\documentstyle[referee]{laa}

\begin{document}
\thesaurus{07  
       (07.09.1; 
	07.13.1;  
       )}

\title{ELECTROMAGNETIC RADIATION AND MOTION OF A PARTICLE}
\author{J.~Kla\v{c}ka}
\institute{Astronomical Institute,
   Faculty of Mathematics, Physics and Informatics, \\
   Comenius University,
   Mlynsk\'{a} dolina, 842~48 Bratislava, Slovak Republic \\
   e-mail: klacka@fmph.uniba.sk \\
\\
I would like to dedicate this paper to the memory of my sister \\
Zuzka Kla\v{c}kov\'{a} (* 5. 7. 1966 \dag 27. 3. 1991)}

\date{}
\maketitle

\begin{abstract}
We consider the motion of uncharged dust grains of arbitrary shape
including the effects of electromagnetic radiation and thermal emission.
The resulting relativistically covariant equation of motion is expressed in terms
of standard optical parameters.
Explicit expressions for secular changes of osculating orbital elements
are derived in detail for the special case of the
Poynting-Robertson effect. Two subcases are considered:
(i) central acceleration due to gravity and the radial component of radiation
pressure independent of the particle velocity, (ii)
central acceleration given by gravity and the radiation force as
the disturbing force. The latter case yields results which may be
compared with secular orbital evolution in terms of orbital elements for
an arbitrarily shaped dust particle. The effects of solar wind are also presented.

\keywords{cosmic dust, electromagnetic radiation, thermal emission,
relativity theory,
equation of motion, orbital motion, celestial mechanics}

\end{abstract}

\section{Introduction}
Poynting (1904) formulated a problem of finding equation of
motion for a perfectly absorbing spherical particle under the action
of electromagnetic radiation. Poynting did not succeed in finding
correct solution.
The second case, closely connected with the relativistic equation
of motion for a free particle under the action of electromagnetic
radiation, was presented by Einstein (1905), who calculated the
change of energy and light pressure at arbitrary angle of
incidence on a plane mirror. Robertson (1937) subsequently derived
a correct equation of motion for a perfectly absorbing spherical
particle. This result has been applied to astronomical problems
for several decades and is known as the Poynting-Robertson (P-R)
effect.
Robertson's case was relativistically generalized by
Kla\v{c}ka (1992a), who showed that the generalized P-R effect
holds only for the special case where the total momentum of the
outgoing radiation per unit time is colinear with the incident
radiation (in the proper/rest frame of reference of the particle),
which may include radiation normally incident on a plane mirror.
Since real particles interact with electromagnetic radiation in a
complicated manner and particles of various optical properties
exist (e. g., Mishchenko {\it et al.} 2002),
it is essential to have an equation of motion sufficiently
general to cover a wide range of optical parameters, not just the
limited cases previously investigated. As an attempt of formulating
such a general equation of motion, we can mention Lyttleton (1976),
Kla\v{c}ka (1994a), Kla\v{c}ka and Kocifaj (1994),
Kocifaj {\it et al.} (2000). The last three
presentations are of hypothetical character only, since none of them proves
correctness of the equation of motion. However,
Kla\v{c}ka (2000a) succeeded in deriving
relativistically covariant equation of motion. This equation of motion was
derived in other possible ways by Kla\v{c}ka (2000b), Kla\v{c}ka and
Kocifaj (2001), within the accuracy corresponding to Kla\v{c}ka
(1994a). Finally, knowing the results obtained by Kla\v{c}ka and Kocifaj (1994)
and Kocifaj {\it et al.} (2000) and having in disposal papers by
Kla\v{c}ka (2000a, 2000b, 2000c), Kimura {\it et al.} (2002) presented a
repetition of the equation of motion corresponding to Kla\v{c}ka
and Kocifaj (1994) and Kocifaj {\it et al.} (2000).

The equations of motion for a moving particle have been constructed under
the assumption that thermal emission from a particle is isotropic and does
not exert a radiation pressure force on the particle in the particle
frame of reference.
Recently, Mishchenko (2001) has formulated the radiation pressure on
arbitrarily shaped particles arising from an anisotropy of thermal emission.
In this paper, we derive the equation of motion for an arbitrarily shaped
particle moving in a radiation field taking into account the radiation pressure
caused by an anisotropy of thermal emission as well as scattering and
absorption of light. Relativistically covariant equation of motion is presented.
Derivation of the equation of motion is physically fully reasoned.

We begin by reviewing in Sec. 2 and 3 the basic physical processes in proper
and stationary frames. The equation of motion for simultaneous action of
gravity and electromagnetic radiation
is presented in Sec. 4. Sec. 5 then applies these results to the
calculation of osculating orbital elements, including the special
case of the P-R effect, which may be treated analytically. The
calculation is carried out in Sec. 6 to first order in $\vec{v}/c$
and applied to the ejection of a
dust particle from a parent body such as a comet or asteroid. Two
cases are considered: (i) the disturbing acceleration is given in
terms of velocity (Robertson 1937, Wyatt and Whipple 1950), and
(ii) the electromagnetic radiation itself is a disturbing
function. However, these authors (including Lyttleton 1976)
did not obtain the correct expression for the secular change of longitude
of pericenter (perihelion). In Sec. 7 we use the equation of
motion to second order in $\vec{v}/c$ for the P-R effect. In
particular we obtain a correct expression for the secular change
of longitude of pericenter (perihelion). Sec. 8 then finds the
secular change in the advancement of perihelion to first order in
$\vec{v}/c$, with gravity as a central acceleration. Next, we
briefly discuss, in Sec. 9, the secular evolution of orbital
elements for the P-R effect and nearly-circular orbits. Sec. 10
then treats the effect of the solar wind on the secular changes in
the orbital elements to the second order in $\vec{v}/u$, where $u$
is the speed of solar wind particles. Finally, Sec. 11 summarizes
our results.

Other theoretical papers on the basic properties of the P-R effect
were written during the last decades: Burns {\it et al.} (1979),
Mediavilla and Buitrago (1989),
Mignard (1992), Srikanth (1999), Williams (2002). Some others
will be mentioned within the context of the discussed problems.
Since some confusion exists in presenting derivations and results in
the most cited papers (Robertson 1937; Wyatt and Whipple 1950;
Burns {\it et al.} 1979; Mignard 1992), we present detailed derivation
of the secular changes of orbital elements for heliocentric orbits
and the P-R effect.

\section{Proper reference frame of the particle -- stationary particle}
The term ``stationary particle'' will denote a particle which does
not move in a given inertial frame of reference.
Primed quantities will denote quantities measured in the
proper reference frame of the particle -- rest frame of the particle.

The flux density of photons scattered into an elementary solid
angle $d \Omega ' = \sin \theta ' ~ d \theta ' ~ d \phi '$ is
proportional to  $p' ( \theta ', \phi ') ~ d \Omega '$, where $p'
( \theta ', \phi ')$ is the ``phase function''. The phase function
depends on orientation of the particle with respect to the
direction of the incident radiation and on the particle
characteristics; angles $\theta '$, $\phi '$ correspond to the
direction (and orientation) of travel of the scattered radiation,
$\theta '$ is the polar angle which vanishes for propagation along
the unit vector $\vec{e}_{1} '$ of the incident radiation. The
phase function fulfills the normalisation condition
\begin{equation}\label{1}
\int_{4 \pi} p' ( \theta ', \phi ')~ d \Omega ' = 1 ~.
\end{equation}

The momentum of the incident beam of photons which is lost in the
process of interaction with the particle is proportional to the
cross-section $C'_{ext}$ (extinction). The part proportional to
$C'_{abs}$ (absorption) is emitted in the form of thermal
radiation and the part proportional to
$C'_{ext} ~-~ C'_{abs} = C'_{sca}$ is scattered.
The differential scattering cross section
$dC'_{sca}/d \Omega '$ $\equiv$ $C'_{sca} ~p'(\theta ', \phi ' )$
depends on the polarization state of the incident light as well as
on the incidence and scattering directions (e. g., Mishchenko {\it
et al.} 2002).

The momentum (per unit time) of the scattered photons into an elementary
solid angle $d \Omega '$ is
\begin{equation}\label{2}
d \vec{p'}_{sca} = \frac{1}{c} ~ S' ~ C'_{sca} ~
	   p' ( \theta ', \phi ')~ \vec{K'} ~ d \Omega ' ~,
\end{equation}
where the unit vector in the direction of scattering is
\begin{equation}\label{3}
\vec{K'} = \cos \theta '~\vec{e}_{1} ' ~+~
	 \sin \theta ' ~ \cos \phi ' ~ \vec{e}_{2} ' ~+~
	 \sin \theta ' ~ \sin \phi ' ~ \vec{e}_{3} ' ~.
\end{equation}
$S'$ is the flux density of radiation energy
(energy flow through unit area perpendicular to the ray per unit time).
The system of unit vectors
used on the RHS of the last equation forms an orthogonal basis.
The total momentum (per unit time) of the scattered photons is
\begin{equation}\label{4}
\vec{p'}_{sca} = \frac{1}{c} ~ S' ~ C'_{sca} ~ \int_{4 \pi} ~
	 p' ( \theta ', \phi ')~ \vec{K'} ~ d \Omega ' ~.
\end{equation}

The momentum (per unit time) obtained by the particle due to the interaction
with radiation -- radiation force acting on the particle -- is
\begin{equation}\label{5}
\frac{d~ \vec{p'}}{d~ t'} = \frac{1}{c} ~ S' ~ \left \{
		  C'_{ext} ~\vec{e}_{1} '
		  ~-~ C'_{sca} ~ \int_{4 \pi} ~
		  p' ( \theta ', \phi ')~ \vec{K'} ~
		  d \Omega ' \right \} ~+~ \vec{F}'_{e} ( T' ) ~,
\end{equation}
where the emission component of the radiation force acting on the particle
of absolute temperature $T'$ is (Mishchenko {\it et al.} 2002, pp. 63-66)
\begin{equation}\label{6}
\vec{F}'_{e} ( T' ) = -~ \frac{1}{c} ~ \int_{0}^{\infty} ~ d \omega' ~
	      \int_{4 \pi} ~ \hat{\vec{r}}' ~
	      K'_{e} \left ( \hat{\vec{r}}', T', \omega ' \right ) ~
	      d \hat{\vec{r}}' ~.
\end{equation}
The unit vector $\hat{\vec{r}}' = \vec{r}' / r'$ is given by position vector
$\vec{r}'$ of the observation point with origin inside the particle
(the emitted radiation in the far-field zone of the particle propagates
in the radial direction, i. e., along the unit vector $\hat{\vec{r}}'$),
$\omega '$ is (angular) frequency of radiation,
\begin{equation}\label{7}
K'_{e} \left ( \hat{\vec{r}}', T', \omega ' \right ) =
       I'_{b} \left ( T', \omega ' \right )  \left [
       K'_{11} \left ( \hat{\vec{r}}', \omega ' \right ) -
       \int_{4 \pi}
       Z'_{11} \left ( \hat{\vec{r}}', \hat{\vec{r}}'', \omega ' \right )
	      d \hat{\vec{r}}'' \right ] ~,
\end{equation}
where $K'_{11}$ is the (1,1) element of the particle extinction matrix,
$Z'_{11}$ is the (1,1) element of the phase matrix and
the Planck blackbody energy distribution is given by the well-known relation
\begin{equation}\label{8}
I'_{b} \left ( T', \omega ' \right ) =
       \frac{\hbar ~\omega'^{3}}{4~ \pi ^{3} ~c^{2}}
       \left \{ \exp \left ( \frac{\hbar ~\omega'}{k~T'} \right )
       ~-~ 1 \right \}^{-1} ~.
\end{equation}
Thermal emission has to be included in the total interaction of the particle
with electromagnetic radiation:
if the particle's absolute temperature is above zero, it can emit as well
as scatter and absorb electromagnetic radiation.

For the sake of brevity, we will use dimensionless efficiency factors $Q'_{x}$
instead of cross sections $C'_{x}$:
$C'_{x} = Q'_{x} ~ A'$, where $A'$ is geometrical
cross section of a sphere of volume equal to the volume of the
particle. Equation (5) can be rewritten to the form
\begin{eqnarray}\label{9}
\frac{d ~\vec{p'}}{d~ \tau} &=& \frac{1}{c} ~ S'~A'~ ~ \left \{ \left [
       Q'_{ext} ~-~ < \cos \theta'> ~ Q'_{sca} \right ] ~
       \vec{e}_{1} ' ~+~ \left [ ~-~ < \sin \theta' ~ \cos \phi ' > ~ Q'_{sca}
       \right ] ~ \vec{e}_{2} ' ~+~
       \right .
\nonumber \\
& &  \left .
       \left [ ~-~ < \sin \theta' ~ \sin \phi ' > ~ Q'_{sca}
       \right ] ~ \vec{e}_{3} ' \right \} ~+~
       \sum_{j=1}^{3} ~F'_{ej} ~ \vec{e}'_{j} ~,
\end{eqnarray}
where $< x' > \equiv  \int_{4 \pi} ~ x' ~p' ( \theta ', \phi ') ~
d \Omega '$ and $F'_{ej} \equiv \vec{F}'_{e} ( T' ) \cdot \vec{e}'_{j}$.
As for the energy, we assume that it is conserved: the energy (per
unit time) of the incoming radiation $E'_{i}$, equals to the
energy (per unit time) of the outgoing radiation (after
interaction with the particle) $E'_{o}$. We will use the fact that
time $t' = \tau$, where $\tau$ is proper time.

Summarizing important equations, we can write them in a short form
\begin{equation}\label{10}
\frac{d~ \vec{p}'}{d~ \tau} = \sum_{j=1}^{3} \left ( \frac{S'~A'}{c} ~
	 Q'_{j} ~+~ F'_{ej}  \right ) ~\vec{e}_{j} ' ~;~~
\frac{d ~E'}{d~ \tau} = 0 ~,
\end{equation}
where $Q'_{1} \equiv Q'_{ext} ~-~ < \cos \theta'> ~ Q'_{sca}$,
$Q'_{2} \equiv  -~ < \sin \theta' ~ \cos \phi ' > ~ Q'_{sca}$,
$Q'_{3} \equiv -~ < \sin \theta' ~ \sin \phi ' > ~ Q'_{sca}$.
We have added an assumption of equilibrium state when the particle's
mass does not change.

\subsection{Summary of the important equations}
Using the text concerning energy below Eq. (9) and the last Eq. (10), we
may describe the total process of interaction in the form of the
following equations (energies and momenta per unit time):
\begin{eqnarray}\label{11}
E_{o} ' &=& E_{i} ' = A'~S' ~,
\nonumber \\
\vec{p}_{o} ' &=& ( 1 ~-~ Q_{1} ' ) ~ \vec{p}_{i} ' ~-~
	  (  Q_{2} ' ~ \vec{e}_{2} ' ~+~
	     Q_{3} ' ~ \vec{e}_{3} ' ) ~ E_{o} ' / c ~-~
       \sum_{j=1}^{3} ~F'_{ej} ~ \vec{e}'_{j} ~,
\nonumber \\
\vec{p}_{i} ' &=& ( E_{i} ' / c ) ~ \vec{e}_{1} ' ~,
\end{eqnarray}
The index $"i"$ represents the incoming radiation, beam of photons, the index
$"o"$ represents the outgoing radiation. The relation for
$\vec{p}_{o} '$ represents a generalization of the following cases: \\
i) $Q_{1} ' =$ 1,  $Q_{2} ' = Q_{3} ' = F'_{e1} = F'_{e2} = F'_{e3} =$ 0 --
Robertson (1937), Robertson and Noonan (1968), Srikanth (1999); \\
ii) $Q_{1} '$ arbitrary, $Q_{2} ' = Q_{3} ' = F'_{e1} = F'_{e2} = F'_{e3} =$ 0
-- Kla\v{c}ka (1992a); \\
iii) $Q_{1} '$, $Q_{2} '$, $Q_{3} '$ arbitrary,
$F'_{e1} = F'_{e2} = F'_{e3} =$ 0 -- Kla\v{c}ka (2000a).

The changes of energy and momentum of the particle due to the interaction
with electromagnetic radiation are
\begin{eqnarray}\label{12}
\frac{d ~E'}{d~ \tau} &=& E_{i} ' ~-~ E_{o} ' = 0 ~,
\nonumber \\
\frac{d ~\vec{p'}}{d~ \tau} &=& \vec{p}_{i} ' ~-~ \vec{p}_{o} ' ~.
\end{eqnarray}

\section{Stationary frame of reference}
By the term ``stationary frame of reference''
(laboratory frame) we mean a frame of reference
in which particle moves with a velocity vector $\vec{v} = \vec{v} (t)$.
The physical quantities measured in the stationary frame of reference
will be denoted by unprimed symbols.

Our aim is to derive equation of motion for the particle in the
stationary frame of reference. We will use the fact that we know
this equation in the proper frame of reference -- see Eqs. (11) and
(12).

If we have a four-vector $A^{\mu} = ( A^{0}, \vec{A} )$, where
$A^{0}$ is its time component and $\vec{A}$ is its spatial component,
generalized special Lorentz transformation yields
\begin{eqnarray}\label{13}
A^{0 '}  &=& \gamma ~ ( A^{0} ~-~ \vec{v} \cdot \vec{A} / c ) ~,
\nonumber \\
\vec{A} ' &=& \vec{A} ~+~ [ ( \gamma ~-~ 1 ) ~ \vec{v} \cdot \vec{A}  /
	  \vec{v} ^{2} ~-~ \gamma ~ A^{0} / c ] ~ \vec{v}  ~,
\end{eqnarray}
with inverse
\begin{eqnarray}\label{14}
A^{0}  &=& \gamma ~ ( A^{0 '} ~+~ \vec{v} \cdot \vec{A} ' / c ) ~,
\nonumber \\
\vec{A}  &=& \vec{A} ' ~+~ [ ( \gamma ~-~ 1 ) ~ \vec{v} \cdot \vec{A} ' /
	  \vec{v} ^{2} ~+~ \gamma ~ A^{0 '} / c ] ~ \vec{v}  ~,
\end{eqnarray}
where
$\gamma = 1 / \sqrt{1~-~\vec{v} ^{2} / c ^{2} }$ ~.

As for four-vectors we immediately introduce four-momentum:
\begin{equation}\label{15}
p^{\mu} = ( p^{0}, \vec{p} ) \equiv ( E / c, \vec{p} ) ~.
\end{equation}

\subsection{Incoming radiation}
Applying Eqs. (14) and (15) to quantity $( E_{i} ' / c, \vec{p}_{i} ')$
(four-momentum per unit time -- proper time is a scalar quantity) and
taking into account also Eqs. (11), we can write
\begin{eqnarray}\label{16}
E_{i}  &=& E_{i} ' ~ \gamma ~ ( 1 ~+~ \vec{v} \cdot \vec{e}_{1} ' / c ) ~,
\nonumber \\
\vec{p}_{i}  &=& \frac{E_{i} '}{c}  ~ \left \{ \vec{e}_{1} ' ~+~
	 \left [ \left ( \gamma ~-~ 1 \right ) ~
	 \vec{v} \cdot \vec{e}_{1} '  /
	 \vec{v} ^{2} ~+~ \gamma / c \right ] ~ \vec{v} \right \} ~.
\end{eqnarray}

Using the fact that $p^{\mu} = ( h ~\nu , h ~\nu ~ \vec{e}_{1} )$
for photons, we have
\begin{eqnarray}\label{17}
\nu ' &=& \nu  ~w_{1} ~,
\nonumber \\
\vec{e}_{1} ' &=& \frac{1}{w_{1}}  ~ \left \{ \vec{e}_{1} ~+~
	 \left [ \left ( \gamma ~-~ 1 \right ) ~
	 \vec{v} \cdot \vec{e}_{1}  /
	 \vec{v} ^{2} ~-~ \gamma / c \right ] ~ \vec{v} \right \} ~,
\end{eqnarray}
where
\begin{equation}\label{18}
w_{1} \equiv \gamma ~ ( 1 ~-~ \vec{v} \cdot \vec{e}_{1} / c ) ~.
\end{equation}

Inserting the second of Eqs. (17) into Eqs. (18), one obtains
\begin{eqnarray}\label{19}
E_{i} &=& ( 1 / w_{1} ) ~ E_{i} ' ~,
\nonumber \\
\vec{p}_{i} &=& ( 1 / w_{1} ) ~ ( E_{i} ' / c ) ~ \vec{e}_{1} ~.
\end{eqnarray}
We have four-vector
$p_{i}^{\mu} = ( E_{i} / c, \vec{p}_{i} ) = ( 1, \vec{e}_{1} ) E_{i} / c$
$= ( 1 / w_{1}, \vec{e}_{1} / w_{1} ) ~ w_{1} ~ E_{i} / c$
$\equiv$ $b_{1}^{\mu} ~ w_{1} ~ E_{i} / c$.

For monochromatic radiation the flux density of radiation energy
becomes
\begin{equation}\label{20}
S' = n' ~h~ \nu ' ~ c ~; ~~~ S = n ~h~ \nu  ~ c ~,
\end{equation}
where $n$ and $n'$ are concentrations of photons (photon number
densities) in the corresponding reference frames. We also have
continuity equation
\begin{equation}\label{21}
\partial _{\mu} ~ j^{\mu} = 0 ~, ~~~j^{\mu} = ( c~n, c~n~ \vec{e}_{1} ) ~,
\end{equation}
with current density $j^{\mu}$. Application of Eq. (13) then
yields
\begin{equation}\label{22}
n' = w_{1} ~ n ~.
\end{equation}
Using Eqs. (17), (20) and (22) we finally obtain
\begin{equation}\label{23}
S' = w_{1}^{2} ~ S ~.
\end{equation}
Eqs. (11), (19) and (23) then together give $E_{i} = w_{1} ~ S
~A'$, $\vec{p}_{i} = w_{1} ~ S ~A'~\vec{e}_{1} / c$.

\subsection{Outgoing radiation}
The situation is analogous to that of the preceding subsection. It
is only a little more algebraically complicated, since radiation
may also spread out in directions given by unit vectors
$\vec{e}_{2}$, $\vec{e}_{3}$. How can we find transformations for
the unit vectors $\vec{e}_{2} '$ and $\vec{e}_{3} '$? The crucial
point is what physics do these unit vectors describe. The vectors
$\vec{e}_{2} '$ and $\vec{e}_{3} '$ can be used to describe directions
of propagation of radiation scattered by the particle.
Thus,  aberration of light also exists for each of these unit
vectors. The relations between $\vec{e}_{2} '$ and $\vec{e}_{2}$,
$\vec{e}_{3} '$ and $\vec{e}_{3}$, are analogous to that presented
by the second of Eq. (17):
\begin{equation}\label{24}
\vec{e}_{j} ' = \frac{1}{w_{j}}  ~ \left \{ \vec{e}_{j} ~+~
	 \left [ \left ( \gamma ~-~ 1 \right ) ~
	 \vec{v} \cdot \vec{e}_{j}  /
	 \vec{v} ^{2} ~-~ \gamma / c \right ] ~ \vec{v} \right \} ~,
	 ~~ j = 1, 2, 3 ~,
\end{equation}
where
\begin{equation}\label{25}
w_{j} \equiv \gamma ~ ( 1 ~-~ \vec{v} \cdot \vec{e}_{j} / c )
	  ~, ~~ j = 1, 2, 3 ~.
\end{equation}
It is worth mentioning that vectors
$\left \{ \vec{e}_{j} ' ; j = 1, 2, 3 \right \}$ form an orthonormal
set of vectors, and, unit vectors
$\left \{ \vec{e}_{j} ; j = 1, 2, 3 \right \}$ are not orthogonal unit
vectors.

Applying Eqs. (14) and (15) to the quantity $( E_{o} ' / c,
\vec{p}_{o} ')$ (four-momentum per unit time -- proper time is a
scalar quantity), we can write
\begin{eqnarray}\label{26}
E_{o}  &=&  \gamma ~ ( E_{o} ' ~+~ \vec{v} \cdot \vec{p}_{o} ' ) ~,
\nonumber \\
\vec{p}_{o}  &=& \vec{p}_{o} ' ~+~
	 \left [ \left ( \gamma ~-~ 1 \right ) ~
	 \vec{v} \cdot \vec{p}_{o} ' /
	 \vec{v} ^{2} ~+~ \gamma ~ \frac{E_{o} '}{c^{2}}
	~\right ] ~ \vec{v}  ~.
\end{eqnarray}
Using also $\vec{p}_{i} '= E_{i} ' ~\vec{e}_{1} '/ c$ and Eqs. (11), (24),
(26),
\begin{eqnarray}\label{27}
E_{o}  &=& Q_{1} ' ~ w_{1} ~ E_{i} ~ \gamma  ~+~ ( 1 ~-~ Q_{1} ' ) ~ E_{i} ~+~
\nonumber \\
& &    w_{1} ~ E_{i} ~ ( Q_{2} ' ~+~ Q_{3} ' ) ~ \gamma  ~-~
       w_{1} ~ E_{i} ~ ( Q_{2} ' / w_{2} ~+~ Q_{3} ' / w_{3} )  ~-~
\nonumber \\
& & \sum_{j=1}^{3} F'_{ej} ~ \left ( c / w_{j} - \gamma ~c \right ) ~,
\nonumber \\
\vec{p}_{o}  &=& ( 1 ~-~ Q_{1} ' ) ~ \frac{E_{i}}{c} ~\vec{e}_{1}  ~+~
	 Q_{1} ' ~ \frac{w_{1} ~E_{i}}{c^{2}} ~\gamma ~ \vec{v} ~-~
	 \sum_{j=2}^{3} ~ Q_{j} ' ~\frac{w_{1} ~E_{i}}{c^{2}} ~ \left (
	 c ~ \vec{e}_{j} / w_{j}
	 ~-~ \gamma ~ \vec{v}  \right )  ~-~
\nonumber \\
& &	\frac{1}{c} \sum_{j=1}^{3} ~F'_{ej} ~ \left ( c ~\vec{e}_{j} / w_{j} -
	\gamma	~ \vec{v} \right ) ~.
\end{eqnarray}

\subsection{Equation of motion}
In analogy with Eqs. (12), we have for the changes of energy and momentum of the
particle due to the interaction with electromagnetic radiation
\begin{eqnarray}\label{28}
\frac{d ~E}{d~ \tau} &=& E_{i}	~-~ E_{o} ~,
\nonumber \\
\frac{d ~\vec{p}}{d~ \tau} &=& \vec{p}_{i}  ~-~ \vec{p}_{o}  ~.
\end{eqnarray}
Putting Eqs. (27) into Eqs. (28), using also
$\vec{p}_{i} = ( E_{i} / c ) ~\vec{e}_{1}$,
one easily obtains
\begin{eqnarray}\label{29}
\frac{d ~E / c}{d~ \tau} &=&  \sum_{j=1}^{3} \left ( Q_{j} ' ~
		  \frac{w_{1} ~ E_{i}}{c^{2}} ~+~ \frac{1}{c} ~F'_{ej} \right )
	      \left ( c ~ \frac{1}{w_{j}}  ~-~ \gamma ~c \right ) ~,
\nonumber \\
\frac{d ~\vec{p}}{d~ \tau} &=& \sum_{j=1}^{3} \left ( Q_{j} ' ~
		  \frac{w_{1} ~ E_{i}}{c^{2}} ~+~ \frac{1}{c} ~F'_{ej} \right )
		   ~\left ( c ~\frac{\vec{e}_{j}}{w_{j}}
		  ~-~ \gamma \vec{v}  \right )	~.
\end{eqnarray}

Eq. (29) may be rewritten in terms of four-vectors:
\begin{eqnarray}\label{30}
\frac{d ~p^{\mu}}{d~ \tau} &=& \sum_{j=1}^{3} \left ( Q_{j} ' ~
	      \frac{w_{1} ~E_{i}}{c^{2}} ~+~ \frac{1}{c} ~F'_{ej} \right )
	      \left ( c ~ b_{j}^{\mu} ~-~ u^{\mu}  \right ) ~,
\end{eqnarray}
where $p^{\mu}$ is four-vector of the particle of mass $m$
\begin{equation}\label{31}
p^{\mu} = m~ u^{\mu} ~,
\end{equation}
four-vector of the world-velocity of the particle is
\begin{equation}\label{32}
u^{\mu} = ( \gamma ~c, \gamma ~ \vec{v} ) ~.
\end{equation}
We have also other four-vectors
\begin{equation}\label{33}
b_{j}^{\mu} = ( 1 / w_{j} , \vec{e}_{j} / w_{j} ) ~, ~~ j= 1, 2, 3 ~.
\end{equation}

It can be easily verified that:\\
i) the quantity $w~E_{i}$ is a scalar quantity
-- see first of Eqs. (19); \\
ii) Eq. (30) reduces to Eq. (10) for the case
of proper inertial frame of reference of the particle; \\
iii) Eq. (30) yields $d ~m / d~ \tau =$ 0.

We introduce
\begin{eqnarray}\label{34}
b_{j}^{0} &\equiv& 1 / w_{j} = \gamma ~( 1 ~+~ \vec{v} \cdot \vec{e}'_{j} ~/~c ) ~,
\nonumber \\
\vec{b_{j}} &\equiv& \vec{e_{j}} ~/~ w_{j} = \vec{e}'_{j} ~+~ [ ( \gamma ~-~ 1 )~
	   \vec{v} \cdot \vec{e}'_{j} ~/~ \vec{v}^{2} ~+~
	   \gamma ~/~c ] ~ \vec{v} ~, ~~j = 1,2,3
\end{eqnarray}
for the purpose of practical calculations. Physics of these relations
corresponds to aberration of light.

We have derived an equation of motion for real dust particle under
the action of electromagnetic radiation (including thermal emission).
It is supposed that the
equation of motion is represented by Eqs. (11) and (12) in the
proper frame of reference of the particle. The final covariant
form is represented by Eq. (30), or using $E_{i} = w_{1} ~ S ~A'$
(see Eqs. (11), (19) and (23)),
\begin{eqnarray}\label{35}
\frac{d ~p^{\mu}}{d~ \tau} &=& \sum_{j=1}^{3} \left (
		  \frac{w_{1}^{2} ~S ~A'}{c^{2}} ~Q_{j} ' ~+~
		  \frac{1}{c} ~F'_{ej} \right )
		  \left ( c ~ b_{j}^{\mu} ~-~ u^{\mu}  \right )  ~.
\end{eqnarray}

Another form of equation of motion is presented in Appendix A. There is also
explained why considerations presented by Kimura {\it et al.} (2002) are
incorrect.

To first order in $\vec{v} / c$, Eqs. (34)-(35) yield
\begin{eqnarray}\label{36}
\frac{d~ \vec{v}}{d ~t} &=&  \frac{S ~A'}{m~c} ~ \sum_{j=1}^{3} ~Q_{j} '
	      ~\left [	\left ( 1~-~ 2~
	  \vec{v} \cdot \vec{e}_{1} / c ~+~
	  \vec{v} \cdot \vec{e}_{j} / c \right ) ~ \vec{e}_{j}
	  ~-~ \vec{v} / c \right ]  ~+~
\nonumber   \\
& &	  \frac{1}{m} ~ \sum_{j=1}^{3} ~F'_{ej} ~\left [  \left ( 1~+~
	  \frac{\vec{v} \cdot \vec{e}_{j}}{c} \right ) ~ \vec{e}_{j}
	  ~-~ \frac{\vec{v}}{c} \right ]  ~,
\nonumber \\
\vec{e}_{j} &=& ( 1 ~-~ \vec{v} \cdot \vec{e}'_{j} / c ) ~ \vec{e}'_{j} ~+~
	  \vec{v} / c ~~~, ~~j = 1, 2, 3~.
\end{eqnarray}
It is worth mentioning to stress that the values of
$Q'-$coefficients depend on particle's orientation with respect to
the incident radiation -- their values are time dependent. A little
different derivation of the equation of motion within the accuracy to the
first order in $\vec{v} / c$ is presented in Kla\v{c}ka and Kocifaj (2001).
Inspite of knowing the correct theoretical results, Kimura {\it et al.} (2002)
have succeeded in publishing a paper where different set of unit vectors is used
(in reality, it is repetition of the older results of Kla\v{c}ka and Kocifaj
1994, Kocifaj {\it et al.} 2000). The relation between our set of unit vectors
$\{ \vec{e}_{j} ; j = 1, 2, 3 \}$ and the set of unit vectors
$\{ \vec{k}_{j} ; j = 1, 2, 3 \}$ used by Kimura {\it et al.} (2002) is
following: $\vec{e}_{1} = \vec{k}_{1}$,
$\vec{e}_{2} = \vec{k}_{2} + ( \vec{v} \cdot \vec{k}_{2} / c ) ~
( \vec{k}_{1} - \vec{k}_{2} ) + \vec{v} / c$,
$\vec{e}_{3} = \vec{k}_{3} - ( \vec{v} \cdot \vec{k}_{3} / c ) ~
\vec{k}_{3} + \vec{v} / c$
and definition $\vec{v} \cdot \vec{k}_{3} =$ 0 is used -- this transformation
yields Eqs. (1) -- (3) in Kimura {\it et al.} (2002):
$\vec{e}'_{1} = ( 1 + \vec{v} \cdot \vec{k}_{1} / c ) \vec{k}_{1} - \vec{v} / c$,
$\vec{e} '_{2} = \vec{k}_{2} + ( \vec{v} \cdot \vec{k}_{2} / c ) \vec{k}_{1}$,
$\vec{e} '_{3} = \vec{k}_{3}$.
As for physics of the paper Kimura {\it et al.} (2002), we refer
to Appendix C.

It can be verified that Eq. (35) (or Eqs. (36)
within the accuracy to the first order in $\vec{v} / c$) yields as special
cases the situations discussed in Einstein (1905) and Robertson (1937) --
Robertson's case is obtained simply by substituting $Q'_{1} =$ 1,
$Q'_{2}$ $=$ $Q'_{3}$ $=$ 0, $F'_{e1} = F'_{e2} = F'_{e3} = 0$,
Einstein's results require a little more calculations (see Appendix B).

\subsection{Heuristic derivation}
Since we completely understand the physics of Eq. (30), we are
able to present a short simple derivation of Eq. (36). We have: \\
i) $d \vec{v}' / dt = \sum_{i=1}^{3} \left \{ [ S' A' / ( m c ) ] ~
      Q_{i} '~+~ F'_{ej} / m \right \} ~ \vec{e}_{i} '$ (see Eq. (10));
$d m / d t =$ 0 is supposed; \\
ii) $S' = S ( 1 - 2 \vec{v} \cdot \vec{e}_{1} / c )$ (see Eq. (23)),
due to the change of concentration of photons
$n' = n ( 1 - \vec{v} \cdot \vec{e}_{1} / c )$ -- Eq. (22) --
and Doppler effect
$\nu' = \nu ( 1 - \vec{v} \cdot \vec{e}_{1} / c )$ -- Eq. (17); \\
iii) $\vec{e}_{j} ' = ( 1 + \vec{v} \cdot \vec{e}_{j} / c ) \vec{e}_{j} -
\vec{v} / c$, $j =$ 1, 2, 3 (aberration of light -- see Eq. (24)). \\
Taking into account these physical phenomena, we finally obtain Eq. (36).
However, only relativistic covariant formulation proves that
$\{ Q_{j} ' , F'_{ej} ; j = 1, 2, 3 \}$ behave
as relativistically invariant quantities.

A reader may compare this heuristic derivation of the general case with
heuristic derivation of the special case
($Q'_{1} = 1$, $Q'_{2} = Q'_{3} = 0$, $\vec{F}'_{e} = 0$)
presented by Burns {\it et al.} (1979, pp. 5-6).

Another instructive text enabling better understanding of physics of
the P-R effect can be found in (http://xxx.lanl.gov,astro-ph/0108210),
which deals with explanation presented in Harwit (1988; pp. 176-177).

A little different derivation of the relation between $S'$ and
$S$, based on: a) stress-energy tensor (energy-momentum tensor),
b) transformations of electric and magnetic fields, may be found
in Kla\v{c}ka (1992a: sections 2.3 and 2.5).

\subsection{Continuous distribution of density flux of energy}
For a continuous frequency distribution of density flux of energy, we can write
\begin{eqnarray}\label{37}
\frac{d ~\vec{p'}}{d~ \tau} &=& \sum_{j=1}^{3} \left \{ \frac{A'}{c} ~
     \int_{0}^{\infty} ~c~h ~\nu ' ~
     \frac{\partial n'}{\partial \nu '} ~ Q_{j} '( \nu ') ~ d \nu ' ~+~
     F'_{ej}  \right \} ~\vec{e}_{j} ' \equiv
\nonumber \\
&\equiv& \sum_{j=1}^{3} \left ( \frac{S'~A'}{c} ~ \bar{Q}'_{j} ~+~
     F'_{ej}  \right ) ~\vec{e'}_{j} ~.
\end{eqnarray}
Taking into account that concentration of photons fulfills
$n' = w_{1} ~n$ (Eq. (22)) and that Doppler effect yields
$\nu ' = w_{1}~ \nu$ (Eq. (17)), we have $\partial n' / \partial \nu '$ $=$
$\partial n / \partial \nu$. Lorentz transformation finally yields
\begin{eqnarray}\label{38}
\frac{d ~p^{\mu}}{d~ \tau} &=& \sum_{j=1}^{3} \left \{
     \frac{w_{1}^{2}~A'}{c^{2}} ~ \int_{0}^{\infty} ~
     c~h ~ \frac{\partial n}{\partial \nu} ~\nu~ Q_{j} '( w_{1} ~\nu ) ~
     d \nu ~+~ \frac{1}{c} ~F'_{ej} \right \}
     \left ( c ~ b_{j}^{\mu} ~-~ u^{\mu}  \right )
\nonumber \\
&\equiv& \sum_{j=1}^{3} \left ( \frac{w_{1}^{2}~S~A'}{c^{2}} ~ \bar{Q}'_{j} ~
     ~+~ \frac{1}{c} ~F'_{ej} \right )
     \left ( c ~ b_{j}^{\mu} ~-~ u^{\mu}  \right ) ~.
\end{eqnarray}
As a consequence, $d m / d \tau =$ 0 (this corresponds to the condition
$d E' / d \tau =$ 0).

If the particle is not irradiated, then one has to use
\begin{equation}\label{39}
\left ( \frac{d ~p^{\mu}}{d~ \tau} \right ) _{e} =
	     \sum_{j=1}^{3} ~F_{ej} ' ~b_{j}^{\mu} ~,
\end{equation}
instead of Eq. (38).
As a consequence, Eq. (39) yields
$( d m / d \tau ) _{e} =
	     \sum_{j=1}^{3} ~F_{ej} ' ~/~c$.
Mass of the particle decreases due to the thermal emission, alone.

\section{Gravitation and electromagnetic radiation}
The generally covariant equation of motion can be immediately written on the
basis of Eq. (38) (see e. g., Landau and Lifshitz 1975):
\begin{equation}\label{40}
\frac{D ~p^{\mu}}{d~ \tau} = \sum_{j=1}^{3} \left ( \frac{w_{1}^{2} ~S~A'}{c} ~
		 \bar{Q}'_{j} ~+~ F_{ej} '  \right ) ~
		 \left ( b_{j}^{\mu} ~-~ u^{\mu} ~/~c  \right ) ~.
\end{equation}
where the operator $D ~/~d \tau$ is the ``total'' covariant derivative
in the general relativistic sense and includes gravitational effects.

\subsection{Gravity and radiation -- equation of motion to the first order in
$v/c$}
We can write, on the basis of Eqs. (40), (25) and (33)
\begin{eqnarray}\label{41}
\frac{d~ \vec{v}}{d ~t} &=& -~ \frac{G~M_{\odot}}{r^{2}} ~ \vec{e}_{1} ~+~
\nonumber   \\
& &	  \frac{G~M_{\odot}}{r^{2}} ~
	  \sum_{j=1}^{3} ~\beta_{j} ~\left [  \left ( 1~-~ 2~
	  \frac{\vec{v} \cdot \vec{e}_{1}}{c} ~+~
	  \frac{\vec{v} \cdot \vec{e}_{j}}{c} \right ) ~ \vec{e}_{j}
	  ~-~ \frac{\vec{v}}{c} \right ] ~+~
\nonumber   \\
& &	  \frac{1}{m} ~ \sum_{j=1}^{3} ~F'_{ej} ~\left [  \left ( 1~+~
	  \frac{\vec{v} \cdot \vec{e}_{j}}{c} \right ) ~ \vec{e}_{j}
	  ~-~ \frac{\vec{v}}{c} \right ]  ~,
\end{eqnarray}
where
\begin{eqnarray}\label{42}
\vec{e}_{j} &=& ( 1 ~-~ \vec{v} \cdot \vec{e}'_{j} / c ) ~ \vec{e}'_{j}  ~+~
	  \vec{v} / c ~, ~~j = 1, 2, 3 ~,
\nonumber   \\
\beta_{1} &=& \frac{\pi ~R_{\odot}^{2}}{G~M_{\odot}~m~c}
	  \int_{0}^{\infty} B_{\odot} ( \lambda ) \left \{
	  C'_{ext} ( \lambda / w ) - C'_{sca} ( \lambda / w ) ~
	  g'_{1} ( \lambda / w ) \right \}  d \lambda ~,
\nonumber   \\
\beta_{2} &=& \frac{\pi ~R_{\odot}^{2}}{G~M_{\odot}~m~c}
	  \int_{0}^{\infty} B_{\odot} ( \lambda ) \left \{
	  ~-~ C'_{sca} ( \lambda / w ) ~
	  g'_{2} ( \lambda / w ) \right \}  d \lambda ~,
\nonumber   \\
\beta_{3} &=& \frac{\pi ~R_{\odot}^{2}}{G~M_{\odot}~m~c}
	  \int_{0}^{\infty} B_{\odot} ( \lambda ) \left \{
	  ~-~ C'_{sca} ( \lambda / w ) ~
	  g'_{3} ( \lambda / w ) \right \}  d \lambda ~,
\nonumber   \\
w &=& 1~-~\vec{v} \cdot \vec{e}_{1} ~/~ c ~,
\end{eqnarray}
if we use Sun as a source of gravitation and radiation.
$R_{\odot}$ denotes the radius of the Sun and $B_{\odot} ( \lambda )$ is
the solar radiance at a wavelength of $\lambda$; $G$, $M_{\odot}$, and
$r$ are the gravitational constant, the mass of the Sun, and the distance
of the particle from the center of the Sun, respectively.
The asymmetry parameter vector $\vec{g}'$ is defined by $\vec{g}'$ $=$
$( 1 / C'_{sca} ) \int \vec{n}' ( d C'_{sca} / d \Omega ') d \Omega '$, where
$\vec{n}'$ is a unit vector in the direction of scattering;
$\vec{g}'$ $=$ $g'_{1} ~ \vec{e}'_{1}$ $+$ $g'_{2} ~ \vec{e}'_{2}$ $+$
$g'_{3} ~ \vec{e}'_{3}$, $\vec{e}'_{j}$ $=$
$( 1 ~+~ \vec{v} \cdot \vec{e}_{j} / c ) ~ \vec{e}_{j}$  $-$
$\vec{v} / c$, $\vec{e}'_{i} \cdot \vec{e}'_{j} = \delta _{ij}$.
We may mention that $\beta_{1} \equiv \beta$,
$\beta_{2} \equiv \beta ~ \bar{Q}'_{2} / \bar{Q}'_{1}$,
$\beta_{3} \equiv \beta ~ \bar{Q}'_{3} / \bar{Q}'_{1}$ correspond to
quantities used in Kla\v{c}ka and Kocifaj (2001).

\section{Motion and orbital elements}
Equation of motion is given by Eq. (40), or, by Eq. (41) to the first order
in $\vec{v}/c$.
We have to take into account that the non-dimensional parameter, for Solar System
\begin{equation}\label{43}
\beta_{1} \equiv \beta = \frac{r^{2} ~S ~A'}{G ~M_{\odot} ~m ~c} ~\bar{Q}'_{1}
      \equiv \frac{L_{\odot} ~A'}{4~ \pi ~G ~M_{\odot} ~m ~c} ~\bar{Q}'_{1}
	    = \frac{0.02868}{12 ~\pi} ~\bar{Q}'_{1} ~
	    \frac{A' \left [ m^{2} \right ]}{m \left [ kg \right ]}
\end{equation}
may change during the motion. Here $L_{\odot}$ is the rate of energy outflow
from the Sun, the solar luminosity. $\beta$ may change and at each point of
the orbit all three parameters $\beta_{1}$, $\beta_{2}$ and $\beta_{3}$ have to
be numerically calculated (except for a very special cases, when
the $\beta$ parameters are constant during the motion).

If we are interested in orbital evolution in terms of osculating orbital
elements, we come to the crucial point: Which type of osculating
orbital elements is correct? Mainly, if the ``radial radiation pressure''
(the dominant term containing parameter $\beta \equiv \beta_{1}$ in Eq. (75))
has to be considered together with the central gravitational force, or not
(compare Kla\v{c}ka 1992b).

As for a definition of osculating orbital elements, we refer to any textbook
on celestial mechanics.
Brouwer and Clemence (1961) write (p. 273): ``As the motion progresses
under the influence of the various attracting bodies, the coordinates and
velocity components at any instant may be used to obtain a set of six
orbital elements. These are precisely the elements of the ellipse that the
body would follow if from that particular instant on, the accelerations
caused by all ``perturbing'' bodies ceased to exist.'' As for our purposes,
it is sufficient to make a small change: ``attracting bodies'' are replaced
by ``forces''.

In reality, our perturbing physical force corresponds to the total electromagnetic
radiation force. Thus, physics incites us to use gravitational force alone,
as the central acceleration. However, it may occur that the dominant term
containing parameter $\beta_{1} \equiv \beta$ in Eq. (41) is comparable to
central gravity term. As a consequence, orbital elements will change
very rapidly during the motion. This would suggest to divide physical
force into two parts and to use ``radial radiation pressure''
as a part of central acceleration, i. e. together with central gravitational
acceleration. The final problem concerns the changes of $\beta$ (almost random
changes) which prevent us to use the term ``Keplerian orbit'' for an unperturbed
problem.

Thus, we have to use gravitational acceleration of the central body (Sun) as an
acceleration defining Keplerian orbits -- complete electromagnetic
radiative acceleration is a disturbing acceleration -- in Eq. (41).
As for long-term orbital evolution, we may use mean values of orbital
elements -- they may be defined as time mean values of orbital elements
when true anomaly changes in 2 $\pi$ radians.

Let us consider that dust particle is ejected from a parent body. Orbital
elements of the parent body, at the moment of ejection, are:
$a_{P}$, $e_{P}$, $i_{P}$, $\Omega_{P}$, $\omega_{P}$ and $\Theta_{P}$,
i. e., semi-major axis, eccentricity, inclination, longitude of
the ascending node, longitude of pericenter -- longitude of perihelion
for the case of Solar System -- and position angle of the body measured
from the ascending node in the direction of the motion of the body.
The velocity vector of the parent body, in
a given coordinate system with the origin in the dominant central body
(Sun in the case of the Solar System) is:
\begin{eqnarray}\label{44}
\vec{v}_{P} &=& v_{R} ~\vec{e}_{PR} ~+~  v_{T} ~\vec{e}_{PT} ~,
\nonumber   \\
v_{R} &=& \sqrt{\frac{G ~M_{\odot}}{a_{P} ~
      \left ( 1 ~-~ e_{P}^{2} \right )}} ~
      e_{P} ~ \sin f_{P} ~,
\nonumber   \\
v_{T} &=& \sqrt{\frac{G ~M_{\odot}}{a_{P} ~
      \left ( 1 ~-~ e_{P}^{2} \right )}} ~
      \left ( 1 ~+~e_{P} ~ \cos f_{P} \right ) ~,
\nonumber   \\
\vec{r}_{P} &=& \frac{a_{P} ~ \left ( 1 ~-~ e_{P}^{2} \right )}{1 ~+~
      e_{P} ~ \cos f_{P}} ~\vec{e}_{PR}
\nonumber   \\
\vec{e}_{PR} &=& ( \cos \Omega_{P} ~ \cos \Theta_{P} ~-~
	  \sin \Omega_{P} ~ \sin \Theta_{P} ~ \cos i_{P} ~,
\nonumber   \\
& &	\sin \Omega_{P} ~ \cos \Theta_{P} ~+~
	\cos \Omega_{P} ~ \sin \Theta_{P} ~ \cos i_{P} ~,~
	\sin \Theta_{P} ~ \sin i_{P} ) ~,
\nonumber   \\
\vec{e}_{PT} &=& ( -~ \cos \Omega_{P} ~ \sin \Theta_{P} ~-~
	      \sin \Omega_{P} ~ \cos \Theta_{P} ~ \cos i_{P} ~,
\nonumber   \\
& &	  -~ \sin \Omega_{P} ~ \sin \Theta_{P} ~+~
	      \cos \Omega_{P} ~ \cos \Theta_{P} ~ \cos i_{P} ~,~
	      \cos \Theta_{P} ~ \sin i_{P} ) ~,
\nonumber   \\
f_{P} &=& \Theta_{P} ~-~ \omega _{P} ~.
\end{eqnarray}
Radial and transversal velocity components are $v_{R}$ and $v_{T}$,
unit vector $\vec{e}_{PT}$ is
normal to the radial vector and oriented in orientation of motion. True anomaly
$f_{P}$ equals 0 for pericenter/perihelion and $\pi$ for apocenter/aphelion.

The initial conditions for the particle ejected with velocity $\vec{\Delta}$
from the parent body are:
\begin{eqnarray}\label{45}
\vec{r}_{in} &=& \vec{r}_{P} ~,
\nonumber \\
\vec{v}_{in} &=& \vec{v}_{P} ~+~ \vec{\Delta} ~.
\end{eqnarray}

Eqs. (44)-(45) immediately yield that initial orbital elements of the
particle are equal to those of the parent body for $\vec{\Delta} =$ 0:
\begin{eqnarray}\label{46}
a_{in} &=& a_{P} ~, ~~ e_{in} = e_{P} ~, ~~ i_{in} = i_{P} ~,
\nonumber \\
\Omega_{in} &=& \Omega_{P} ~, ~~ \omega_{in} = \omega_{P} ~, ~~
\Theta_{in} = \Theta_{P} ~.
\end{eqnarray}
(Initial orbital elements for general case represented by Eq. (45)
can be calculated from the relations presented in Eq. (47) below.)

Complete equation of motion for dust particle orbiting the central body of
mass $M_{\odot}$ under central's body electromagnetic radiation and
gravity interaction is given by Eq. (41).
We can find osculating orbital elements for dust particle according to
the following equations:
\begin{eqnarray}\label{47}
E &=&  \frac{1}{2} ~\vec{v} ^{2} -~
       \frac{G ~M_{\odot}}{r} ~,
\nonumber \\
\vec{H} &=&  \vec{r} \times ~\vec{v} ~,
\nonumber \\
p &=&  \frac{| \vec{H} | ^{2}}{G ~M_{\odot}} ~,
\nonumber \\
e &=&  1 ~+~ \frac{2 ~p ~E}{G ~M_{\odot}} ~,
\nonumber \\
a &=&  \frac{p}{1 ~-~ e^{2}} ~, ~~ q = a ~ \left ( 1 ~-~ e  \right ) ~,
\nonumber \\
i &=& \arccos ( H_{z} / | \vec{H} | ) ~,
\nonumber \\
\sin \Omega ~ \sin i &=&  H_{x} / | \vec{H} |  ~, ~~
- ~\cos \Omega ~ \sin i =  H_{y} / | \vec{H} |	~,
\nonumber \\
\vec{e}_{R} &=& \vec{r} / | \vec{r} | = ( x, y, z ) / r ~,
\nonumber \\
\vec{e}_{N} &=& \vec{H} / | \vec{H} | ~,
\nonumber \\
\vec{e}_{T} &=& \vec{e}_{N} \times \vec{e}_{R} ~,
\nonumber \\
\sin \Theta ~ \sin i &=&  z  / r ~,
\nonumber \\
\cos \Theta ~ \sin i &=&  \left ( \vec{e}_{T} \right ) _{z}
     \equiv \left ( y ~ H_{x} ~-~ x ~ H_{y} \right )  /
     \left ( r	| \vec{H} | \right ) ~,
\nonumber \\
\sin \left ( \Theta ~-~ \omega \right ) &=& \vec{v} \cdot \vec{e}_{R} /
       \left ( e \sqrt{G ~ M_{\odot} / p} \right ) ~,
\nonumber \\
\cos \left ( \Theta ~-~ \omega \right ) &=& \vec{v} \cdot \vec{e}_{T} /
       \left ( e \sqrt{G ~ M_{\odot} / p} \right ) ~-~ 1 ~,
\end{eqnarray}
where $q$ is perihelion distance.
If $i =$ 0, then $\Theta$ has to be obtained from
$\cos ( \Omega + \Theta ) = x / r$,
$\sin ( \Omega + \Theta ) = y / r$, and, we may take $\Omega$ in an
arbitrary way (e. g., the requirement that $\Omega$ is a continuous function of
time may define $\Omega$ for $i =$ 0). As for the value
of inclination, we repeat the well-known fact: prograde orbits exist for
$i \in \langle 0, \pi / 2 )$, retrograde orbits exist for
$i \in ( \pi / 2, \pi \rangle$.
The case $e > 1$ describes the case when the particle is ejected from the
Solar System.

\subsection{Poynting-Robertson effect}
Putting $\bar{Q}'_{2} = \bar{Q}'_{3} = F'_{e1} = F'_{e2} =$
$F'_{e3} \equiv$ 0 in Eq. (40), we obtain Poynting-Robertson
effect (Robertson 1937, Kla\v{c}ka 1992a):
\begin{equation}\label{48}
\frac{D ~p^{\mu}}{d~ \tau} = \frac{w_{1}^{2} ~S~A'}{c} ~
		 \bar{Q}'_{1}
		 \left ( b_{j}^{\mu} ~-~ u^{\mu} ~/~c  \right ) ~.
\end{equation}

Eq. (48) enables us some analytical
calculations for changes of osculating orbital elements: it is supposed
that $\bar{Q}'_{1}$ is a constant.
We will present them, in the following sections.
Section 6 will consider the first order in $\vec{v}/c$ of Eq. (48),
section 7 will consider also the second order in $\vec{v}/c$ of Eq. (48).

\section{P-R effect -- equation of motion to the first order in $v/c$}
We can write, on the basis of Eqs. (48), (41) -- (43)
\begin{equation}\label{49}
\frac{d~ \vec{v}}{d ~t} = -~ \frac{\mu}{r^{2}} ~ \vec{e}_{R} ~+~
	  \frac{\mu}{r^{2}} ~
	  \beta ~\left \{  \left ( 1~-~
	  \frac{\vec{v} \cdot \vec{e}_{R}}{c} \right ) ~ \vec{e}_{R}
	  ~-~ \frac{\vec{v}}{c} \right \} ~,
\end{equation}
where $\mu \equiv G~M_{\odot}$, $\vec{e}_{R} \equiv \vec{e}_{1}$ and
$\beta \equiv \beta_{1}$is a non-dimensional parameter
(``the ratio of radiation pressure force to the gravitational force'';
see also Eq. (43)).
Eq. (43) reduces to
$\beta = 5.7 \times 10^{-5} \bar{Q}'_{1} / ( \varrho [g/cm^{3}] ~s [cm] )$,
for homogeneous spherical particle:
$\varrho$ is mass density and $s$ is radius of the sphere.

\subsection{Secular changes of orbital elements -- radiation pressure
as a part of central acceleration}
We have to use $-~\mu ~( 1~-~\beta )~\vec{e}_{R}~/~r^{2}$ as a central
acceleration determining osculating orbital elements if we want to take a time
average ($T$ is time interval between passages through two following
pericenters) in an analytical way
\begin{eqnarray}\label{50}
\langle g \rangle &\equiv& \frac{1}{T}	\int_{0}^{T} g(t) dt =
\frac{\sqrt{\mu ~( 1 - \beta )}}{a_{\beta}^{3/2}} ~
 \frac{1}{2 \pi}  \int_{0}^{2 \pi} g(f_{\beta})
\left ( \frac{df_{\beta}}{dt} \right )^{-1} df_{\beta}
\nonumber \\
&=& \frac{\sqrt{\mu ~( 1 - \beta )}}{a_{\beta}^{3/2}} ~ \frac{1}{2 \pi} \int_{0}^{2 \pi}
g(f_{\beta}) ~\frac{r^{2}}{\sqrt{\mu ~( 1 - \beta ) ~p_{\beta}}}~ df_{\beta}
\nonumber \\
&=& \frac{1}{a_{\beta}^{2}~ \sqrt{1 - e_{\beta}^{2}}} ~\frac{1}{2 \pi} ~
  \int_{0}^{2 \pi} ~ g(f_{\beta}) ~r^{2} ~ df_{\beta} ~,
\end{eqnarray}
assuming non-pseudo-circular orbits and the fact that orbital elements exhibit
only small changes during the time interval $T$; $a_{\beta}$ is semi-major axis,
$e_{\beta}$ is eccentricity, $f_{\beta}$ is true anomaly,
$p_{\beta} = a_{\beta}(1 - e_{\beta}^{2})$;
the second and the third Kepler's laws were used:
$r^{2} ~df_{\beta}/dt = \sqrt{\mu ( 1 - \beta ) p_{\beta}}$ --
conservation of angular momentum,
$a_{\beta}^{3}/T^{2} = \mu ( 1 - \beta ) / (4 \pi^{2})$.

Rewriting Eq. (49) into the form
\begin{equation}\label{51}
\frac{d\vec{v}}{dt} = - \frac{\mu ~\left ( 1 - \beta \right )}{r^{2}} \vec{e}_{R}
	      - \beta \frac{\mu}{r^{2}}
    \left \{
    \left ( \frac{\vec{v} \cdot \vec{e}_{R}}{c} \right ) \vec{e}_{R}
    + \frac{\vec{v}}{c} \right \} ~,
\end{equation}
we can immediately write for components of perturbation acceleration
to Keplerian motion:
\begin{equation}\label{52}
F_{\beta ~R} = -~2~ \beta ~\frac{\mu}{r^{2}} ~\frac{v_{\beta~R}}{c} ~,~~
F_{\beta ~T} = -~ \beta ~\frac{\mu}{r^{2}} ~\frac{v_{\beta~T}}{c} ~,~~
F_{\beta ~N} = 0 ~,
\end{equation}
where $F_{\beta ~R}$, $F_{\beta ~T}$ and $F_{\beta ~N}$ are radial, transversal
and normal components of perturbation acceleration, and
two-body problem yields
\begin{eqnarray}\label{53}
v_{\beta~R} &=& \sqrt{\frac{\mu ~( 1 - \beta )}{p_{\beta}}} ~
	e_{\beta} \sin f_{\beta} ~,~
\nonumber \\
v_{\beta~T} &=& \sqrt{\frac{\mu ~( 1 - \beta )}{p_{\beta}}} ~
\left ( 1 + e_{\beta} \cos f_{\beta} \right ) ~.
\end{eqnarray}
The important fact that perturbation acceleration is proportional to
$v/c$ ($\ll$ 1) ensures the above metioned small changes of orbital elements
during the time interval $T$.

Perturbation equations of celestial mechanics yield for osculating orbital
elements ($a_{\beta}$ -- semi-major axis; $e_{\beta}$ -- eccentricity;
$i_{\beta}$ -- inclination
(of the orbital plane to the reference frame);
$\Omega_{\beta}$ -- longitude of the ascending node; $\omega_{\beta}$ --
longitude of pericenter; $\Theta_{\beta}$
is the position angle of the particle on the orbit, when measured
from the ascending node in the direction of the particle's motion,
$\Theta_{\beta} = \omega_{\beta} + f_{\beta}$):
\begin{eqnarray}\label{54}
\frac{d a_{\beta}}{d t} &=& \frac{2~a_{\beta}}{1~-~e_{\beta}^{2}} ~
      \sqrt{\frac{p_{\beta}}{\mu \left ( 1 ~-~ \beta \right )}} ~
      \left \{
      F_{\beta ~R} ~e_{\beta}~ \sin f_{\beta} +
      F_{\beta ~T} \left ( 1~+~e_{\beta}~ \cos f_{\beta} \right ) \right \} ~,
\nonumber \\
\frac{d e_{\beta}}{d t} &=&
      \sqrt{\frac{p_{\beta}}{\mu \left ( 1 ~-~ \beta \right )}} ~ \left \{
      F_{\beta ~R} ~ \sin f_{\beta} +
      F_{\beta ~T} \left [ \cos f_{\beta} ~+~
     \frac{e_{\beta} +	\cos f_{\beta}}{1 + e_{\beta} \cos f_{\beta}}
	  \right ] \right \} ~,
\nonumber \\
\frac{d i_{\beta}}{d t} &=& \frac{r}{\sqrt{\mu \left ( 1 ~-~ \beta \right ) p_{\beta}}} ~
	    F_{\beta ~N} ~ \cos \Theta_{\beta} ~,
\nonumber \\
\frac{d \Omega_{\beta}}{d t} &=&
      \frac{r}{\sqrt{\mu \left ( 1 ~-~ \beta \right ) p_{\beta}}} ~
      F_{\beta ~N} ~ \frac{\sin \Theta_{\beta}}{\sin i_{\beta}} ~,
\nonumber \\
\frac{d \omega_{\beta}}{d t} &=& -~ \frac{1}{e_{\beta}} ~
      \sqrt{\frac{p_{\beta}}{\mu \left ( 1 ~-~ \beta \right )}} ~ \left \{
      F_{\beta ~R} \cos f_{\beta} - F_{\beta ~T}
      \frac{2 + e_{\beta} \cos f_{\beta}}{1 + e_{\beta} \cos f_{\beta}}
      \sin f_{\beta} \right \} ~-~
\nonumber \\
& &   \frac{r}{\sqrt{\mu \left ( 1 ~-~ \beta \right ) p_{\beta}}} ~
      F_{\beta ~N} ~ \frac{\sin \Theta_{\beta}}{\sin i_{\beta}} ~\cos i_{\beta} ~,
\nonumber \\
\frac{d \Theta_{\beta}}{d t} &=&
      \frac{\sqrt{\mu \left ( 1 ~-~ \beta \right ) p_{\beta}}}{r^{2}} ~-~
      \frac{r}{\sqrt{\mu \left ( 1 ~-~ \beta \right ) p_{\beta}}} ~
      F_{\beta ~N} ~ \frac{\sin \Theta_{\beta}}{\sin i_{\beta}} ~\cos i_{\beta} ~,
\end{eqnarray}
where $r = p_{\beta} / (1 + e_{\beta} \cos f_{\beta})$.

Inserting Eqs. (52) -- (53) into Eq. (54), one easily obtains
\begin{eqnarray}\label{55}
\frac{da_{\beta}}{dt} &=& -~\beta \frac{\mu}{r^{2}} \frac{2 a_{\beta}}{c}
    \frac{1 + e_{\beta}^{2} + 2 e_{\beta} \cos f_{\beta}
    + e_{\beta}^{2}  \sin^{2} f_{\beta}}{1~-~e_{\beta}^{2}} ~,
\nonumber \\
\frac{de_{\beta}}{dt} &=& -~ \beta \frac{\mu}{r^{2}} \frac{1}{c}  \left (
      2 e_{\beta} + e_{\beta}  \sin^{2} f_{\beta} + 2 \cos f_{\beta} \right ) ~,
\nonumber \\
\frac{d i_{\beta}}{dt} &=& 0 ~,
\nonumber \\
\frac{d\Omega_{\beta}}{dt} &=& 0 ~,
\nonumber \\
\frac{d\omega_{\beta}}{dt} &=& -~ \beta \frac{\mu}{r^{2}} \frac{1}{c}
    \frac{1}{e_{\beta}} ~ \left (
    2  - e_{\beta} \cos f_{\beta} \right ) \sin f_{\beta} ~,
\nonumber \\
\frac{d \Theta_{\beta}}{dt} &=& \frac{\sqrt{\mu \left ( 1 - \beta \right )
		p_{\beta}}}{r^{2}} ~.
\end{eqnarray}
It is worth mentioning that $da_{\beta}/dt <$ 0 for any time $t$.

The set of differential equations Eqs. (55) has to be complemented
with initial conditions. If the subscript ``$P$'' denotes orbital elements
of the parent body (see Eq. (44)), then
\begin{eqnarray}\label{56}
\vec{r}_{\beta~in} &=& \vec{r}_{P} \equiv
	 \frac{p_{P}}{1 ~+~ e_{P} ~ \cos f_{P}} ~ \vec{e}_{PR}
\nonumber \\
\vec{v}_{\beta~in} &=& \vec{v}_{P} ~+~ \vec{\Delta} ~,
\end{eqnarray}
where $\vec{\Delta}$ is
velocity with respect to the nucleus of the parent body, and
\begin{eqnarray}\label{57}
\vec{r}_{\beta~in} &=&
    \frac{p_{\beta ~in}}{1 ~+~ e_{\beta~in} ~ \cos f_{\beta~in}} ~
    \vec{e}_{\beta ~R ~in} ~,
\nonumber \\
\vec{v}_{\beta~in} &=& v_{\beta~R~in} ~\vec{e}_{\beta ~R~in} ~+~
	       v_{\beta~T~in} ~\vec{e}_{\beta ~T~in} ~,
\nonumber \\
v_{\beta~R~in} &=& \sqrt{\frac{\mu ~( 1 - \beta )}{p_{\beta~in}}} ~
	e_{\beta~in} \sin f_{\beta~in} ~,~
\nonumber \\
v_{\beta~T~in} &=& \sqrt{\frac{\mu ~( 1 - \beta )}{p_{\beta~in}}} ~
\left ( 1 + e_{\beta~in} \cos f_{\beta~in} \right ) ~.
\nonumber   \\
\vec{e}_{\beta ~R ~in} &=& ( \cos \Omega_{\beta ~in} ~ \cos \Theta_{\beta ~in}
	 ~-~ \sin \Omega_{\beta ~in} ~ \sin \Theta_{\beta ~in} ~
	 \cos i_{\beta ~in} ~,
\nonumber   \\
& &	\sin \Omega_{\beta ~in} ~ \cos \Theta_{\beta ~in} ~+~
	\cos \Omega_{\beta ~in} ~ \sin \Theta_{\beta ~in}
	\cos i_{\beta ~in} ~,~
	\sin \Theta_{\beta ~in} ~ \sin i_{\beta ~in} ) ~,
\nonumber   \\
\vec{e}_{\beta ~T ~in} &=& ( -~ \cos \Omega_{\beta ~in} ~
	\sin \Theta_{\beta ~in} ~-~
	\sin \Omega_{\beta ~in} ~ \cos \Theta_{\beta ~in} ~
	\cos i_{\beta ~in} ~,
\nonumber   \\
& &	-~ \sin \Omega_{\beta ~in} ~ \sin \Theta_{\beta ~in} ~+~
	\cos \Omega_{\beta ~in} ~ \cos \Theta_{\beta ~in} ~
	\cos i_{\beta ~in} ~,~
	\cos \Theta_{\beta ~in} ~ \sin i_{\beta ~in} ) ~,
\nonumber   \\
f_{\beta ~in} &=& \Theta_{\beta ~in} ~-~ \omega _{\beta ~in} ~.
\end{eqnarray}
We write ($\vec{e}_{PN} = \vec{e}_{PR} \times \vec{e}_{PT}$)
\begin{equation}\label{58}
\vec{\Delta} = \Delta v_{R} ~\vec{e}_{PR} ~+~  \Delta v_{T} ~\vec{e}_{PT} ~+~
	   \Delta v_{N} ~\vec{e}_{PN} ~.
\end{equation}
Using Eqs. (44), (56) -- (58), we finally obtain
(compare Gajdo\v{s}\'{\i}k and Kla\v{c}ka 1999):
\begin{eqnarray}\label{59}
p_{\beta ~in} &=& \frac{p_{P}}{1 ~-~ \beta} ~
      \frac{\left ( v_{T} ~+~ \Delta v_{T} \right ) ^{2} ~+~
      \left ( \Delta v_{N} \right ) ^{2}}
      {v_{T}^{2}} ~, \nonumber \\
e_{\beta ~in}^{2} &=& \left ( \frac{1 ~+~ e_{P} ~ \cos f_{P}}{1 ~-~ \beta} ~
  \frac{v_{TS}^{2}}{v_{T}^{2}} ~-~ 1 \right ) ^{2} ~+~
  \nonumber \\
& & \left ( \frac{1 ~+~ e_{P} ~ \cos f_{P}}{1 ~-~ \beta} ~
  \frac{v_{TS}}{v_{T}}	\right ) ^{2}
    \left ( \frac{v_{R} ~+~ \Delta v_{R}}{v_{T}} \right ) ^{2} ~,
  \nonumber \\
\cos i_{\beta ~in} &=& \frac{v_{T} ~+~ \Delta v_{T}}{v_{TS}} \cos i_{P} ~-~
\frac{\Delta v_{N}}{v_{TS}} ~ \cos \Theta_{P} ~ \sin i_{P} ~, \nonumber \\
\sin i_{\beta ~in} ~\cos \Omega_{\beta ~in} &=&
\frac{v_{T} ~+~ \Delta v_{T}}{v_{TS}} \sin i_{P}
~ \cos \Omega_{P} ~+~ \nonumber \\
& & \frac{\Delta v_{N}}{v_{TS}} ~  \left ( \cos \Theta_{P} ~
\cos i_{P} ~ \cos \Omega_{P}  ~-~ \sin \Theta_{P} ~ \sin \Omega_{P} \right ) ~,
\nonumber \\
\sin i_{\beta ~in} ~\sin \Omega_{\beta ~in} &=& \frac{v_{T} ~+~
	  \Delta v_{T}}{v_{TS}} \sin i_{P}
	  ~ \sin \Omega_{P} ~+~ \nonumber \\
& & \frac{\Delta v_{N}}{v_{TS}} ~  \left ( \cos \Theta_{P} ~
\cos i_{P} ~ \sin \Omega_{P}  ~+~ \sin \Theta_{P} ~ \cos \Omega_{P} \right ) ~,
\nonumber \\
e_{\beta ~in} ~ \cos f_{\beta ~in} &=& \frac{p_{\beta ~in}}{p_{P}} ~
		  \left ( 1 ~+~ e_{P} ~\cos f_{P} \right ) ~-~ 1 ~,
\nonumber \\
e_{\beta ~in} ~ \sin f_{\beta ~in} &=& \frac{1 ~+~ e_{P} ~\cos f_{P}}{1 ~-~ \beta} ~
    \frac{v_{TS} ~ \left ( v_{R} ~+~ \Delta v_{R} \right )}{v_{T}^{2}} ~,
\nonumber \\
\sin i_{\beta ~in} ~\cos \Theta_{\beta ~in} &=& \frac{v_{T} ~+~ \Delta v_{T}}{v_{TS}} ~\sin i_{P}
~ \cos \Theta_{P} ~+~ \nonumber \\
& & \frac{\Delta v_{N}}{v_{TS}} ~ \cos i_{P} ~,
\nonumber \\
\sin i_{\beta ~in} ~\sin \Theta_{\beta ~in} &=& \sin i_{P} ~ \sin \Theta_{P} ~,
\nonumber \\
v_{TS}^{2} &\equiv& ( v_{T} ~+~ \Delta v_{T} )^{2} ~+~ ( \Delta v_{N} )^{2} ~,
\end{eqnarray}
where $v_{R}$ and $v_{T}$ are given by expressions presented in Eq. (78) --
$p_{P} = a_{P} ~ \left ( 1 ~-~ e_{P}^{2} \right )$.

For the special case $\vec{\Delta} =$ 0 Eq. (59) reduces to
\begin{equation}\label{60}
a_{\beta ~in} = a_{P} \left ( 1 - \beta \right ) \left ( 1 - 2 \beta
       \frac{1 + e_{P} \cos f_{P}}{1~-~e_{P}^{2}}
       \right ) ^{-1} ~,
\end{equation}
\begin{equation}\label{61}
e_{\beta ~in} = \sqrt{1 - \frac{1 - e_{P}^{2} - 2 \beta \left (
      1 + e_{P} \cos f_{P} \right )}{
      \left ( 1 - \beta \right )^{2}}} ~,
\end{equation}
where $f_{P} \equiv \Theta_{P} - \omega_{P}$,
$\omega _{\beta ~in}$ has to be obtained from
\begin{eqnarray}\label{62}
e_{\beta~ in} ~ \cos \left ( \Theta _{P} - \omega _{\beta ~in} \right ) &=&
  \frac{\beta + e_{P} \cos f_{P}}{1 - \beta} ~,
\nonumber \\
e_{\beta ~in} ~ \sin \left ( \Theta _{P} - \omega _{\beta ~in} \right ) &=&
  \frac{e_{P} \sin f_{P}}{1 - \beta} ~,
\end{eqnarray}
\begin{equation}\label{63}
\Omega_{\beta~ in} = \Omega_{P} ~, ~~i_{\beta~ in} = i_{P} ~,~~
\Theta_{\beta ~in} = \Theta_{P} ~.
\end{equation}
Physics of Eqs. (60) -- (61) is following: meteoroid escapes from
the Solar System when the orbital energy becomes positive and this
can happen when the energy due to the radial component of the
radiation force is included, without it being necessary for this
force to exceed the gravitational attraction (Harwit 1963). Some
figures may be found in Kla\v{c}ka (1992b). As an example we may
mention that particle of $\beta = ( 1 - e_{P} ) / 2$ moves in
parabolic orbit if ejected at perihelion of the parent body, and,
in an orbit with eccentricity $e_{\beta ~in} = | 1 - 3 e_{P} | / (
1 + e_{P} )$ if released at aphelion; $e_{\beta ~in} =$ 0 for
$\beta = e_{P} = 1 / 3$ and aphelion ejection.

By inserting Eqs. (55) into Eq. (50), taking into account that $e_{\beta ~in} <$ 1
(see Eq. (61)), one can easily obtain for the secular changes of orbital elements:
\begin{equation}\label{64}
\langle  \frac{d a_{\beta}}{d t} \rangle = -~ \beta ~ \frac{\mu}{c} ~
       \frac{2 + 3 e_{\beta}^{2}}{a_{\beta}~
     \left ( 1 - e_{\beta}^{2} \right )^{3/2}} ~,
\end{equation}
\begin{equation}\label{65}
\langle  \frac{d e_{\beta}}{d t} \rangle = -~ \frac{5}{2}~ \beta ~ \frac{\mu}{c} ~
       \frac{e_{\beta}}{a_{\beta}^{2}~
     \left ( 1 - e_{\beta}^{2} \right )^{1/2}} ~,
\end{equation}
\begin{equation}\label{66}
\langle  \frac{d \Theta_{\beta}}{d t} \rangle =
    \frac{\sqrt{\mu ~ \left ( 1 - \beta \right )}}{a_{\beta}^{3/2}} ~,
\end{equation}
\begin{equation}\label{67}
\langle  \frac{d i_{\beta}}{d t} \rangle =
\langle  \frac{d \Omega_{\beta}}{d t} \rangle =
\langle  \frac{d \omega_{\beta}}{d t} \rangle = 0 ~.
\end{equation}
As an remark we may mention that a little more simple set of differential
equations than that represented by Eqs. (64) -- (65) is obtained when
semi-major axis $a_{\beta}$ is replaced by the quantity
$p_{\beta}$ $=$ $a_{\beta}$ ( $1 - e_{\beta}^{2}$ ). It can be easily
verified that a new set of differential equations for secular evolution
is given by the following equations: \\
i)  $d p_{\beta} / dt = - 2 \beta ( \mu / c ) ( 1 - e_{\beta}^{2} ) ^{3/2} / p_{\beta}$, \\
ii) $d e_{\beta} / dt = - ( 5 / 2 ) \beta ( \mu / c ) [ ( 1 - e_{\beta}^{2} ) ^{3/2} / p_{\beta} ]
e_{\beta} / p_{\beta}$. \\
These two equations immediately yield $p_{\beta} = p_{\beta in}$
$( e_{\beta} / e_{\beta in} ) ^{4/5}$. \\
The last relation corresponds to the relation
between orbital angular momentum and eccentricity of the orbit:
$H_{\beta} = H_{\beta in}$ $( e_{\beta} / e_{\beta in} ) ^{2/5}$. \\
Analogously, for secular evolution of osculating orbital energy (per unit mass)
holds: $E_{\beta} = \vec{v} ^{2} / 2 - \mu ( 1 - \beta ) / r$ $\equiv$
$- \mu ( 1 - \beta ) ( 1 - e_{\beta}^{2} ) / ( 2 p_{\beta} )$,	\\
$E_{\beta} = E_{\beta in} [ ( 1 - e_{\beta}^{2} ) / ( 1 - e_{\beta in}^{2} ) ]$
$( e_{\beta in} / e_{\beta} ) ^{4/5}$. \\
Equations i) and ii) enable to write: \\
i)  $d p_{\beta} / dt = - 2 \beta ( \mu / c )$
$[ 1 - e_{\beta in}^{2} ( p_{\beta} / p_{\beta in} ) ^{5/2} ] ^{3/2}$ $/$
$p_{\beta}$, \\
ii) $d e_{\beta} / dt = - ( 5 / 2 ) \beta ( \mu / c )$
$( e_{\beta in}^{8/5} / p_{\beta in} ^{2} )$
$( 1 - e_{\beta}^{2} ) ^{3/2}$ $/$ $e_{\beta}^{3/5}$. \\
These equations point out that near-circular / pseudo-circular orbits
($e_{\beta} \approx$ 0) and orbits with small values of $p_{\beta}$
are not described by the discussed secular changes of orbital elements
and by the considered $v/c$ approximation for the P-R effect.

\subsection{Secular changes of orbital elements --
gravitation as a central acceleration}
As it was discussed in section 5,
it is not wise to use $-~\mu ~( 1~-~\beta_{1} )~\vec{e}_{R}~/~r^{2}$
as a central acceleration determining osculating orbital elements
for the general case represented by Eqs. (40) or (41), since
$\beta_{1}$ changes almost randomly during a motion.
In order to have a comparable results in disposal, we have to find secular
changes of orbital elements when central central acceleration
is given by $-~\mu ~\vec{e}_{R}~/~r^{2}$.
Thus, we will use $-~\mu ~\vec{e}_{R}~/~r^{2}$ as a central
acceleration determining osculating orbital elements.

Taking into account Eq. (49), we take the action of electromagnetic radiation
as a perturbation to the two-body problem.
We can immediately write for components of perturbation acceleration:
\begin{equation}\label{68}
F_{R} = \beta ~\frac{\mu}{r^{2}} ~-
~2~ \beta ~\frac{\mu}{r^{2}} ~\frac{v_{Rd}}{c} ~,~~
F_{T} = -~ \beta ~\frac{\mu}{r^{2}} ~\frac{v_{Td}}{c} ~,~~
F_{N} = 0 ~,
\end{equation}
where $F_{R}$, $F_{T}$ and $F_{N}$ are radial, transversal
and normal components of perturbation acceleration, and
two-body problem yields
\begin{eqnarray}\label{69}
v_{Rd} &=& \sqrt{\frac{\mu}{p}} ~ e \sin f ~,~
\nonumber \\
v_{Td} &=& \sqrt{\frac{\mu}{p}} ~ \left ( 1 + e \cos f \right ) ~.
\end{eqnarray}

Perturbation equations of celestial mechanics yield for osculating orbital
elements ($a$ -- semi-major axis; $e$ -- eccentricity; $i$ -- inclination
(of the orbital plane to the reference frame);
$\Omega$ -- longitude of the ascending node; $\omega$ --
longitude of pericenter; $\Theta$ --
$\Theta = \omega + f$ is the position angle of the particle on the orbit,
when measured from the ascending node in the direction of the particle's
motion):
\begin{eqnarray}\label{70}
\frac{d a}{d t} &=& \frac{2~a}{1~-~e^{2}} ~
      \sqrt{\frac{p}{\mu}} ~ \left \{
      F_{R} ~e~ \sin f +
      F_{T} \left ( 1~+~e~ \cos f \right ) \right \} ~,
\nonumber \\
\frac{d e}{d t} &=& \sqrt{\frac{p}{\mu}} ~ \left \{ F_{R} ~ \sin f +
	    F_{T} \left [ \cos f ~+~ \frac{e +	\cos f}{1 + e \cos f}
	    \right ] \right \} ~,
\nonumber \\
\frac{d i}{d t} &=& \frac{r}{\sqrt{\mu ~p}} ~
	    F_{N} ~ \cos \Theta ~,
\nonumber \\
\frac{d \Omega}{d t} &=& \frac{r}{\sqrt{\mu ~p}} ~
	     F_{N} ~ \frac{\sin \Theta}{\sin i} ~,
\nonumber \\
\frac{d \omega}{d t} &=& -~ \frac{1}{e} ~\sqrt{\frac{p}{\mu}} ~ \left \{
	     F_{R} \cos f - F_{T} \frac{2 + e \cos f}{1 + e \cos f}
	     \sin f \right \} ~-~
\nonumber \\
& &	     \frac{r}{\sqrt{\mu ~p}} ~
	     F_{N} ~ \frac{\sin \Theta}{\sin i} ~\cos i ~,
\nonumber \\
\frac{d \Theta}{d t} &=& \frac{\sqrt{\mu ~p}}{r^{2}} ~-~
	     \frac{r}{\sqrt{\mu ~p}} ~
	     F_{N} ~ \frac{\sin \Theta}{\sin i} ~\cos i ~,
\end{eqnarray}
where $r = p / (1 + e \cos f)$.

Inserting Eqs. (68) -- (69) into Eq. (70), one easily obtains
\begin{eqnarray}\label{71}
\frac{da}{dt} &=& \frac{2~ \beta}{r^{2}} ~\sqrt{\frac{\mu~ a^{3}}{1~-~e^{2}}} ~
    e ~\sin f ~-~
    \beta \frac{\mu}{r^{2}} ~\frac{2~a}{c}~
    \frac{1 + e^{2} + 2 e \cos f + e^{2}  \sin^{2} f}{1~-~e^{2}} ~,
\nonumber \\
\frac{de}{dt} &=& \beta ~ \frac{\sqrt{\mu~p}}{r^{2}} ~ \sin f ~-~
      \beta ~ \frac{\mu}{r^{2}} \frac{1}{c}  \left (
      2 e + e  \sin^{2} f + 2 \cos f \right ) ~,
\nonumber \\
\frac{d i}{dt} &=& 0 ~,
\nonumber \\
\frac{d\Omega}{dt} &=& 0 ~,
\nonumber \\
\frac{d\omega}{dt} &=& -~ \beta ~ \frac{\sqrt{\mu~p}}{r^{2}} ~\frac{1}{e} ~
	       \cos f ~-~ \beta ~ \frac{\mu}{r^{2}} ~ \frac{1}{c}
	       \frac{1}{e} ~ \left ( 2	- e \cos f \right ) ~ \sin f ~,
\nonumber \\
\frac{d \Theta}{dt} &=& \frac{\sqrt{\mu ~p}}{r^{2}} ~.
\end{eqnarray}
It is worth mentioning that $da/dt <$ 0 for any time $t$ does not hold
(perturbation corresponds to complete nongravitational acceleration, and, thus,
Eqs. (55) do not hold).

The set of differential equations Eqs. (71) has to be complemented
with initial conditions. If the subscript ``$P$'' denotes orbital elements
of the parent body, then the initial orbital elements for the particle
ejected with velocity $\vec{\Delta}$ are:
\begin{eqnarray}\label{72}
p_{in} &=& p_{P} ~
      \frac{\left ( v_{T} ~+~ \Delta v_{T} \right ) ^{2} ~+~
      \left ( \Delta v_{N} \right ) ^{2}}
      {v_{T}^{2}} ~, \nonumber \\
e_{in}^{2} &=& \left [ \left ( 1 ~+~ e_{P} ~ \cos f_{P} \right ) ~
  \frac{v_{TS}^{2}}{v_{T}^{2}} ~-~ 1 \right ] ^{2} ~+~
  \nonumber \\
& & \left [ \left ( 1 ~+~ e_{P} ~ \cos f_{P} \right ) ~
  \frac{v_{TS}}{v_{T}}	\right ] ^{2}
    \left ( \frac{v_{R} ~+~ \Delta v_{R}}{v_{T}} \right ) ^{2} ~,
  \nonumber \\
\cos i_{in} &=& \frac{v_{T} ~+~ \Delta v_{T}}{v_{TS}} \cos i_{P} ~-~
\frac{\Delta v_{N}}{v_{TS}} ~ \cos \Theta_{P} ~ \sin i_{P} ~, \nonumber \\
\sin i_{in} ~\cos \Omega_{in} &=&
\frac{v_{T} ~+~ \Delta v_{T}}{v_{TS}} \sin i_{P}
~ \cos \Omega_{P} ~+~ \nonumber \\
& & \frac{\Delta v_{N}}{v_{TS}} ~  \left ( \cos \Theta_{P} ~
\cos i_{P} ~ \cos \Omega_{P}  ~-~ \sin \Theta_{P} ~ \sin \Omega_{P} \right ) ~,
\nonumber \\
\sin i_{in} ~\sin \Omega_{in} &=& \frac{v_{T} ~+~
	  \Delta v_{T}}{v_{TS}} \sin i_{P}
	  ~ \sin \Omega_{P} ~+~ \nonumber \\
& & \frac{\Delta v_{N}}{v_{TS}} ~  \left ( \cos \Theta_{P} ~
\cos i_{P} ~ \sin \Omega_{P}  ~+~ \sin \Theta_{P} ~ \cos \Omega_{P} \right ) ~,
\nonumber \\
e_{in} ~ \cos f_{in} &=& \frac{p_{in}}{p_{P}} ~
	     \left ( 1 ~+~ e_{P} ~\cos f_{P} \right ) ~-~ 1 ~,
\nonumber \\
e_{in} ~ \sin f_{in} &=& \left ( 1 ~+~ e_{P} ~\cos f_{P} \right ) ~
    \frac{v_{TS} ~ \left ( v_{R} ~+~ \Delta v_{R} \right )}{v_{T}^{2}} ~,
\nonumber \\
\sin i_{in} ~\cos \Theta_{in} &=& \frac{v_{T} ~+~ \Delta v_{T}}{v_{TS}} ~\sin i_{P}
~ \cos \Theta_{P} ~+~ \nonumber \\
& & \frac{\Delta v_{N}}{v_{TS}} ~ \cos i_{P} ~,
\nonumber \\
\sin i_{in} ~\sin \Theta_{in} &=& \sin i_{P} ~ \sin \Theta_{P} ~,
\nonumber \\
v_{TS}^{2} &\equiv& ( v_{T} ~+~ \Delta v_{T} )^{2} ~+~ ( \Delta v_{N} )^{2} ~,
\end{eqnarray}
where $v_{R}$ and $v_{T}$ are given by expressions presented in Eq. (44) --
$p_{P} = a_{P} ~ \left ( 1 ~-~ e_{P}^{2} \right )$.

For the special case $\vec{\Delta} =$ 0 Eq. (72) reduces to a simple
fact: initial osculating orbital elements of the particle are identical with
those of the parent body:
\begin{eqnarray}\label{73}
a_{in} &=& a_{P} ~, ~~e_{in} = e_{P} ~,~~i_{in} = i_{P} ~,
\nonumber \\
\Omega_{in} &=& \Omega_{P} ~,~~\omega_{in} = \omega_{P} ~,~~
\Theta_{in} = \Theta_{P} ~.
\end{eqnarray}

The important fact is that Eqs. (71) contain also terms not proportional
to $v/c$ ($\ll$ 1). These important terms protect us to use procedure
analogous to that represented by Eq. (50). While dispersion of osculating orbital
elements is very small during a time interval $T$ for the case when
$\mu ~ ( 1 - \beta )$ is used in central acceleration, the dispersion of
osculating orbital elements may be large during the same time interval
for the case when $\mu$ is used in central acceleration
(compare Figs. 1 and 2 in Kla\v{c}ka 1994b). Thus, any formal
averaging of Eqs. (71) leading to equations analogous to Eqs. (64)-(67)
is not correct.

We have explained that it is not allowed to make a simple time averaging
analogous to that described by Eq. (50), when $-~\mu ~\vec{e}_{R}~/~r^{2}$ is used
as a central acceleration determining osculating orbital elements. However, it
is of interest if we have to numerically solve Newtonian vectorial equation of
motion (Eq. (49)) and make numerical time averaging (over a time interval between
passages through two following pericenters), or if some analytical simplifications
can be done -- something analogous to Eqs. (64)-(67).

Fortunately, we can make analytical calculations for the purpose of
obtaining secular changes of semi-major axis and eccentricity even
when $-~\mu ~\vec{e}_{R}~/~r^{2}$ is used
as a central acceleration determining osculating orbital elements. We will
derive the correct equations in the following two subsections.

\subsubsection{Radial forces and orbital elements}
We will proceed according to Kla\v{c}ka (1994), in this subsection.

Let us consider a gravitational system of two bodies
\begin{equation}\label{74}
\dot{\vec{v}} = -~\frac{\mu}{r^{2}} ~ \vec{e}_{R} ~.
\end{equation}
Let perturbation acceleration exists in the form
\begin{equation}\label{75}
\vec{a} = \beta ~ \frac{\mu}{r^{2}} ~ \vec{e}_{R} ~,
\end{equation}
$0 \le \beta <$ 1. Thus, the final equation of motion is
\begin{equation}\label{76}
\dot{\vec{v}} = -~\frac{\mu ~ \left ( 1 ~-~ \beta \right )}{r^{2}} ~ \vec{e}_{R} ~.
\end{equation}
Eq. (76) yields as a solution the well-known Keplerian motion and the orbit
is given by
\begin{equation}\label{77}
r = \frac{p_{c}}{1 ~+~ e_{c} ~\cos \left ( \Theta ~-~ \omega_{c} \right )} ~,
\end{equation}
where
\begin{equation}\label{78}
p_{c} = a_{c} ~ \left ( 1 ~-~ e_{c}^{2} \right ) ~.
\end{equation}
The subscript ``$c$'' denotes that orbital elements are constants of motion.
If we write
\begin{equation}\label{79}
\vec{v} = v_{cR} ~ \vec{e}_{R} ~+~ v_{cT} ~ \vec{e}_{T} ~,
\end{equation}
where $\vec{e}_{T}$ is a unit vector transverse to the radial vector
$\vec{e}_{R}$ in the plane of the trajectory (positive in the direction of motion),
we have
\begin{equation}\label{80}
v_{cR} = \sqrt{\mu ~ \left ( 1 ~-~ \beta \right )~p_{c}^{-1}} ~e_{c} ~
    \sin \left ( \Theta ~-~ \omega_{c} \right ) ~,
\end{equation}
\begin{equation}\label{81}
v_{cT} = \sqrt{\mu ~ \left ( 1 ~-~ \beta \right )~p_{c}^{-1}} ~\left [
    1 ~+~ e_{c} ~
    \cos \left ( \Theta ~-~ \omega_{c} \right ) \right ] ~.
\end{equation}

In principle, we may consider also a new set of orbital elements, which
are defined by the central gravitational acceleration. Eqs. (77) --
(81) are then of the form
\begin{equation}\label{82}
r = \frac{p}{1 ~+~ e ~\cos \left ( \Theta ~-~ \omega \right )} ~,
\end{equation}
\begin{equation}\label{83}
p = a ~ \left ( 1 ~-~ e^{2} \right ) ~,
\end{equation}
\begin{equation}\label{84}
\vec{v} = v_{dR} ~ \vec{e}_{R} ~+~ v_{dT} ~ \vec{e}_{T} ~,
\end{equation}
\begin{equation}\label{85}
v_{dR} = \sqrt{\mu ~p^{-1}} ~e ~ \sin \left ( \Theta ~-~ \omega \right ) ~,
\end{equation}
\begin{equation}\label{86}
v_{dT} = \sqrt{\mu ~p^{-1}} ~\left [ 1 ~+~ e ~
    \cos \left ( \Theta ~-~ \omega \right ) \right ] ~;
\end{equation}
the fact that $\Theta$ is unchanged in both sets of orbital elements is used;
$\vec{e}_{R}$ $=$ ( $\cos \Theta, \sin \Theta$ ),
$\vec{e}_{T}$ $=$ ( $ - \sin \Theta, \cos \Theta$ ).

Position vector and velocity vector define a state of the body at any time.
Equations (77) and (82) yield then
\begin{equation}\label{87}
\frac{p_{c}}{1 ~+~ e_{c} ~\cos \left ( \Theta ~-~ \omega_{c} \right )}
= \frac{p}{1 ~+~ e ~\cos \left ( \Theta ~-~ \omega \right )} ~.
\end{equation}
Analogously, the other two pairs of equations (Eqs. (80) and (85), and,
Eqs. (81) and (86)) give
\begin{equation}\label{88}
\sqrt{\left ( 1 ~-~ \beta \right )~p_{c}^{-1}} ~e_{c} ~
    \sin \left ( \Theta ~-~ \omega_{c} \right )
= \sqrt{p^{-1}} ~e ~ \sin \left ( \Theta ~-~ \omega \right ) ~,
\end{equation}
\begin{equation}\label{89}
\sqrt{\left ( 1 ~-~ \beta \right )~p_{c}^{-1}} ~\left [ 1 ~+~ e_{c} ~
    \cos \left ( \Theta ~-~ \omega_{c} \right ) \right ]
= \sqrt{p^{-1}} ~\left [ 1 ~+~ e ~
    \cos \left ( \Theta ~-~ \omega \right ) \right ] ~.
\end{equation}
One can easily obtain, using Eqs. (87) and (89),
\begin{equation}\label{90}
p_{c} ~ \left ( 1 ~-~ \beta \right ) = p ~,
\end{equation}
and Eqs. (88)-(89) yield then
\begin{equation}\label{91}
\left ( 1 ~-~ \beta \right ) ~e_{c} ~
    \sin \left ( \Theta ~-~ \omega_{c} \right ) =
    e~ \sin \left ( \Theta ~-~ \omega \right ) ~,
\end{equation}
\begin{equation}\label{92}
\left ( 1 ~-~ \beta \right ) ~\left [
    1 ~+~ e_{c} ~
    \cos \left ( \Theta ~-~ \omega_{c} \right ) \right ] =
    1 ~+~ e ~ \cos \left ( \Theta ~-~ \omega \right ) ~.
\end{equation}
Eq. (92) may be written in the form
\begin{equation}\label{93}
\left ( 1 ~-~ \beta \right ) ~ e_{c} ~
    \cos \left ( \Theta ~-~ \omega_{c} \right ) ~-~ \beta  =
    e~ \cos \left ( \Theta ~-~ \omega \right ) ~.
\end{equation}
Eqs. (91) and (93) yield
\begin{equation}\label{94}
e^{2} = \left ( 1 ~-~ \beta \right ) ^{2} ~ e_{c}^{2} ~+~ \beta ^{2} ~-~
    2~ \beta ~ \left ( 1 ~-~ \beta \right ) ~ e_{c} ~
    \cos \left ( \Theta ~-~ \omega_{c} \right ) ~.
\end{equation}
Equation for $\omega$ is given by Eqs. (91) and (93), using also Eq. (94).
Finally, Eqs. (78), (83), (79) and (94) yield
\begin{equation}\label{95}
a = a_{c} ~ \left \{ 1 ~+~ \beta ~ \frac{1 ~+~ e_{c}^{2} ~+~ 2~e_{c} ~
    \cos \left ( \Theta ~-~ \omega_{c} \right )}{1~-~e_{c}^{2}} \right \}
    ^{-1}  ~.
\end{equation}
Eqs. (94)-(95) show that orbital osculating elements $a$ and $e$ change in time,
during an orbital revolution -- the larger $\beta$, the larger change of
$a$ and $e$.

The osculating orbital elements $e$ and $a$ obtain values between their
maxima and minima, which can be easily found from Eqs. (94)-(95). One can
easily verify that these relations hold:
\begin{equation}\label{96}
e_{min} = | \left ( 1 ~-~ \beta \right ) ~e_{c} ~-~ \beta | \le e \le
      \left ( 1 ~-~ \beta \right ) ~e_{c} ~+~ \beta =
      e_{max} ~,
\end{equation}
\begin{equation}\label{97}
\frac{a_{min}}{a_{c}} =
     \frac{1~-~e_{c}}{1~-~e_{c} ~+~ \beta ~\left ( 1 ~+~ e_{c} \right )}
\le \frac{a}{a_{c}} \le
     \frac{1~+~e_{c}}{1~+~e_{c} ~+~ \beta ~\left ( 1 ~-~ e_{c} \right )}
= \frac{a_{max}}{a_{c}} ~.
\end{equation}

\subsubsection{Mean values of semi-major axis and eccentricity}
Eqs. (96) and (97) represent interval of values for eccentricity and semi-major
axis, when $-~\mu ~\vec{e}_{R}~/~r^{2}$ is used
as a central acceleration determining osculating orbital elements. However,
we can make time averaging during a period $T$, which was decribed by Eq. (50):
\begin{eqnarray}\label{98}
\langle e \rangle &=& \frac{1}{T} ~ \int_{0}^{T} e(t) ~dt ~, ~~
\langle a \rangle = \frac{1}{T} ~ \int_{0}^{T} a(t) ~dt ~,
\nonumber \\
r^{2} ~ \frac{d f_{c}}{d t} &=&
\sqrt{\mu ~\left ( 1 ~-~ \beta \right )~p_{c}} ~, ~~
\frac{a_{c}^{3}}{T^{2}} = \frac{\mu ~ \left ( 1 ~-~ \beta \right )}{4~\pi^{2}} ~, ~~
r = \frac{p_{c}}{1 ~+~ e_{c} ~\cos f_{c}} ~.
\end{eqnarray}
Eqs. (98) yield
\begin{eqnarray}\label{99}
\langle e \rangle &=& \left ( 1 ~-~ e_{c}^{2} \right ) ^{3/2} ~
     \frac{1}{2~\pi} ~ \int_{0}^{2 \pi}
\frac{e \left ( f_{c} \right )}{ \left ( 1 ~+~ e_{c} ~\cos f_{c} \right )^{2}}
~df_{c} ~,
\nonumber \\
\langle a \rangle &=& \left ( 1 ~-~ e_{c}^{2} \right ) ^{3/2} ~
     \frac{1}{2~\pi} ~ \int_{0}^{2 \pi}
\frac{a \left ( f_{c} \right )}{ \left ( 1 ~+~ e_{c} ~\cos f_{c} \right )^{2}}
~df_{c} ~.
\end{eqnarray}
Using Eqs. (94) and (95), we finally obtain
\begin{eqnarray}\label{100}
\langle e \rangle &=& \left ( 1 - e_{c}^{2} \right ) ^{3/2} ~
     \frac{1}{2 \pi} ~ \int_{0}^{2 \pi}
\frac{\sqrt{\left ( 1 - \beta \right ) ^{2}  e_{c}^{2} + \beta ^{2} -
    2 \beta \left ( 1 - \beta \right )	e_{c}
    \cos f_{c}}}{\left ( 1 + e_{c} \cos f_{c} \right )^{2}}
    ~df_{c} ~,
\nonumber \\
\langle a \rangle &=& a_{c}  \left ( 1 - e_{c}^{2} \right ) ^{3/2} ~
     \frac{1}{2 \pi} ~ \int_{0}^{2 \pi}
    \frac{\left [ 1 + \beta \left ( 1 + e_{c}^{2} + 2 e_{c}
    \cos f_{c} \right ) / \left ( 1 - e_{c}^{2} \right ) \right ]^{-1}}
    {\left ( 1 + e_{c} \cos f_{c} \right )^{2}}
    ~df_{c} ~.
\end{eqnarray}


The following properties can be verified: \\
i) $\langle e \rangle \ge \beta$,
$\langle e \rangle = \beta$ for $e_{c} =$ 0; ~
ii) $\langle e \rangle / e_{c} \ge$ 1,
$\langle e \rangle = e_{c}$ for $\beta =$ 0; \\
iii) $\partial \langle e \rangle / \partial e_{c} >$ 0;~
iv) $\partial \langle e \rangle / \partial \beta >$ 0; \\
v) $\langle a \rangle \ge a_{c} / ( 1 + \beta )$,
$\langle a \rangle = a_{c} / ( 1 + \beta )$ for $e_{c} =$ 0; \\
vi) $\langle a \rangle / a_{c} \le$ 1,
$\langle a \rangle = a_{c}$ for $\beta =$ 0; \\
vii) $\partial \langle a \rangle / \partial e_{c} >$ 0;~
viii) $\partial \langle a \rangle / \partial a_{c} >$ 0;~
ix) $\partial \langle a \rangle / \partial \beta <$ 0.

\subsubsection{Secular changes of semi-major axis and eccentricity}
Summarizing our results, it is possible to calculate secular evolution of
eccentricity and semi-major axis, according to the following prescription.

At first, initial conditions for $a_{\beta}$ and $e_{\beta}$ are calculated:
\begin{eqnarray}\label{101}
\left ( a_{\beta} \right )_{in} &=& a_{P} \left ( 1 - \beta \right )
       \left ( 1 - 2 \beta \frac{1 + e_{P} \cos f_{P}}{1~-~e_{P}^{2}}
       \right ) ^{-1} ~,
\nonumber \\
\left ( e_{\beta} \right )_{in} &=& \sqrt{1 - \frac{1 - e_{P}^{2} - 2 \beta \left (
      1 + e_{P} \cos f_{P} \right )}{\left ( 1 - \beta \right )^{2}}} ~,
\end{eqnarray}
supposing that particle was ejected with zero ejection velocity
from a parent body -- quantities
with subscript ``$P$'' belongs to the parent body's trajectory; as for
more general case, Eqs. (56), (58) and (59) have to be used,
$( a_{\beta} )_{in}$ $=$ $p_{\beta ~in}$ $/$ $( 1 - e^{2}_{\beta ~in} )$,
$( e_{\beta} )_{in}$ $\equiv$ $e_{\beta ~in}$.

As the second step, the set of the following differential equations must
be solved for the above presented initial conditions:
\begin{eqnarray}\label{102}
\frac{d a_{\beta}}{d t} &=& -~ \beta ~ \frac{\mu}{c} ~
       \frac{2 + 3 e_{\beta}^{2}}{a_{\beta}~ \left ( 1 - e_{\beta}^{2} \right )^{3/2}} ~,
\nonumber \\
\frac{d e_{\beta}}{d t} &=& -~ \frac{5}{2}~ \beta ~ \frac{\mu}{c} ~
       \frac{e_{\beta}}{a_{\beta}^{2}~ \left ( 1 - e_{\beta}^{2} \right )^{1/2}} ~.
\end{eqnarray}

Finally, semi-major axis and eccentricity are calculated from:
\begin{eqnarray}\label{103}
a &=& a_{\beta} \left ( 1 - e_{\beta}^{2} \right ) ^{3/2} ~
     \frac{1}{2 \pi} ~ \int_{0}^{2 \pi}
    \frac{\left [ 1 + \beta \left ( 1 + e_{\beta}^{2} + 2 e_{\beta}
    \cos x \right ) / \left ( 1 - e_{\beta}^{2} \right ) \right ]^{-1}}
    {\left ( 1 + e_{\beta} \cos x \right )^{2}}
    ~dx ~,
\nonumber \\
e &=& \left ( 1 - e_{\beta}^{2} \right ) ^{3/2} ~
     \frac{1}{2 \pi} ~ \int_{0}^{2 \pi}
\frac{\sqrt{\left ( 1 - \beta \right ) ^{2}  e_{\beta}^{2} + \beta ^{2} -
    2 \beta \left ( 1 - \beta \right )	e_{\beta}
    \cos x}}{\left ( 1 + e_{\beta} \cos x \right )^{2}}
    ~dx ~.
\end{eqnarray}

The set of equations represented by Eqs. (101)-(103) fully corresponds to
detailed numerical calculations of vectorial equation of motion, if
we are interested in secular evolution of eccentricity and semi-major
axis (supposing $(e_{\beta})_{in} <$ 1 and $e_{\beta}$ does not correspond to
pseudo-circular orbit)
for the case when central acceleration is defined by gravity alone.

It is worth mentioning that instantaneous time derivatives of semi-major axis
and eccentricity may be both positive and negative,
while secular evolution yields that semi-major axis and
eccentricity are decreasing functions of time.

We may mention that a little more simple procedure is obtained when
semi-major axis $a_{\beta}$ is replaced by the quantity
$p_{\beta}$ $=$ $a_{\beta}$ ( $1 - e_{\beta}^{2}$ ). As it was already
mentioned behind Eq. (67), \\
i)  $d p_{\beta} / dt = - 2 \beta ( \mu / c )$
$[ 1 - e_{\beta ~in}^{2} ( p_{\beta} / p_{\beta ~in} ) ^{5/2} ] ^{3/2}$ $/$
$p_{\beta}$, \\
ii) $d e_{\beta} / dt = - ( 5 / 2 ) \beta ( \mu / c )$
$( e_{\beta ~in}^{8/5} / p_{\beta ~in} ^{2} )$
$( 1 - e_{\beta}^{2} ) ^{3/2}$ $/$ $e_{\beta}^{3/5}$. \\
Initial condition for $p_{\beta}$ is given by the first relation of Eq. (59),
e. g., $p_{\beta ~in}$ $=$ $p_{P}$ $/$ $( 1 - \beta )$ for zero ejection
velocity.
The first integral of Eq. (103) is replaced by a simple relation:
$p = p_{\beta} ~ \left ( 1 ~-~ \beta \right )$ (Eq. (90)). \\
(It is important to stress that quantities $ p \equiv \langle p \rangle$, and
$ a \equiv \langle a \rangle$, $ e \equiv \langle e \rangle$, present in Eq. (137),
do not fulfil relation $p = a ( 1 - e^{2} )$.)

\section{P-R effect -- equation of motion to the second order in $v/c$}
We can write, on the basis of Eqs. (48) (Balek and Kla\v{c}ka 2002)
\begin{eqnarray}\label{104}
\frac{d~ \vec{v}}{d ~t} &=& -~ \frac{\mu}{r^{2}} ~ \vec{e}_{R} ~+~
	  \beta ~ \frac{\mu}{r^{2}} ~
	  \left \{  \left [ 1~-~
	  \frac{\vec{v} \cdot \vec{e}_{R}}{c} ~-~ \frac{1}{2} ~
	  \left ( \frac{\vec{v}}{c} \right ) ^{2} \right ] ~ \vec{e}_{R}
	  ~-~ \left ( 1 ~-~ \frac{\vec{v} \cdot \vec{e}_{R}}{c} \right )
	  \frac{\vec{v}}{c} \right \}
\nonumber \\
& & ~-~ \frac{\mu}{r^{2}} ~\left \{ \left [
    \left ( \frac{\vec{v}}{c} \right ) ^{2} ~-~ 4~
    \frac{\mu}{c^{2} r} \right ]  ~\vec{e}_{R}
    ~-~ 4 ~\frac{\vec{v} \cdot \vec{e}_{R}}{c} ~
    \frac{\vec{v}}{c} \right \}  ~-~ 7~ \beta ~
    \frac{\mu}{r^{2}} ~
    \frac{\mu}{c^{2} r} ~\vec{e}_{R} ~,
\end{eqnarray}
where $\mu \equiv G~M_{\odot}$, $\vec{e}_{R} \equiv \vec{e}_{1}$ and
$\beta \equiv \beta_{1}$is a non-dimensional parameter
(``the ratio of radiation pressure force to the gravitational force'';
see also Eq. (43)). The first term in Eq. (104) is Newtonian gravity
for two-body problem, the second term is the Poynting-Robertson effect
in flat spacetime, the third term corresponds to Einstein's correction
to Newtonian gravity and the last term is a sort of interference term
describing the mixing between the effects of gravity and radiation pressure.

\subsection{Secular changes of orbital elements -- radiation pressure
as a part of central acceleration}
We have to use $-~\mu ~( 1~-~\beta )~\vec{e}_{R}~/~r^{2}$ as a central
acceleration determining osculating orbital elements if we want to use
a fact that the elements do not change rapidly during
a motion described by Eq. (104) --
during a time interval $T$, where
$T$ is time interval between passages through two following pericenters /
perihelia:
\begin{eqnarray}\label{105}
\frac{d~ \vec{v}}{d ~t} &=& -~
	  \frac{\mu \left ( 1 - \beta \right )}{r^{2}} ~ \vec{e}_{R} ~-~
	  \beta ~ \frac{\mu}{r^{2}} ~
	  \left \{ \left [
	  \frac{\vec{v} \cdot \vec{e}_{R}}{c} ~+~ \frac{1}{2} ~
	  \left ( \frac{\vec{v}}{c} \right ) ^{2} \right ] ~ \vec{e}_{R}
	  ~+~ \left ( 1 ~-~ \frac{\vec{v} \cdot \vec{e}_{R}}{c} \right )
	  \frac{\vec{v}}{c} \right \}
\nonumber \\
& & ~-~ \frac{\mu}{r^{2}} ~\left \{ \left [
    \left ( \frac{\vec{v}}{c} \right ) ^{2} ~-~ 4~
    \frac{\mu}{c^{2} r} \right ]  ~\vec{e}_{R}
    ~-~ 4 ~\frac{\vec{v} \cdot \vec{e}_{R}}{c} ~
    \frac{\vec{v}}{c} \right \}  ~-~ 7~ \beta ~
    \frac{\mu}{r^{2}} ~
    \frac{\mu}{c^{2} r} ~\vec{e}_{R} ~.
\end{eqnarray}
We can immediately write for components of perturbation acceleration
to Keplerian motion, on the basis of Eq. (105) -- $\beta$ is considered to
be a constant during the motion:
\begin{eqnarray}\label{106}
F_{\beta ~R} &=& -~2~ \beta ~\frac{\mu}{r^{2}} ~\frac{v_{\beta~R}}{c} ~+~
      \beta ~\frac{\mu}{r^{2}} \left \{ \frac{1}{2} ~ \left [
      \left ( \frac{v_{\beta~R}}{c} \right ) ^{2} ~-~
      \left ( \frac{v_{\beta~T}}{c} \right ) ^{2} \right ] ~-~
      7 ~ \frac{\mu}{c^{2} r} \right \} ~+~
\nonumber \\
& &   \frac{\mu}{r^{2}} \left \{ 3 ~
      \left ( \frac{v_{\beta~R}}{c} \right ) ^{2} ~-~
      \left ( \frac{v_{\beta~T}}{c} \right ) ^{2} ~+~
      4~ \frac{\mu}{c^{2} r} \right \} ~, ~~
\nonumber \\
F_{\beta ~T} &=& -~ \beta ~\frac{\mu}{r^{2}} ~\frac{v_{\beta~T}}{c} ~+~
	 \frac{\mu}{r^{2}} ~ \left ( 4 ~+~ \beta \right ) ~
	 \frac{v_{\beta~R} ~ v_{\beta~T}}{c^{2}} ~, ~~
\nonumber \\
F_{\beta ~N} &=& 0 ~,
\end{eqnarray}
where $F_{\beta ~R}$, $F_{\beta ~T}$ and $F_{\beta ~N}$ are radial, transversal
and normal components of perturbation acceleration, and
the two-body problem yields
\begin{eqnarray}\label{107}
v_{\beta~R} &=& \sqrt{\frac{\mu ~( 1 - \beta )}{p_{\beta}}} ~
	e_{\beta} \sin f_{\beta} ~,~
\nonumber \\
v_{\beta~T} &=& \sqrt{\frac{\mu ~( 1 - \beta )}{p_{\beta}}} ~
\left ( 1 + e_{\beta} \cos f_{\beta} \right ) ~,
\end{eqnarray}
where $e_{\beta}$ is osculating eccentricity of the orbit,
$f_{\beta}$ is true anomaly,  $p_{\beta} = a_{\beta} ( 1 - e_{\beta}^{2} )$,
$a_{\beta}$ is semi-major axis.
The important fact that perturbation acceleration is proportional to
$v/c$ ($\ll$ 1) ensures the above metioned small changes of orbital elements
during the time interval $T$.

Perturbation equations of celestial mechanics yield for osculating orbital
elements ($a_{\beta}$ -- semi-major axis; $e_{\beta}$ -- eccentricity;
$i_{\beta}$ -- inclination
(of the orbital plane to the reference frame);
$\Omega_{\beta}$ -- longitude of the ascending node; $\omega_{\beta}$ --
longitude of pericenter / perihelion; $\Theta_{\beta}$
is the position angle of the particle on the orbit, when measured
from the ascending node in the direction of the particle's motion,
$\Theta_{\beta} = \omega_{\beta} + f_{\beta}$):
\begin{eqnarray}\label{108}
\frac{d a_{\beta}}{d t} &=& \frac{2~a_{\beta}}{1~-~e_{\beta}^{2}} ~
      \sqrt{\frac{p_{\beta}}{\mu \left ( 1 ~-~ \beta \right )}} ~
      \left \{
      F_{\beta ~R} ~e_{\beta}~ \sin f_{\beta} +
      F_{\beta ~T} \left ( 1~+~e_{\beta}~ \cos f_{\beta} \right ) \right \} ~,
\nonumber \\
\frac{d e_{\beta}}{d t} &=&
      \sqrt{\frac{p_{\beta}}{\mu \left ( 1 ~-~ \beta \right )}} ~ \left \{
      F_{\beta ~R} ~ \sin f_{\beta} +
      F_{\beta ~T} \left [ \cos f_{\beta} ~+~
     \frac{e_{\beta} +	\cos f_{\beta}}{1 + e_{\beta} \cos f_{\beta}}
	  \right ] \right \} ~,
\nonumber \\
\frac{d i_{\beta}}{d t} &=& \frac{r}{\sqrt{\mu \left ( 1 ~-~ \beta \right ) p_{\beta}}} ~
	    F_{\beta ~N} ~ \cos \Theta_{\beta} ~,
\nonumber \\
\frac{d \Omega_{\beta}}{d t} &=&
      \frac{r}{\sqrt{\mu \left ( 1 ~-~ \beta \right ) p_{\beta}}} ~
      F_{\beta ~N} ~ \frac{\sin \Theta_{\beta}}{\sin i_{\beta}} ~,
\nonumber \\
\frac{d \omega_{\beta}}{d t} &=& -~ \frac{1}{e_{\beta}} ~
      \sqrt{\frac{p_{\beta}}{\mu \left ( 1 ~-~ \beta \right )}} ~ \left \{
      F_{\beta ~R} \cos f_{\beta} - F_{\beta ~T}
      \frac{2 + e_{\beta} \cos f_{\beta}}{1 + e_{\beta} \cos f_{\beta}}
      \sin f_{\beta} \right \} ~-~
\nonumber \\
& &   \frac{r}{\sqrt{\mu \left ( 1 ~-~ \beta \right ) p_{\beta}}} ~
      F_{\beta ~N} ~ \frac{\sin \Theta_{\beta}}{\sin i_{\beta}} ~\cos i_{\beta} ~,
\nonumber \\
\frac{d \Theta_{\beta}}{d t} &=&
      \frac{\sqrt{\mu \left ( 1 ~-~ \beta \right ) p_{\beta}}}{r^{2}} ~-~
      \frac{r}{\sqrt{\mu \left ( 1 ~-~ \beta \right ) p_{\beta}}} ~
      F_{\beta ~N} ~ \frac{\sin \Theta_{\beta}}{\sin i_{\beta}} ~\cos i_{\beta} ~,
\end{eqnarray}
where $r = p_{\beta} / (1 + e_{\beta} \cos f_{\beta})$.

Inserting Eqs. (106) -- (107) into Eq. (108), one easily obtains
\begin{eqnarray}\label{109}
\frac{da_{\beta}}{dt} &=& -~\beta \frac{\mu}{r^{2}} \frac{2 a_{\beta}}{c}
    \frac{1 + e_{\beta}^{2} + 2 e_{\beta} \cos f_{\beta}
    + e_{\beta}^{2}  \sin^{2} f_{\beta}}{1~-~e_{\beta}^{2}} ~+~
\nonumber \\
& & \frac{\mu}{r^{2}} ~ \frac{a_{\beta}}{c^{2}} ~
    \sqrt{\frac{\mu \left ( 1 - \beta \right )}{p_{\beta}}} ~
    \frac{\beta \times X_{a1} ~+~ 6 \times X_{a2}}{1 - e_{\beta}^{2}} ~,
\nonumber \\
X_{a1} &=& \left ( 1 - \frac{14}{1 - \beta} \right ) e_{\beta} \sin f_{\beta} +
       \left ( 1 - \frac{7}{1 - \beta} \right )
       e_{\beta}^{2} \sin \left ( 2 f_{\beta} \right ) +
       e_{\beta}^{3} \sin f_{\beta} ~,
\nonumber \\
X_{a2} &=&
    \left ( 1 + \frac{4 / 3}{1 - \beta} \right )
    e_{\beta} \sin f_{\beta} + \left ( 1 + \frac{2 / 3}{1 - \beta} \right )
    e_{\beta} ^{2} \sin \left ( 2 f_{\beta} \right ) +
    e_{\beta} ^{3} \sin f_{\beta} ~,
\nonumber \\
\frac{de_{\beta}}{dt} &=& -~ \beta \frac{\mu}{r^{2}} \frac{1}{c}  \left (
      2 e_{\beta} + e_{\beta}  \sin^{2} f_{\beta} + 2 \cos f_{\beta} \right ) ~+~
\nonumber \\
& & \frac{\mu}{r^{2}} ~ \frac{1}{c^{2}} ~
    \sqrt{\frac{\mu \left ( 1 - \beta \right )}{p_{\beta}}} ~
    \left ( \frac{1}{2} ~\beta \times X_{e1} ~+~ X_{e2} \right ) ~,
\nonumber \\
X_{e1} &=& - \left ( 1 + \frac{14}{1 - \beta} \right ) \sin f_{\beta} +
       \left ( 1 - \frac{7}{1 - \beta} \right )
       e_{\beta} \sin \left ( 2 f_{\beta} \right ) +
       3 e_{\beta}^{2} \sin f_{\beta} ~,
\nonumber \\
X_{e2} &=& -
    \left ( 1 - \frac{4}{1 - \beta} \right )
    \sin f_{\beta} + \left ( 3 + \frac{2}{1 - \beta} \right )
    e_{\beta} \sin \left ( 2 f_{\beta} \right ) +
    7 e_{\beta} ^{2} \sin f_{\beta} ~,
\nonumber \\
\frac{d i_{\beta}}{dt} &=& 0 ~,
\nonumber \\
\frac{d\Omega_{\beta}}{dt} &=& 0 ~,
\nonumber \\
\frac{d\omega_{\beta}}{dt} &=& -~ \beta \frac{\mu}{r^{2}} \frac{1}{c}
    \frac{1}{e_{\beta}} ~ \left (
    2  - e_{\beta} \cos f_{\beta} \right ) \sin f_{\beta} ~+~
\nonumber \\
& & \frac{\mu}{r^{2}} ~ \frac{1}{c^{2}} ~\frac{1}{e_{\beta}} ~
    \sqrt{\frac{\mu \left ( 1 - \beta \right )}{p_{\beta}}} ~
    \left ( \frac{1}{2} ~\beta \times X_{\omega 1} ~+~ X_{\omega 2} \right ) ~,
\nonumber \\
X_{\omega 1} &=&
       \left ( 1 + \frac{14}{1 - \beta} \right ) \cos f_{\beta} +
       4 e_{\beta} - \left ( 2 - \frac{14}{1 - \beta} \right )
       e_{\beta} \cos ^{2} f_{\beta} +
       e_{\beta}^{2} \cos f_{\beta} ~,
\nonumber \\
X_{\omega 2} &=&
       \left ( 1 - \frac{4}{1 - \beta} \right ) \cos f_{\beta} +
       8 e_{\beta} - \left ( 6 + \frac{4}{1 - \beta} \right )
       e_{\beta} \cos ^{2} f_{\beta} +
       e_{\beta}^{2} \cos f_{\beta} ~,
\nonumber \\
\frac{d \Theta_{\beta}}{dt} &=& \frac{\sqrt{\mu \left ( 1 - \beta \right ) p_{\beta}}}{r^{2}} ~.
\end{eqnarray}

We want to find secular changes of osculating orbital elements up to
$1/c^{2}$.

As for the terms proportional to $1/c^{2}$ in Eq. (109), we may take a time
average ($T$ is time interval between passages through two following pericenters)
in an analytical way
\begin{eqnarray}\label{110}
\langle g \rangle &\equiv& \frac{1}{T}	\int_{0}^{T} g(t) dt =
\frac{\sqrt{\mu ~( 1 - \beta )}}{a_{\beta}^{3/2}} ~
 \frac{1}{2 \pi}  \int_{0}^{2 \pi} g(f_{\beta})
\left ( \frac{df_{\beta}}{dt} \right )^{-1} df_{\beta}
\nonumber \\
&=& \frac{\sqrt{\mu ~( 1 - \beta )}}{a_{\beta}^{3/2}} ~ \frac{1}{2 \pi} \int_{0}^{2 \pi}
g(f_{\beta}) ~\frac{r^{2}}{\sqrt{\mu ~( 1 - \beta ) ~p_{\beta}}}~ df_{\beta}
\nonumber \\
&=& \frac{1}{a_{\beta}^{2}~ \sqrt{1 - e_{\beta}^{2}}} ~\frac{1}{2 \pi} ~
  \int_{0}^{2 \pi} ~ g(f_{\beta}) ~r^{2} ~ df_{\beta} ~,
\end{eqnarray}
assuming non-pseudo-circular orbits and the fact that orbital elements exhibit
only small changes during the time interval $T$;
the second and the third Kepler's laws were used:
$r^{2} ~df_{\beta}/dt = \sqrt{\mu ( 1 - \beta ) p_{\beta}}$ --
conservation of angular momentum,
$a_{\beta}^{3}/T^{2} = \mu ( 1 - \beta ) / (4 \pi^{2})$.
The result is:
\begin{eqnarray}\label{111}
\langle \frac{d a_{\beta}}{dt} \rangle _{II} &=& 0 ~,
\nonumber \\
\langle \frac{d e_{\beta}}{dt} \rangle _{II}  &=& 0 ~,
\nonumber \\
\langle \frac{d\omega_{\beta}}{dt} \rangle _{II} &=&
 \frac{3 \mu ^{3/2}}{c^{2} a_{\beta} ^{5/2} \left ( 1 - e_{\beta} ^{2} \right )}
 ~ \frac{1 - \beta ^{2} / 2}{\left ( 1 - \beta \right ) ^{1/2}} ~.
\end{eqnarray}

Taking into account terms proportional to $1/c$ in Eq. (109), it is inevitable
to use perturbation theory of the second order. Simple averaging represented by
Eq. (110) is not sufficient: the last equation of Eq. (109) yields
\begin{eqnarray}\label{112}
\frac{d f_{\beta}}{dt} &=& \frac{\sqrt{\mu \left ( 1 - \beta \right ) p_{\beta}}}{r^{2}}
	       ~-~ \frac{d \omega_{\beta}}{dt}
\nonumber \\
	 &\equiv& \frac{H_{\beta}}{r^{2}} ~-~
	      \frac{d \omega_{\beta}}{dt} ~.
\end{eqnarray}
Eq. (146) holds due to the fact, that $F_{\beta ~ N} =$ 0 -- in reality,
equation for $d \Theta _{\beta} / d t$ presented in Eq. (108) has to be used.
Instead of Eq. (110), we have more precise method of averaging, now. We can
write, using Eq. (112):
\begin{eqnarray}\label{113}
\langle g \rangle &\equiv& \frac{1}{T}	\int_{0}^{T} g(t) dt =
\nonumber \\
&=& \left \{ \int_{0}^{2 \pi} \left ( \frac{H_{\beta}}{r^{2}} ~-~
    \frac{d \omega_{\beta}}{dt} \right ) ^{-1} d f_{\beta} \right \} ^{-1}
    \times
    \int_{0}^{2 \pi} \left ( \frac{H_{\beta}}{r^{2}} ~-~
    \frac{d \omega_{\beta}}{dt} \right ) ^{-1} g(f_{\beta})~d f_{\beta} ~.
\end{eqnarray}

We are interested in secular changes of orbital elements up to the
order $1/c^{2}$. Within this accuracy we can write for the terms on the
right-hand side of Eq. (113)
\begin{eqnarray}\label{114}
\left ( \frac{H_{\beta}}{r^{2}} ~-~
    \frac{d \omega_{\beta}}{dt} \right ) ^{-1} &=&
    \frac{r^{2}}{H_{\beta}} \left \{ 1 ~+~
    \frac{r^{2}}{H_{\beta}} ~ \frac{d \omega_{\beta}}{dt} ~+~
    \left ( \frac{r^{2}}{H_{\beta}} ~ \frac{d \omega_{\beta}}{dt} \right )
    ^{2} \right \} ~,
\end{eqnarray}
\begin{eqnarray}\label{115}
LHS &\equiv&
\left \{ \int_{0}^{2 \pi} \left ( \frac{H_{\beta}}{r^{2}} ~-~
\frac{d \omega_{\beta}}{dt} \right ) ^{-1} ~d f_{\beta} \right \} ^{-1} ~,
\nonumber \\
LHS &=& \left ( \int_{0}^{2 \pi}  \frac{r^{2}}{H_{\beta}} d f_{\beta}
    \right ) ^{-1} \times
    \left \{ 1 ~-~
    \left ( \int_{0}^{2 \pi} \frac{r^{2}}{H_{\beta}} d f_{\beta}
    \right ) ^{-1} \times \right .
\nonumber \\
& &  \left . \left [
     \int_{0}^{2 \pi}
     \frac{r^{2}}{H_{\beta}} ~ \left (	\frac{r^{2}}{H_{\beta}} ~
     \frac{d \omega_{\beta}}{dt} \right ) d f_{\beta} ~+~
     \int_{0}^{2 \pi}
     \frac{r^{2}}{H_{\beta}} ~ \left (	\frac{r^{2}}{H_{\beta}} ~
     \frac{d \omega_{\beta}}{dt} \right ) ^{2} d f_{\beta} \right ] ~+~
     \right .
\nonumber \\
& & \left .
     \left ( \int_{0}^{2 \pi} \frac{r^{2}}{H_{\beta}} d f_{\beta}
     \right ) ^{-2} \times \left [
     \int_{0}^{2 \pi}
     \frac{r^{2}}{H_{\beta}} ~ \left (	\frac{r^{2}}{H_{\beta}} ~
     \frac{d \omega_{\beta}}{dt} \right ) d f_{\beta} \right ] ^{2}
     \right \} ~.
\end{eqnarray}

We have to find osculating orbital elements present on the right-hand sides
of Eq. (109) in terms of true anomaly $f_{\beta}$ to the required accuracy
in $1/c^{2}$. Differentiation of the relation
$p_{\beta}$ $=$ $a_{\beta}$ ( $1 - e_{\beta}^{2}$ ) yields, using Eq. (109)
for $d a_{\beta} / dt$ and $d e_{\beta} / dt$,
\begin{eqnarray}\label{116}
\frac{dp_{\beta}}{dt} &=& -~2  ~\beta ~\frac{\mu}{c} ~\frac{p_{\beta}}{r^{2}} ~+~
      \frac{\mu}{c^{2}} ~\frac{a_{\beta}}{r^{2}} ~
      \sqrt{\frac{\mu \left ( 1 - \beta \right )}{p_{\beta}}} \times
\nonumber \\
& &   \left \{ \left ( \beta \times X_{a1} ~+~ 6 \times X_{a2} \right )
      ~-~ 2 e_{\beta}
    \left ( \frac{1}{2} ~\beta \times X_{e1} ~+~ X_{e2} \right ) \right \} ~,
\end{eqnarray}
and, using Eqs. (109), (112) and (114), we obtain
\begin{eqnarray}\label{117}
\frac{dp_{\beta}}{df_{\beta}} &=& -~2  ~\beta ~\frac{\mu}{c} ~
      \frac{p_{\beta}}{H_{\beta}} ~
      \left ( 1 ~-~
      \beta ~ \frac{\mu}{c} ~\frac{1}{H_{\beta}} ~
      \frac{2  - e_{\beta} \cos f_{\beta}}{e_{\beta}} ~
      \sin f_{\beta} \right ) ~+~
      \frac{\mu}{c^{2}} ~\frac{a_{\beta}}{p_{\beta}} ~
      \times
\nonumber \\
& &   \left \{ \left ( \beta \times X_{a1} ~+~ 6 \times X_{a2} \right )
      ~-~ 2 e_{\beta}
    \left ( \frac{1}{2} ~\beta \times X_{e1} ~+~ X_{e2} \right ) \right \} ~,
\end{eqnarray}
\begin{eqnarray}\label{118}
\frac{de_{\beta}}{df_{\beta}} &=& - ~\beta ~\frac{\mu}{c} ~
\frac{2 e_{\beta} + e_{\beta}  \sin^{2} f_{\beta} + 2 \cos f_{\beta}}{H_{\beta}}
   ~ \left ( 1 ~-~ \beta ~ \frac{\mu}{c} ~\frac{1}{H_{\beta}} ~
     \frac{2  - e_{\beta} \cos f_{\beta}}{e_{\beta}} ~
     \sin f_{\beta} \right ) ~+~
\nonumber \\
& &  \frac{\mu}{c^{2}} ~\frac{1}{p_{\beta}} ~
    \left ( \frac{1}{2} ~\beta \times X_{e1} ~+~ X_{e2} \right ) ~.
\end{eqnarray}

Now, if we restrict ourselves to the first order in $1/c$, Eqs. (117)-(118)
reduce to
\begin{eqnarray}\label{119}
\frac{dp_{\beta}}{df_{\beta}} &=& -~2  ~\frac{\beta}{\sqrt{1 - \beta}} ~
      \frac{\sqrt{\mu}}{c} ~\sqrt{p_{\beta}} ~,
\end{eqnarray}
\begin{eqnarray}\label{120}
\frac{de_{\beta}}{df_{\beta}} &=& - ~\frac{\beta}{\sqrt{1 - \beta}} ~
      \frac{\sqrt{\mu}}{c} ~ \frac{2 e_{\beta} + e_{\beta}
      \sin^{2} f_{\beta} + 2 \cos f_{\beta}}{\sqrt{p_{\beta}}} ~.
\end{eqnarray}
Eqs. (119)-(120) can be easily solved:
\begin{eqnarray}\label{121}
\sqrt{p_{\beta}} &=& \sqrt{p_{\beta 0}}
      ~-~ \frac{\beta}{\sqrt{1 - \beta}} ~\frac{\sqrt{\mu}}{c} ~ f_{\beta} ~,
\end{eqnarray}
\begin{eqnarray}\label{122}
e_{\beta} &=& e_{\beta 0} ~-~ \frac{\beta}{\sqrt{1 - \beta}} ~
      \frac{\sqrt{\mu}}{c} ~ \frac{1}{\sqrt{p_{\beta 0}}} ~
      \left \{ 2 \sin f_{\beta} + \frac{e_{\beta 0}}{2} ~
      \left ( 5 f_{\beta} - \frac{1}{2} \sin  2 f_{\beta}
      \right ) \right \} ~,
\end{eqnarray}
where initial values $p_{\beta 0}$, $e_{\beta 0}$ correspond to
$f_{\beta 0} =$ 0.

On the basis of Eq. (121), we can immediately write solution
of Eq. (117):
\begin{eqnarray}\label{123}
p_{\beta} &=& p_{\beta 0}
    ~-~ 2 ~\frac{\beta}{\sqrt{1 - \beta}} ~\frac{\sqrt{\mu}}{c} ~
    \sqrt{p_{\beta 0}} ~ f_{\beta} ~+~ \frac{\mu}{c^{2}} ~
    \left ( \frac{\beta ^{2}}{1 - \beta} \times X_{p1} + X_{p2} \right ) ~,
\nonumber \\
X_{p1} &=& f_{\beta}^{2} ~+~ \frac{4}{e_{\beta 0}} ~ \left ( 1 -
       \cos f_{\beta} \right ) ~-~ \frac{1}{2} ~ \left ( 1 -
       \cos 2 f_{\beta} \right ) ~,
\nonumber \\
X_{p2} &=& 2~ \left ( 4 ~+~ \beta \right ) ~ e_{\beta 0}~
       \left ( 1 - \cos f_{\beta} \right ) ~.
\end{eqnarray}

On the basis of Eqs. (121) -- (122), we can rewrite Eq. (118) to the form
\begin{eqnarray}\label{124}
\frac{de_{\beta}}{df_{\beta}} &=& - ~\frac{\beta}{\sqrt{1 - \beta}} ~
      \frac{\sqrt{\mu}}{c} ~
      \frac{\left ( 5 - \cos 2 f_{\beta} \right ) e_{\beta 0} / 2 +
      2 \cos f_{\beta}}{\sqrt{p_{\beta 0}}} ~-~
      \frac{\beta ^{2}}{1 - \beta} ~ \frac{\mu}{c^{2}} ~\frac{1}{p_{\beta 0}} ~
      \times X_{e3} ~+~
\nonumber \\
& &  \frac{\mu}{c^{2}} ~\frac{1}{p_{\beta 0}} ~
    \left ( \frac{1}{2} ~\beta \times X_{e1} ~+~ X_{e2} \right ) ~,
\nonumber \\
X_{e3} &=& - ~ \frac{15}{4} ~ e_{\beta 0} ~f_{\beta} ~+~
       \frac{3}{4} ~ e_{\beta 0} ~f_{\beta} \cos ( 2 f_{\beta} ) ~+~
       2 f_{\beta} \sin f_{\beta} ~-~\frac{21}{2} ~\sin f_{\beta} ~+~
\nonumber \\
& &    \left ( \frac{7}{4} ~ e_{\beta 0} - \frac{2}{e_{\beta 0}} \right )
       \sin ( 2 f_{\beta} ) ~+~ \frac{3}{2} ~ \sin ( 3 f_{\beta} ) ~-~
       \frac{5}{16} ~ e_{\beta 0} ~ \sin ( 4 f_{\beta} ) ~,
\end{eqnarray}
where $e_{\beta 0}$ has to be inserted into expressions for
$X_{e1}$ and $X_{e2}$ (see Eq. (109)) instead of $e_{\beta}$. Solution
of Eq. (124) is:
\begin{eqnarray}\label{125}
e_{\beta} &=& e_{\beta 0} ~-~ \frac{\beta}{\sqrt{1 - \beta}} ~
      \frac{\sqrt{\mu}}{c} ~ \frac{1}{\sqrt{p_{\beta 0}}} ~
      \left \{ 2 \sin f_{\beta} + \frac{e_{\beta 0}}{2} ~
      \left ( 5 f_{\beta} - \frac{1}{2} \sin  2 f_{\beta}
      \right ) \right \} ~+~
\nonumber \\
& & \frac{\mu}{c^{2}} ~\frac{1}{p_{\beta 0}} ~
    \left ( \frac{1}{2} ~\beta \times X_{inte1} ~+~ X_{inte2} \right )
    ~-~ \frac{\beta ^{2}}{1 - \beta} ~ \frac{\mu}{c^{2}} ~\frac{1}{p_{\beta 0}}
    \times X_{inte3} ~,
\nonumber \\
X_{inte1} &=& - \left ( 1 + \frac{14}{1 - \beta} \right )
	  \left ( 1 - \cos f_{\beta} \right ) ~+~
\nonumber \\
& &    \frac{1}{2}
       \left ( 1 - \frac{7}{1 - \beta} \right )
       e_{\beta 0} \left ( 1 - \cos 2 f_{\beta} \right ) ~+~
       3 e_{\beta 0}^{2} \left ( 1 - \cos f_{\beta} \right ) ~,
\nonumber \\
X_{inte2} &=& -
       \left ( 1 - \frac{4}{1 - \beta} \right )
       \left ( 1 - \cos f_{\beta} \right )  ~+~
\nonumber \\
& &    \frac{1}{2} \left ( 3 + \frac{2}{1 - \beta} \right )
       e_{\beta 0} \left ( 1 - \cos 2 f_{\beta} \right ) ~+~
       7 e_{\beta 0} ^{2} \left ( 1 - \cos f_{\beta} \right )  ~,
\nonumber \\
X_{inte3} &=& - ~ \frac{15}{8} ~ e_{\beta 0} ~f_{\beta}^{2} ~-~
       \frac{3}{16} ~ e_{\beta 0} ~\left ( 1 -
       \cos 2 f_{\beta} \right ) ~+~
       \frac{3}{8} ~ e_{\beta 0} ~f_{\beta} \sin ( 2 f_{\beta} ) ~+~
       2 \sin f_{\beta}
\nonumber \\
& &    ~-~ 2 f_{\beta} \cos f_{\beta} ~-~
       \frac{21}{2} ~\left ( 1 - \cos f_{\beta} \right ) ~+~
       \left ( \frac{7}{8} ~ e_{\beta 0} - \frac{1}{e_{\beta 0}} \right )
       \left ( 1 - \cos 2 f_{\beta} \right )
\nonumber \\
& &    ~+~ \frac{1}{2} ~ \left ( 1 - \cos 3 f_{\beta} \right ) ~-~
       \frac{5}{64} ~ e_{\beta 0} ~
       \left ( 1 - \cos 4 f_{\beta} \right ) ~.
\end{eqnarray}
Semi-major axis $a_{\beta}$ can be obtained from the relation
$p_{\beta}$ $=$ $a_{\beta}$ ( $1 - e_{\beta}^{2}$ ), using
Eqs. (123) and (125).

Let us calculate secular change of $\omega _{\beta}$ to the order
$1/c^{2}$, generated by osculating
orbital change proportional to $1/c$ -- compare Eqs. (109) and (111).
Putting Eq. (114) into Eq. (113), we can write to the required
accuracy,
\begin{eqnarray}\label{126}
\langle \frac{d \omega_{\beta}}{dt} \rangle _{I}
&=& \left \{ \int_{0}^{2 \pi} \left ( \frac{H_{\beta}}{r^{2}} ~-~
    \frac{d \omega_{\beta}}{dt} \right ) ^{-1} d f_{\beta} \right \} ^{-1}
    \int_{0}^{2 \pi}
    \frac{r^{2}}{H_{\beta}} \left ( 1 ~+~
    \frac{r^{2}}{H_{\beta}} ~ \frac{d \omega_{\beta}}{dt} \right ) ~
    \frac{d \omega_{\beta}}{dt} ~d f_{\beta} ~,
\end{eqnarray}
since time derivative of $\omega _{\beta}$ is proportional to $1/c$, and,
thus, the third term on the right-hand side of Eq. (148) would yield a higher
order than $1/c^{2}$ in the last integral.

The last integral in Eq. (126) consists of two terms:
\begin{eqnarray}\label{127}
I_{\omega} &\equiv& \int_{0}^{2 \pi}
    \frac{r^{2}}{H_{\beta}} \left ( 1 ~+~
    \frac{r^{2}}{H_{\beta}} ~ \frac{d \omega_{\beta}}{dt} \right ) ~
    \frac{d \omega_{\beta}}{dt} ~d f_{\beta} \equiv
    I_{\omega 1} ~+~ I_{\omega 2} ~,
\nonumber \\
I_{\omega 1} &\equiv&
    \int_{0}^{2 \pi} \frac{r^{2}}{H_{\beta}}
    \frac{r^{2}}{H_{\beta}} ~ \frac{d \omega_{\beta}}{dt} ~
    \frac{d \omega_{\beta}}{dt} ~d f_{\beta} ~,
\nonumber \\
I_{\omega 2} &\equiv&
    \int_{0}^{2 \pi} \frac{r^{2}}{H_{\beta}}
    \frac{d \omega_{\beta}}{dt} ~d f_{\beta} ~.
\end{eqnarray}
Inserting relation for $d \omega_{\beta} / dt$ from Eq. (109),
we can immediately write for $I_{\omega 1}$, within the required accuracy:
\begin{eqnarray}\label{128}
I_{\omega 1} &=& \left ( -~ \beta ~\frac{\mu}{c} \right ) ^{2} ~
    \int_{0}^{2 \pi} \left \{ \frac{\left ( 2  - e_{\beta} \cos f_{\beta}
    \right ) \sin f_{\beta}}{H_{\beta} e_{\beta}} \right \} ^{2}
    ~d f_{\beta}
\nonumber \\
&=& \left ( -~ \beta ~\frac{\mu}{c} \right ) ^{2} ~
    \int_{0}^{2 \pi} \left \{ \frac{\left ( 2  - e_{\beta 0} \cos f_{\beta}
    \right ) \sin f_{\beta}}{H_{\beta 0} e_{\beta 0}} \right \} ^{2}
    ~d f_{\beta}
\nonumber \\
&=& \frac{\beta ^{2}}{1 - \beta} ~\frac{\mu}{c^{2}} ~\frac{2 \pi}{p_{\beta 0}} ~
    \left ( \frac{2}{e_{\beta 0}^{2}} ~+~ \frac{1}{8} \right ) ~,
\end{eqnarray}
where results of Eqs. (121) and (122) were used, and,
$H_{\beta}$ $=$ $\sqrt{\mu ( 1 - \beta ) p_{\beta}}$.
\begin{eqnarray}\label{129}
I_{\omega 2} &=& -~ \beta ~\frac{\mu}{c} ~
    \int_{0}^{2 \pi} \frac{\left ( 2  - e_{\beta} \cos f_{\beta}
    \right ) \sin f_{\beta}}{H_{\beta} e_{\beta}}
    ~d f_{\beta}
\equiv I_{\omega 21} ~+~ I_{\omega 22} ~,
\nonumber \\
I_{\omega 21} &\equiv& \beta ~\frac{\mu}{c} ~
    \int_{0}^{2 \pi} \frac{\cos f_{\beta} \sin f_{\beta}}{H_{\beta}}
    ~d f_{\beta} ~,
\nonumber \\
I_{\omega 22} &\equiv&	-~ 2 ~ \beta ~\frac{\mu}{c} ~
    \int_{0}^{2 \pi} \frac{\sin f_{\beta}}{H_{\beta} e_{\beta}}
    ~d f_{\beta} ~.
\end{eqnarray}
As for calculation of $I_{\omega 21}$, we need to consider that
$H_{\beta}$ $=$ $\sqrt{\mu ( 1 - \beta ) p_{\beta}}$ and Eq. (121).
We obtain:
\begin{eqnarray}\label{130}
I_{\omega 21} &=& \beta ~\frac{\mu}{c} ~
    \int_{0}^{2 \pi} \frac{\cos f_{\beta} \sin f_{\beta}}
    {\sqrt{\mu ( 1 - \beta ) p_{\beta}}} ~d f_{\beta}
\nonumber \\
&=& \beta ~\frac{\mu}{c} ~ \frac{1}{H_{\beta 0}} ~
    \int_{0}^{2 \pi} \left ( 1 ~+~ \frac{\beta}{\sqrt{1 - \beta}} ~
    \frac{\sqrt{\mu}}{c \sqrt{p_{\beta 0}}} ~f_{\beta} \right )
    \cos f_{\beta} \sin f_{\beta} ~d f_{\beta} =
\nonumber \\
&=& -~ \frac{\beta ^{2}}{1 - \beta} ~\frac{\mu}{c^{2}}~
    \frac{\pi / 2}{p_{\beta 0}} ~.
\end{eqnarray}
Calculation of $I_{\omega 22}$ is analogous to calculation of $I_{\omega 21}$,
but we have to use Eqs. (121) and (122) simultaneously. The result is:
\begin{eqnarray}\label{131}
I_{\omega 22} &=& \frac{\beta ^{2}}{1 - \beta} ~\frac{\mu}{c^{2}}~
    \frac{4 \pi}{p_{\beta 0}} ~
    \left ( \frac{7}{2~ e_{\beta 0}} ~-~ \frac{1}{e_{\beta 0} ^{2}} \right ) ~.
\end{eqnarray}
Eqs. (129)-(131) yield
\begin{eqnarray}\label{132}
I_{\omega 2} &=& \frac{\beta ^{2}}{1 - \beta} ~\frac{\mu}{c^{2}}~
    \frac{4 \pi}{p_{\beta 0}} ~\left ( -~ \frac{1}{8} ~+~
    \frac{7}{2~ e_{\beta 0}} ~-~ \frac{1}{e_{\beta 0} ^{2}} \right ) ~.
\end{eqnarray}
Eqs. (127), (128) and (132) yield
\begin{eqnarray}\label{133}
I_{\omega} &=& \frac{\beta ^{2}}{1 - \beta} ~\frac{\mu}{c^{2}}~
    \frac{4 \pi}{p_{\beta 0}} ~\left ( -~ \frac{1}{16} ~+~
    \frac{7}{2~ e_{\beta 0}} \right ) ~.
\end{eqnarray}

We have calculated the last integral in Eq. (126) -- it is represented
by Eq. (133). Since it is proportional to $1/c^{2}$, the first
integral in Eq. (126) has to be proportional to $1/c^{0}$. Thus,
the first integral in Eq. (126) reduces to
\begin{eqnarray}\label{134}
I_{\omega D} &\equiv& \int_{0}^{2 \pi} \left ( \frac{H_{\beta}}{r^{2}} ~-~
    \frac{d \omega_{\beta}}{dt} \right ) ^{-1} d f_{\beta} ~,
\nonumber \\
I_{\omega D} &\rightarrow& \int_{0}^{2 \pi} \frac{r^{2}}{H_{\beta 0}} d f_{\beta}
  \equiv \frac{1}{H_{\beta 0}} ~\int_{0}^{2 \pi} \left \{
  \frac{p_{\beta 0}}{1 + e_{\beta 0} \cos f_{\beta}} \right \} ^{2} d f_{\beta} =
\nonumber \\
&=& \frac{p_{\beta 0}^{3/2}}{\sqrt{\mu \left ( 1 - \beta \right )}} ~
    \frac{2 ~\pi}{\left ( 1 - e_{\beta 0}^{2} \right )^{3/2}} ~.
\end{eqnarray}

Finally, Eqs. (126), (133) and (134) lead to
\begin{eqnarray}\label{135}
\langle \frac{d \omega_{\beta}}{dt} \rangle _{I} &=&
    \frac{\beta ^{2}}{\sqrt{1 - \beta}} ~\frac{\mu ^{3/2}}{c^{2}} ~
    \frac{\left ( 1 - e_{\beta 0}^{2} \right )^{3/2}}{p_{\beta 0} ^{5/2}} ~
    \left ( -~ \frac{1}{8} ~+~
    \frac{7}{e_{\beta 0}} \right ) ~.
\end{eqnarray}

Total shift of pericenter / perihelion is given as a sum of
Eqs. (111) and (135). Changing the index $\beta 0$ into $\beta$ and using
$p_{\beta}$ $=$ $a_{\beta}$ $( 1 - e_{\beta}^{2} )$,
we finally receive:
\begin{eqnarray}\label{136}
\langle \frac{d\omega_{\beta}}{dt} \rangle &=&
 \frac{3 \mu ^{3/2}}{c^{2} a_{\beta} ^{5/2} \left ( 1 - e_{\beta} ^{2} \right )}
 ~ \frac{1 + \beta ^{2} \left ( - 13 / 8 + 7 / e_{\beta} \right ) / 3}
   {\left ( 1 - \beta \right ) ^{1/2}} ~.
\end{eqnarray}

It can be easily verified that $< d \omega_{\beta} ~/~ d t >$ is \\
i) an increasing function of $\beta$, \\
ii) the perihelion circulates in a positive direction, \\
iii) the rate of the advancement of perihelion is not bounded for
$\beta \rightarrow$ 1.

Let us calculate secular change of $a_{\beta}$ to the order
$1/c^{2}$, generated by osculating
orbital change proportional to $1/c$ -- compare Eqs. (109) and (111).
Putting Eq. (114) into Eq. (113), we can write to the required
accuracy,
\begin{eqnarray}\label{137}
\langle \frac{d a_{\beta}}{dt} \rangle _{I}
&=& \left \{ \int_{0}^{2 \pi} \left ( \frac{H_{\beta}}{r^{2}} ~-~
    \frac{d \omega_{\beta}}{dt} \right ) ^{-1} d f_{\beta} \right \} ^{-1}
    \int_{0}^{2 \pi}
    \frac{r^{2}}{H_{\beta}} \left ( 1 ~+~
    \frac{r^{2}}{H_{\beta}} ~ \frac{d \omega_{\beta}}{dt} \right ) ~
    \frac{d a_{\beta}}{dt} ~d f_{\beta} ~,
\end{eqnarray}
since time derivatives of $a _{\beta}$ and $\omega _{\beta}$ are
proportional to $1/c$, and,
thus, the third term on the right-hand side of Eq. (114) would yield a higher
order than $1/c^{2}$ in the last integral.

The last integral in Eq. (137) consists of two terms:
\begin{eqnarray}\label{138}
I_{a} &\equiv& \int_{0}^{2 \pi}
    \frac{r^{2}}{H_{\beta}} \left ( 1 ~+~
    \frac{r^{2}}{H_{\beta}} ~ \frac{d \omega_{\beta}}{dt} \right ) ~
    \frac{d a_{\beta}}{dt} ~d f_{\beta} \equiv
    I_{a 1} ~+~ I_{a 2} ~,
\nonumber \\
I_{a 1} &\equiv&
    \int_{0}^{2 \pi} \frac{r^{2}}{H_{\beta}}
    \frac{r^{2}}{H_{\beta}} ~ \frac{d \omega_{\beta}}{dt} ~
    \frac{d a_{\beta}}{dt} ~d f_{\beta} ~,
\nonumber \\
I_{a 2} &\equiv&
    \int_{0}^{2 \pi} \frac{r^{2}}{H_{\beta}}
    \frac{d a_{\beta}}{dt} ~d f_{\beta} ~.
\end{eqnarray}
Inserting relations for $d a_{\beta} / dt$ and $d \omega_{\beta} / dt$
from Eq. (109),
we can immediately write for $I_{a 1}$, within the required accuracy:
\begin{eqnarray}\label{139}
I_{a 1} &=& \left ( \beta ~\frac{\mu}{c} \right ) ^{2} ~
    \int_{0}^{2 \pi} \frac{\left ( 2  - e_{\beta} \cos f_{\beta}
    \right ) \sin f_{\beta}}{H_{\beta} e_{\beta}} ~
    \frac{2 a_{\beta} \left (
    1 + e_{\beta}^{2} + 2 e_{\beta} \cos f_{\beta}
    + e_{\beta}^{2}  \sin^{2} f_{\beta} \right )}{H_{\beta}
    \left ( 1~-~e_{\beta}^{2} \right )}
    ~d f_{\beta}
\nonumber \\
&=& \left ( \beta ~\frac{\mu}{c} \right ) ^{2} ~
    \int_{0}^{2 \pi} \frac{\left ( 2  - e_{\beta 0} \cos f_{\beta}
    \right ) \sin f_{\beta}}{H_{\beta 0} e_{\beta 0}} ~
    \frac{1 + e_{\beta 0}^{2} + 2 e_{\beta 0} \cos f_{\beta}
    + e_{\beta 0}^{2}  \sin^{2} f_{\beta}}{H_{\beta 0}
    \left ( 1~-~e_{\beta 0}^{2} \right ) / \left ( 2 a_{\beta 0}  \right )}
    ~d f_{\beta}
\nonumber \\
&=& 0 ~,
\end{eqnarray}
where results of Eqs. (121) and (122) were used.

\begin{eqnarray}\label{140}
I_{a 2} &=& -~ \beta ~\frac{\mu}{c} ~ \int_{0}^{2 \pi}
    \frac{2 a_{\beta} \left (
    1 + e_{\beta}^{2} + 2 e_{\beta} \cos f_{\beta}
    + e_{\beta}^{2}  \sin^{2} f_{\beta} \right )}{H_{\beta}
    \left ( 1~-~e_{\beta}^{2} \right )} ~d f_{\beta} ~.
\end{eqnarray}
Considering that $H_{\beta}$ $=$ $\sqrt{\mu ( 1 - \beta ) p_{\beta}}$ and
Eqs. (121) and (122), we have
\begin{eqnarray}\label{141}
\frac{1}{H_{\beta}} &=& \frac{1}{H_{\beta 0}} ~
    \left ( 1 ~+~ \frac{\beta}{\sqrt{1 - \beta}} ~
    \frac{\sqrt{\mu}}{c \sqrt{p_{\beta 0}}} ~f_{\beta} \right ) ~,
\nonumber \\
e_{\beta}^{2} &=& e_{\beta 0}^{2} ~-~ 2 e_{\beta 0} ~
      \frac{\beta}{\sqrt{1 - \beta}} ~
      \frac{\sqrt{\mu}}{c} ~ \frac{1}{\sqrt{p_{\beta 0}}}
      \left \{ 2 \sin f_{\beta} + \frac{e_{\beta 0}}{2} ~
      \left ( 5 f_{\beta} - \frac{1}{2} \sin  2 f_{\beta}
      \right ) \right \} ~,
\nonumber \\
\frac{1}{1~-~e_{\beta}^{2}} &=& \frac{1}{1~-~e_{\beta 0}^{2}} ~\left \{ 1 ~-~
      \frac{2 e_{\beta 0}}{1~-~e_{\beta 0}^{2}} ~
      \frac{\beta}{\sqrt{1 - \beta}} ~
      \frac{\sqrt{\mu}}{c} ~ \frac{1}{\sqrt{p_{\beta 0}}} \times \right .
\nonumber \\
& &   \left .	\left [ 2 \sin f_{\beta} + \frac{e_{\beta 0}}{2} ~
      \left ( 5 f_{\beta} - \frac{1}{2} \sin  2 f_{\beta}
      \right ) \right ] \right \} ~,
\nonumber \\
a_{\beta} &=& p_{\beta} ~ \frac{1}{1~-~e_{\beta}^{2}} =
	  \left ( p_{\beta 0} ~-~ 2 ~
	  \frac{\beta}{\sqrt{1 - \beta}} ~\frac{\sqrt{\mu}}{c} ~
	  \sqrt{p_{\beta 0}} ~ f_{\beta} \right ) ~
	  \frac{1}{1~-~e_{\beta}^{2}}
\nonumber \\
&=& a_{\beta 0} ~\left \{ 1 ~-~ 2~
      \frac{\beta}{\sqrt{1 - \beta}} ~
      \frac{\sqrt{\mu}}{c} ~ \frac{1}{\sqrt{p_{\beta 0}}} ~f_{\beta} ~-~
      \right .
\nonumber \\
& &   \left . \frac{2 e_{\beta 0}}{1~-~e_{\beta 0}^{2}} ~
      \frac{\beta}{\sqrt{1 - \beta}} ~
      \frac{\sqrt{\mu}}{c} ~ \frac{1}{\sqrt{p_{\beta 0}}}
      \left [ 2 \sin f_{\beta} + \frac{e_{\beta 0}}{2} ~
      \left ( 5 f_{\beta} - \frac{1}{2} \sin  2 f_{\beta}
      \right ) \right ] \right \} ~.
\end{eqnarray}
Inserting results of Eq. (141) into Eq. (140), one finally obtains
\begin{eqnarray}\label{142}
I_{a 2} &=& -~ \beta ~\frac{\mu}{c} ~
  \frac{4 \pi a_{\beta 0}}{H_{\beta 0} \left ( 1~-~e_{\beta 0}^{2} \right )}
  \left \{ 1 + \frac{3}{2} e_{\beta 0}^{2} -
  \frac{\beta}{\sqrt{1 - \beta}} ~\frac{\sqrt{\mu}}{c \sqrt{p_{\beta 0}}}
  \times \right .
\nonumber \\
& & \left .
    \frac{\pi + \left ( 10 \pi + 3 \right ) e_{\beta 0}^{2} +
    \left ( 9 \pi + 2 \right ) e_{\beta 0}^{4}}{1 - e_{\beta 0}^{2}}
    \right \} ~.
\end{eqnarray}

We have calculated the last integral in Eq. (137) -- it is represented
by Eq. (142). Since its dominant part is proportional to $1/c$, the first
integral in Eq. (137) has to be proportional to $1/c^{1}$. Thus,
the first integral in Eq. (137) is given, according to Eq. (115), as
\begin{eqnarray}\label{143}
LHS &\equiv&
\left \{ \int_{0}^{2 \pi} \left ( \frac{H_{\beta}}{r^{2}} ~-~
\frac{d \omega_{\beta}}{dt} \right ) ^{-1} ~d f_{\beta} \right \} ^{-1} ~,
\nonumber \\
LHS &=& \left ( \int_{0}^{2 \pi}  \frac{r^{2}}{H_{\beta}} d f_{\beta}
    \right ) ^{-1}
    \left \{ 1 ~-~
    \left ( \int_{0}^{2 \pi} \frac{r^{2}}{H_{\beta}} d f_{\beta}
    \right ) ^{-1}
     \int_{0}^{2 \pi}
     \frac{r^{2}}{H_{\beta}} ~ \left (	\frac{r^{2}}{H_{\beta}} ~
     \frac{d \omega_{\beta}}{dt} \right ) d f_{\beta} \right \} ~.
\end{eqnarray}
It can be easily verified that the last integral in Eq. (143) equals zero,
within the required accuracy, and
\begin{eqnarray}\label{144}
LHS &=& \left ( \int_{0}^{2 \pi}  \frac{r^{2}}{H_{\beta}} d f_{\beta}
    \right ) ^{-1} =
    \frac{\sqrt{\mu \left ( 1 - \beta \right )}}{2 \pi a_{\beta 0} ^{3/2}}
    \left \{ 1 + \frac{\beta}{\sqrt{1 - \beta}} \frac{\sqrt{\mu}}{c}
    \frac{1 - e_{\beta 0}^{2}}{\sqrt{a_{\beta 0}}} I_{LHS} \right \} ~,
\nonumber \\
I_{LHS} &=& \frac{5}{2 \pi} \int_{0}^{2 \pi}
	\frac{x}{\left ( 1 + e_{\beta 0} \cos x \right ) ^{3}} ~ d x ~-~
	\frac{2}{2 \pi} \int_{0}^{2 \pi}
	\frac{x}{\left ( 1 + e_{\beta 0} \cos x \right ) ^{2}} ~ d x ~.
\end{eqnarray}
Finally, Eqs. (137) -- (140), (142) -- (144) yield
\begin{eqnarray}\label{145}
\langle \frac{d a_{\beta}}{dt} \rangle _{I} &=& -~ \beta ~\frac{\mu}{c}~
    \frac{2  + 3 e_{\beta 0}^{2}}{a_{\beta 0} \left ( 1~-~e_{\beta 0}^{2}
    \right )^{3/2}} \left \{ 1 +
    \frac{\beta}{\sqrt{1 - \beta}} ~\frac{\sqrt{\mu}}{c}
    \frac{1 - e_{\beta 0}^{2}}{\sqrt{a_{\beta 0}}} \times \right .
\nonumber \\
& & \left . \left [ I_{LHS} - \frac{\pi +
    \left ( 10 \pi + 3 \right ) e_{\beta 0}^{2} +
    \left ( 9 \pi + 2 \right ) e_{\beta 0}^{4}}{\left ( 1
     + 3 e_{\beta 0}^{2} / 2 \right ) \left ( 1 -
    e_{\beta 0}^{2} \right )^{5/2}} \right ] \right \} ~.
\end{eqnarray}

Total secular change of semi-major axis is given as a sum of Eqs. (111) and
(145). Changing the index $\beta 0$ into $\beta$ and using Eq. (144), we
can finally write
\begin{eqnarray}\label{146}
\langle \frac{d a_{\beta}}{dt} \rangle &=& -~ \beta ~\frac{\mu}{c}~
    \frac{2  + 3 e_{\beta}^{2}}{a_{\beta} \left ( 1~-~e_{\beta}^{2}
    \right )^{3/2}} \left \{ 1 +
    \frac{\beta}{\sqrt{1 - \beta}} ~\frac{\sqrt{\mu}}{c}
    \frac{1 - e_{\beta}^{2}}{\sqrt{a_{\beta}}} \times \right .
\nonumber \\
& & \left . \left [ I_{n} - \frac{\pi +
    \left ( 10 \pi + 3 \right ) e_{\beta}^{2} +
    \left ( 9 \pi + 2 \right ) e_{\beta}^{4}}{\left ( 1
     + 3 e_{\beta}^{2} / 2 \right ) \left ( 1 -
    e_{\beta}^{2} \right )^{5/2}} \right ] \right \} ~,
\nonumber \\
I_{n} &=& \frac{5}{2 \pi} \int_{0}^{2 \pi}
	\frac{x}{\left ( 1 + e_{\beta} \cos x \right ) ^{3}} ~ d x ~-~
	\frac{2}{2 \pi} \int_{0}^{2 \pi}
	\frac{x}{\left ( 1 + e_{\beta} \cos x \right ) ^{2}} ~ d x ~.
\end{eqnarray}

Let us calculate secular change of $e_{\beta}$ to the order
$1/c^{2}$, generated by osculating
orbital change proportional to $1/c$ -- compare Eqs. (109) and (111).
Putting Eq. (114) into Eq. (113), we can write to the required
accuracy,
\begin{eqnarray}\label{147}
\langle \frac{d e_{\beta}}{dt} \rangle _{I}
&=& \left \{ \int_{0}^{2 \pi} \left ( \frac{H_{\beta}}{r^{2}} ~-~
    \frac{d \omega_{\beta}}{dt} \right ) ^{-1} d f_{\beta} \right \} ^{-1}
    \int_{0}^{2 \pi}
    \frac{r^{2}}{H_{\beta}} \left ( 1 ~+~
    \frac{r^{2}}{H_{\beta}} ~ \frac{d \omega_{\beta}}{dt} \right ) ~
    \frac{d e_{\beta}}{dt} ~d f_{\beta} ~,
\end{eqnarray}
since time derivatives of $e_{\beta}$ and $\omega _{\beta}$ are
proportional to $1/c$, and,
thus, the third term on the right-hand side of Eq. (114) would yield a higher
order than $1/c^{2}$ in the last integral.

The last integral in Eq. (147) consists of two terms:
\begin{eqnarray}\label{148}
I_{e} &\equiv& \int_{0}^{2 \pi}
    \frac{r^{2}}{H_{\beta}} \left ( 1 ~+~
    \frac{r^{2}}{H_{\beta}} ~ \frac{d \omega_{\beta}}{dt} \right ) ~
    \frac{d e_{\beta}}{dt} ~d f_{\beta} \equiv
    I_{e 1} ~+~ I_{e 2} ~,
\nonumber \\
I_{e 1} &\equiv&
    \int_{0}^{2 \pi} \frac{r^{2}}{H_{\beta}}
    \frac{r^{2}}{H_{\beta}} ~ \frac{d \omega_{\beta}}{dt} ~
    \frac{d e_{\beta}}{dt} ~d f_{\beta} ~,
\nonumber \\
I_{e 2} &\equiv&
    \int_{0}^{2 \pi} \frac{r^{2}}{H_{\beta}}
    \frac{d e_{\beta}}{dt} ~d f_{\beta} ~.
\end{eqnarray}
Inserting relations for $d e_{\beta} / dt$ and $d \omega_{\beta} / dt$
from Eq. (109),
we can immediately write for $I_{e 1}$, within the required accuracy:
\begin{eqnarray}\label{149}
I_{e 1} &=& \left ( \beta ~\frac{\mu}{c} \right ) ^{2} ~
    \int_{0}^{2 \pi} \frac{\left ( 2  - e_{\beta} \cos f_{\beta}
    \right ) \sin f_{\beta}}{H_{\beta} e_{\beta}} ~
    \frac{2 e_{\beta} + e_{\beta} \sin ^{2} f_{\beta}
    + 2 \cos f_{\beta}}{H_{\beta}}
    ~d f_{\beta}
\nonumber \\
&=& \left ( \beta ~\frac{\mu}{c} \right ) ^{2} ~
    \int_{0}^{2 \pi} \frac{\left ( 2  - e_{\beta 0} \cos f_{\beta}
    \right ) \sin f_{\beta}}{H_{\beta 0} e_{\beta 0}} ~
    \frac{2 e_{\beta 0} + e_{\beta 0} \sin ^{2} f_{\beta}
    + 2 \cos f_{\beta}}{H_{\beta 0}}
    ~d f_{\beta}
\nonumber \\
&=& 0 ~,
\end{eqnarray}
where results of Eqs. (121) and (122) were used.
\begin{eqnarray}\label{150}
I_{e 2} &=& -~ \beta ~\frac{\mu}{c} ~ \int_{0}^{2 \pi}
    \frac{2 e_{\beta} + e_{\beta} \sin ^{2} f_{\beta}
    + 2 \cos f_{\beta}}{H_{\beta}} ~d f_{\beta} ~.
\end{eqnarray}
Considering Eq. (122) and the first relation of Eq. (141) for $1/H_{\beta}$,
one finally obtains from Eq. (150):
\begin{eqnarray}\label{151}
I_{e 2} &=& -~ \beta ~\frac{\mu}{c} \frac{5 \pi e_{\beta 0}}{H_{\beta 0}}
    \left \{ 1 -
    \frac{\beta}{\sqrt{1 - \beta}} \frac{\sqrt{\mu}}{c \sqrt{p_{\beta 0}}}
    \pi \left ( 2 + \frac{7}{10} e_{\beta 0} \right ) \right \} ~.
\end{eqnarray}

Putting the results represented by Eqs. (144), (148), (149) and (151)
into Eq. (147), we receive
\begin{eqnarray}\label{152}
\langle \frac{d e_{\beta}}{dt} \rangle _{I} &=& -~ \beta ~\frac{\mu}{c}~
    \frac{5 e_{\beta 0} / 2}{a_{\beta 0} ^{2} \sqrt{1 - e_{\beta 0}^{2}}}
    \left \{ 1 +
    \frac{\beta}{\sqrt{1 - \beta}} ~\frac{\sqrt{\mu}}{c}
    \frac{1 - e_{\beta 0}^{2}}{\sqrt{a_{\beta 0}}} \times \right .
\nonumber \\
& & \left . \left [ I_{LHS} - \pi \frac{2 + 7 e_{\beta 0} / 10}{\left ( 1 -
    e_{\beta 0}^{2} \right )^{3/2}} \right ] \right \} ~.
\end{eqnarray}

Total secular change of eccentricity is given as a sum of Eqs. (111) and
(152). Changing the index $\beta 0$ into $\beta$ in Eq. (152), we
can finally write
\begin{eqnarray}\label{153}
\langle \frac{d e_{\beta}}{dt} \rangle &=& -~ \beta ~\frac{\mu}{c}~
    \frac{5 e_{\beta} / 2}{a_{\beta} ^{2} \sqrt{1 - e_{\beta}^{2}}}
    \left \{ 1 +
    \frac{\beta}{\sqrt{1 - \beta}} ~\frac{\sqrt{\mu}}{c}
    \frac{1 - e_{\beta}^{2}}{\sqrt{a_{\beta}}}
    \left [ I_{n} - \pi \frac{2 + 7 e_{\beta} / 10}{\left ( 1 -
    e_{\beta}^{2} \right )^{3/2}} \right ] \right \} ~,
\nonumber \\
I_{n} &=& \frac{5}{2 \pi} \int_{0}^{2 \pi}
	\frac{x}{\left ( 1 + e_{\beta} \cos x \right ) ^{3}} ~ d x ~-~
	\frac{2}{2 \pi} \int_{0}^{2 \pi}
	\frac{x}{\left ( 1 + e_{\beta} \cos x \right ) ^{2}} ~ d x ~.
\end{eqnarray}

Let us calculate secular change of $p_{\beta}$ (see Eq. (116)) to the order
$1/c^{2}$, generated by osculating
orbital change proportional to $1/c$ -- one can easily verify that
$\langle d p_{\beta} / dt \rangle _{II} =$ 0. Taking into account
Eqs. (109) and (116), we can write to the required
accuracy,
\begin{eqnarray}\label{154}
\langle \frac{d p_{\beta}}{dt} \rangle _{I}
&=& \left \{ \int_{0}^{2 \pi} \left ( \frac{H_{\beta}}{r^{2}} ~-~
    \frac{d \omega_{\beta}}{dt} \right ) ^{-1} d f_{\beta} \right \} ^{-1}
    \int_{0}^{2 \pi}
    \frac{r^{2}}{H_{\beta}} \left ( 1 ~+~
    \frac{r^{2}}{H_{\beta}} ~ \frac{d \omega_{\beta}}{dt} \right ) ~
    \frac{d p_{\beta}}{dt} ~d f_{\beta} ~,
\end{eqnarray}
since time derivatives of $p_{\beta}$ and $\omega _{\beta}$ are
proportional to $1/c$, and,
thus, the third term on the right-hand side of Eq. (114) would yield a higher
order than $1/c^{2}$ in the last integral.

The last integral in Eq. (154) consists of two terms:
\begin{eqnarray}\label{155}
I_{p} &\equiv& \int_{0}^{2 \pi}
    \frac{r^{2}}{H_{\beta}} \left ( 1 ~+~
    \frac{r^{2}}{H_{\beta}} ~ \frac{d \omega_{\beta}}{dt} \right ) ~
    \frac{d p_{\beta}}{dt} ~d f_{\beta} \equiv
    I_{p 1} ~+~ I_{p 2} ~,
\nonumber \\
I_{p 1} &\equiv&
    \int_{0}^{2 \pi} \frac{r^{2}}{H_{\beta}}
    \frac{r^{2}}{H_{\beta}} ~ \frac{d \omega_{\beta}}{dt} ~
    \frac{d p_{\beta}}{dt} ~d f_{\beta} ~,
\nonumber \\
I_{p 2} &\equiv&
    \int_{0}^{2 \pi} \frac{r^{2}}{H_{\beta}}
    \frac{d p_{\beta}}{dt} ~d f_{\beta} ~.
\end{eqnarray}
Inserting relations for $d p_{\beta} / dt$ and $d \omega_{\beta} / dt$
from Eqs. (116) and (109),
we can immediately write for $I_{p 1}$, within the required accuracy:
\begin{eqnarray}\label{156}
I_{p 1} &=& 0 ~,
\nonumber \\
I_{p 2} &=& -~ 2 \beta ~\frac{\mu}{c}
	\frac{1}{\sqrt{\mu \left ( 1 - \beta \right )}} \int_{0}^{2 \pi}
	\sqrt{p_{\beta}} d f_{\beta} =
 -~ 4 \pi  \beta ~\frac{\mu}{c} \left \{
      \sqrt{\frac{p_{\beta 0}}{\mu \left ( 1 - \beta \right )}} -
      \frac{\beta}{1 - \beta} \frac{1}{c} \pi \right \} ~,
\end{eqnarray}
where Eq. (121) was used.
Putting results of Eqs. (143), (144), (155) and (156) into Eq. (154), one obtains
\begin{eqnarray}\label{157}
\langle \frac{d p_{\beta}}{dt} \rangle _{I}
&=& - 2 \beta \frac{\mu}{c} \frac{\left (
    1 - e_{\beta 0}^{2} \right ) ^{3/2}}{p_{\beta 0}}
    \left \{ 1 + \frac{\beta}{\sqrt{1 - \beta}} \frac{\sqrt{\mu}}{c} \left [
    \frac{\left ( 1 - e_{\beta 0}^{2} \right ) ^{3/2}}{\sqrt{p_{\beta 0}}}
    I_{LHS} - \frac{\pi}{\sqrt{p_{\beta 0}}}
    \right ] \right \} ~.
\end{eqnarray}

Total secular change of $p_{\beta}$ is given as a sum of Eqs. (157) and
$\langle d p_{\beta} / dt \rangle _{II} =$ 0.
Changing the index $\beta 0$ into $\beta$, we can finally write
\begin{eqnarray}\label{158}
\langle \frac{d p_{\beta}}{dt} \rangle
&=& - 2 \beta \frac{\mu}{c} \frac{\left (
    1 - e_{\beta}^{2} \right ) ^{3/2}}{p_{\beta}}
    \left \{ 1 + \frac{\beta}{\sqrt{1 - \beta}} \frac{\sqrt{\mu}}{c}
    \frac{\left ( 1 - e_{\beta}^{2} \right ) ^{3/2}}{\sqrt{p_{\beta}}}
    \left [ I_{n} - \frac{\pi}{\left ( 1 - e_{\beta}^{2} \right )^{3/2}}
    \right ] \right \} ~,
\nonumber \\
I_{n} &=& \frac{5}{2 \pi} \int_{0}^{2 \pi}
	\frac{x}{\left ( 1 + e_{\beta} \cos x \right ) ^{3}} ~ d x ~-~
	\frac{2}{2 \pi} \int_{0}^{2 \pi}
	\frac{x}{\left ( 1 + e_{\beta} \cos x \right ) ^{2}} ~ d x ~.
\end{eqnarray}

Comparing Eqs. (153) and (158), one easily obtains
\begin{eqnarray}\label{159}
\frac{d e_{\beta}}{d p_{\beta}} &=& \frac{5}{4}~ \frac{e_{\beta}}{p_{\beta}}
    \left \{ 1 -
    \frac{\beta}{\sqrt{1 - \beta}} ~\frac{\sqrt{\mu}}{c}
    \frac{\pi}{\sqrt{p_{\beta}}}
    \left ( 1 + 7 e_{\beta} / 10 \right ) \right \} ~.
\end{eqnarray}
This can be easily integrated and the results may be written as:
\begin{eqnarray}\label{160}
e_{\beta} &=& e_{\beta in} \left ( \frac{p_{\beta}}{p_{\beta in}} \right )
	  ^{5/4} \left \{ 1 -
    \frac{\beta}{\sqrt{1 - \beta}} ~\frac{\sqrt{\mu}}{c}
    \frac{5 \pi}{3} \frac{1}{\sqrt{p_{\beta}}} \left [ 1 +
    \frac{21}{80} e_{\beta in}
    \left ( \frac{p_{\beta}}{p_{\beta in}} \right ) ^{5/4} \right ] \right \} ~,
\nonumber \\
p_{\beta} &=& p_{\beta in} \left ( \frac{e_{\beta}}{e_{\beta in}} \right )
	  ^{4/5} \left \{ 1 +
    \frac{\beta}{\sqrt{1 - \beta}} ~\frac{\sqrt{\mu}}{c}
    \frac{4 \pi}{3} \frac{1}{\sqrt{p_{\beta in}}}
    \left ( \frac{e_{\beta in}}{e_{\beta}} \right ) ^{2/5}
    \left [ 1 +  \frac{21}{80} e_{\beta}
    \right ] \right \} ~.
\end{eqnarray}

\section{Secular change of advancement of pericenter/perihelion -- gravitation
as a central acceleration}
We will use $-~\mu ~\vec{e}_{R}~/~r^{2}$ as a central
acceleration determining osculating orbital elements. As we have seen
in the previous subsection, corrections of the order $(v/c)^{2}$ represent
very small corrections with respect to the order $v/c$ -- only
$d \omega_{\beta} / dt$ is important, as for secular changes. Thus, the
most important results, for the case when $-~\mu ~\vec{e}_{R}~/~r^{2}$ is used
as a central acceleration, were presented in section 6.2. However, we have
not calculated secular change of $d \omega / d t$ in the section 6.2. We will
do this now, since results of the preceding section 7.1 have to be used.

We will use Eqs. (91), (93) and (94) in the form
\begin{eqnarray}\label{161}
\sin \left ( \Theta ~-~ \omega \right ) &=& \left \{
       \left ( 1 ~-~ \beta \right ) ~e_{c} ~
       \sin \left ( \Theta ~-~ \omega_{c} \right ) \right \} ~/~ e ~,
\nonumber \\
\cos \left ( \Theta ~-~ \omega \right ) &=&
       \left \{ \left ( 1 ~-~ \beta \right ) ~ e_{c} ~
       \cos \left ( \Theta ~-~ \omega_{c} \right ) ~-~ \beta \right \} ~/~  e ~,
\nonumber \\
e &=& \sqrt{\left ( 1 ~-~ \beta \right ) ^{2} ~ e_{c}^{2} ~+~ \beta ^{2} ~-~
       2~ \beta ~ \left ( 1 ~-~ \beta \right ) ~ e_{c} ~
       \cos \left ( \Theta ~-~ \omega_{c} \right )} ~.
\end{eqnarray}

Using the fact that $\Theta - \omega = \Theta - \omega_{c}$ $+$
$\omega_{c} - \omega$ $\equiv$ $f_{c}$ $+$ $\omega_{c} - \omega$, we can write
\begin{eqnarray}\label{162}
\frac{d \cos \left ( \Theta ~-~ \omega \right )}{d t} &=& - ~ \left \{
\frac{d f_{c}}{d t} + \frac{d \omega_{c}}{d t} - \frac{d \omega}{d t}
\right \} \sin \left ( \Theta ~-~ \omega \right ) ~.
\end{eqnarray}
We admit that the subscript "c" may be changed into $\beta$.
Using Eq. (162), one immediately obtains
\begin{eqnarray}\label{163}
\frac{d \omega}{d t} &=&
\frac{d \cos \left ( \Theta ~-~ \omega \right )}{d t} \times
\left \{ \sin \left ( \Theta ~-~ \omega \right ) \right \} ^{-1} ~+~
\frac{d \omega_{c}}{d t} ~+~ \frac{d f_{c}}{d t}  ~.
\end{eqnarray}
Inserting the right-hand side of the second of equations of Eq. (161)
into Eq. (163), one finally, after differentiation, obtains
\begin{eqnarray}\label{164}
\frac{d \omega}{d t} &=& \frac{d \omega_{c}}{d t} ~+~
\frac{\cos f_{c}}{\sin f_{c}} \left ( e_{c}^{-1} ~\frac{d e_{c}}{d t} ~-~
e^{-1} ~\frac{d e}{d t} \right ) ~+~ \frac{\beta}{1 - \beta} ~
\frac{1}{e_{c} \sin f_{c}} ~ e^{-1} ~\frac{d e}{d t} ~.
\end{eqnarray}

Differentiation of the third equation of Eq. (161) yields
\begin{eqnarray}\label{165}
e^{-1} ~\frac{d e}{d t}  &=& \frac{1}{e^{2}} \left \{
\left ( 1 - \beta \right ) ^{2} e_{c} \frac{d e_{c}}{d t} +
\beta \left ( 1 - \beta \right ) \left [ - \frac{d e_{c}}{d t}
\cos f_{c} + \frac{d f_{c}}{d t} e_{c} \sin f_{c} \right ] \right \} ~.
\end{eqnarray}

Inserting Eq. (165) into Eq. (164), we can write
\begin{eqnarray}\label{166}
\frac{d \omega}{d t} &=& \frac{d \omega_{c}}{d t} + \frac{1}{2} \left [ 1 +
\frac{\beta ^{2}  - \left ( 1 - \beta \right ) ^{2} e_{c}^{2}}{e^{2}}
\right ] \frac{d f_{c}}{d t} +
\beta \left ( 1 - \beta \right ) \frac{1}{e^{2}} \frac{d e_{c}}{d t}
\sin f_{c} ~,
\nonumber \\
e^{2} &=& \left ( 1 ~-~ \beta \right ) ^{2} ~ e_{c}^{2} ~+~ \beta ^{2} ~-~
    2~ \beta ~ \left ( 1 ~-~ \beta \right ) ~ e_{c} ~
    \cos f_{c} ~.
\end{eqnarray}
Eq. (166) is the decisive equation which enables us to find secular change
of $\omega$.

As a first approximation, let us consider Keplerian orbit characterized by
conditions $d a_{c} / dt =$ $d p_{c} / dt =$ $d e_{c} / dt =$
$d \omega_{c} / dt =$ 0, $p_{c} = a_{c} ( 1 - e_{c}^{2} )$.
Using averaging of the type
of Eq. (98), or, Eq. (110), we can write for Eq. (166)
(we use $r^{2} d f_{c} / dt = \sqrt{\mu ( 1 - \beta ) p_{c}}$)
\begin{eqnarray}\label{167}
\langle \frac{d \omega}{d t} \rangle &\equiv& \frac{1}{T}  \int_{0}^{T}
\frac{d \omega}{d t} dt =
  \frac{1}{a_{c}^{2}~ \sqrt{1 - e_{c}^{2}}} ~\frac{1}{2 \pi} ~
  \int_{0}^{2 \pi} ~ \frac{d \omega}{d t} (f_{c}) ~r^{2} ~ df_{c} =
  \frac{1}{a_{c}^{2}~ \sqrt{1 - e_{c}^{2}}} ~\frac{1}{2 \pi} \times
\nonumber \\
& & \int_{0}^{2 \pi} \frac{1}{2}
  \left [ 1 +
  \frac{\beta ^{2}  - \left ( 1 - \beta \right ) ^{2} e_{c}^{2}}{
  \left ( 1 - \beta \right ) ^{2}  e_{c}^{2} + \beta ^{2} -
  2 \beta \left ( 1 - \beta \right ) e_{c} \cos f_{c}}
  \right ]
  \sqrt{\mu \left ( 1 - \beta \right ) p_{c}} ~ d f_{c} ~.
\end{eqnarray}
If the result $\int_{0}^{2 \pi} dx / ( 1 + k \cos x ) = 2 \pi /
\sqrt{1 - k^{2}}$ is used, one finally obtains
\begin{equation}\label{168}
 \langle \frac{d \omega}{d t} \rangle =  \left \{ \begin{array}{lll}
    0  & \mbox{if $\beta < e_{c} / ( 1 + e_{c} )$} \\
    \sqrt{\mu \left ( 1 - \beta \right )} / a_{c} ^{3/2} / 2
    & \mbox{if $\beta = e_{c} / ( 1 + e_{c} )$} \\
    \sqrt{\mu \left ( 1 - \beta \right )} / a_{c} ^{3/2}
    & \mbox{if $\beta > e_{c} / ( 1 + e_{c} )$}
	 \end{array}
    \right .  ~.
\end{equation}
Let us remind that $d \Theta _{c} / dt =$
$\sqrt{\mu ( 1 - \beta )} / a_{c} ^{3/2}$ (see Eq. (100)).

Let us consider P-R effect to the first order in $v/c$. We will use
Eq. (55) and also Eq. (112), which we summarize in the following Eq. (169):
\begin{eqnarray}\label{169}
\frac{de_{\beta}}{dt} &=& -~ \beta \frac{\mu}{r^{2}} \frac{1}{c}  \left (
      2 e_{\beta} + e_{\beta}  \sin^{2} f_{\beta} + 2 \cos f_{\beta} \right ) ~,
\nonumber \\
\frac{d\omega_{\beta}}{dt} &=& -~ \beta \frac{\mu}{r^{2}} \frac{1}{c}
    \frac{1}{e_{\beta}} ~ \left (
    2  - e_{\beta} \cos f_{\beta} \right ) \sin f_{\beta} ~,
\nonumber \\
\frac{d f_{\beta}}{dt} &=& \frac{H_{\beta}}{r^{2}} ~-~
	       \frac{d \omega_{\beta}}{dt} ~,
\end{eqnarray}
where $H_{\beta} \equiv \sqrt{\mu \left ( 1 - \beta \right ) p_{\beta}}$.
Changing subscript ``c'' into ``$\beta$'' in Eq. (166) and inserting the
third of Eq. (169) into Eq. (166), we can write
\begin{eqnarray}\label{170}
\frac{d \omega}{d t} &=&
\frac{1}{2} \left [ 1 +
\frac{\beta ^{2}  - \left ( 1 - \beta \right ) ^{2} e_{\beta}^{2}}{e^{2}}
\right ] \frac{H_{\beta}}{r^{2}} +
     \frac{1}{2} \left [ 1 -
\frac{\beta ^{2}  - \left ( 1 - \beta \right ) ^{2} e_{\beta}^{2}}{e^{2}}
\right ] \frac{d \omega_{\beta}}{d t} +
\nonumber \\
& &
\beta \left ( 1 - \beta \right ) \frac{1}{e^{2}} \frac{d e_{\beta}}{d t}
\sin f_{\beta} ~,
\nonumber \\
e^{2} &=& \left ( 1 ~-~ \beta \right ) ^{2} ~ e_{\beta}^{2} ~+~ \beta ^{2} ~-~
    2~ \beta ~ \left ( 1 ~-~ \beta \right ) ~ e_{\beta} ~
    \cos f_{\beta} ~.
\end{eqnarray}
On the basis of Eqs. (113)-(115), (143)-(144), (169)-(170),
we can write within the
required accuracy:
\begin{eqnarray}\label{171}
\langle \frac{d \omega}{d t} \rangle &=& A_{\omega} \int_{0}^{2 \pi}
    \frac{r^{2}}{H_{\beta}} \left ( 1 ~+~
    \frac{r^{2}}{H_{\beta}} ~ \frac{d \omega_{\beta}}{dt} \right )
    \frac{d \omega}{d t} ~d f_{\beta} =
\nonumber \\
&=&
\frac{A_{\omega}}{2} \int_{0}^{2 \pi} \left [ 1 +
\frac{\beta ^{2}  - \left ( 1 - \beta \right ) ^{2} e_{\beta}^{2}}{
    \left ( 1 - \beta \right ) ^{2}  e_{\beta}^{2} + \beta ^{2} -
    2 \beta \left ( 1 - \beta \right ) e_{\beta}
    \cos f_{\beta}}
\right ] ~d f_{\beta} ~,
\nonumber \\
A_{\omega} &\equiv&
    \frac{\sqrt{\mu \left ( 1 - \beta \right )}}{2 \pi a_{\beta 0} ^{3/2}}
    \left \{ 1 + \frac{\beta}{\sqrt{1 - \beta}} \frac{\sqrt{\mu}}{c}
    \frac{1 - e_{\beta 0}^{2}}{\sqrt{a_{\beta 0}}} I_{n 0} \right \} ~,
\nonumber \\
I_{n 0} &=& \frac{5}{2 \pi} \int_{0}^{2 \pi}
	\frac{x}{\left ( 1 + e_{\beta 0} \cos x \right ) ^{3}} ~ d x ~-~
	\frac{2}{2 \pi} \int_{0}^{2 \pi}
	\frac{x}{\left ( 1 + e_{\beta 0} \cos x \right ) ^{2}} ~ d x ~.
\end{eqnarray}
Using also Eq. (122) in the integral of the second line of
Eq. (171), we finally receive
\begin{eqnarray}\label{172}
\langle \frac{d \omega}{d t} \rangle &=&
    \frac{\sqrt{\mu \left ( 1 - \beta \right )}}{a_{\beta} ^{3/2}}
    \left \{ \vartheta _{H} \left ( \beta - \frac{e_{\beta}}{1 + e_{\beta}} \right )
    + \frac{\beta}{\sqrt{1 - \beta}} \frac{\sqrt{\mu}}{c}
    \frac{M_{\omega 1} + M_{\omega 2}}{\sqrt{a_{\beta}
    \left ( 1 - e_{\beta}^{2} \right )}}  \right \} ~,
\nonumber \\
M_{\omega 1} &=&
    \left ( 1 - e_{\beta}^{2} \right ) ^{3/2} \left [
    5~ I_{3} \left ( e_{\beta} \right ) ~-~
    2~ I_{2} \left ( e_{\beta} \right ) \right ] \vartheta _{H}
    \left ( \beta - \frac{e_{\beta}}{1 + e_{\beta}} \right ) ~,
\nonumber \\
M_{\omega 2} &=& \frac{5}{4} \left \{ I_{1} \left ( \xi \right ) -
    \left [ \frac{\beta ^{2} - \left ( 1 - \beta \right ) ^{2}
    e_{\beta}^{2}}{\beta ^{2} + \left ( 1 - \beta \right ) ^{2}
    e_{\beta}^{2}} \right ] ^{2}
    I_{2} \left ( \xi \right ) \right \} ~,
\nonumber \\
\xi &=& -~ \frac{2 \beta \left ( 1 - \beta \right ) e_{\beta}}{
\left ( 1 - \beta \right ) ^{2} e_{\beta}^{2} + \beta ^{2}} ~,
\nonumber \\
I_{\alpha} \left ( \varepsilon \right ) &=& \frac{1}{2 \pi} \int_{0}^{2 \pi}
       \frac{x}{\left ( 1 + \varepsilon \cos x \right ) ^{\alpha}} ~ d x ~,
       ~~~ \alpha = 1, 2, 3 ~,
\end{eqnarray}
where $\vartheta _{H} ( x ) =$ 1 if $x >$ 0,
$\vartheta _{H} ( x ) =$ 0 if $x <$ 0 (Heaviside's step function); it is
assumed that $\beta \ne e_{\beta} / ( 1 + e_{\beta} )$.
We see that advancement of pericenter/perihelion exists already in the first
order in $v/c$ if gravity alone is taken as a central acceleration.

\section{P-R effect and near circular orbits}
When the orbits are near circular (pseudo-circular)
-- central acceleration contains radiation pressure term -- the orbit can not
reduce in semi-major axis without increasing in eccentricity.
Both types of orbital elements, defined by central accelerations,
have been used in detail numerical calculations in papers
by Kla\v{c}ka and Kaufmannov\'{a} (1992, 1993). Due to the property
$\langle e \rangle \ge \beta$ (see section 6.2.2), one must be aware
that also values $\langle e \rangle > \beta$ may correspond to pseudo-circular
orbits. The results for pseudo-circular orbits were analytically confirmed by
Breiter and Jackson (1998). As an advantage, the analytical approach reproduces
known results without detail numerical calculations. However, the analytical
approach paralelly produces nonphysical results which may not be
distinguishable from the correct results. The nonphysical analytical results are
caused by use of the P-R effect in the first order in $v/c$ -- very special
analytical solutions will diminish when higher orders in $v/c$ are used.
The nonphysical analytical results have been discussed in more detail by
Kla\v{c}ka (2001; see also http://xxx.lanl.gov/abs/astro-ph/0004181).

\section{Solar wind effect}
We have calculated secular changes of orbital elements for the P-R effect up to
the second order in $v/c$. The effects of the second order
in $v/c$ seem to be small to play an important role in Solar System studies.
In reality, similar effect coming from the Sun exists and this effect
may play more important role. Although this effect is not connected with
electromagnetic radiation, its secular changes of orbital elements to the
first order in $v/c$ (more correctly, $v/u$, where $u$ is the speed of solar
wind particles) correspond to the secular changes for P-R effect. This effect is
caused by solar wind particles hitting an interplanetary dust particle. The aim
of this section is to obtain secular changes of orbital elements of the
interplanetary dust particle under the action of the solar wind, up to the
second order in $v/u$.

Let us consider equation of motion in the form
(we neglect decrease of particle's mass)
\begin{eqnarray}\label{173}
\frac{d~ \vec{v}_{sw}}{d ~t} &=& \eta ~ \frac{\beta}{\bar{Q}'_{1}} ~
    \frac{\mu}{r^{2}} ~ \frac{u}{c} ~
    \left \{ \left [ \left ( 1~-~ \frac{\vec{v} \cdot \vec{e}_{R}}{u}
    \right ) ~ \vec{e}_{R} ~-~ \frac{\vec{v}}{u} \right ] ~+~  \right .
\nonumber \\
& & \left . \left ( \frac{\vec{v} \cdot \vec{e}_{R}}{u} \right ) ~
    \frac{\vec{v}}{u} ~+~ \frac{1}{2} ~ \left [
    \left ( \frac{\vec{v}}{u} \right ) ^{2} ~-~ \left (
    \frac{\vec{v} \cdot \vec{e}_{R}}{u} \right ) ^{2} \right ]
    \vec{e}_{T} \right \} ~,
\end{eqnarray}
where, as standardly in this paper, $\mu \equiv G~M_{\odot}$ and non-dimensional
parameter $\beta$ is defined in Eq. (43) (``the ratio of radiation
pressure force to the gravitational force''); $\eta \approx$ 1/3. Adding the
right-hand side of Eq. (49) to the right-hand side of Eq. (173), one obtains
\begin{eqnarray}\label{174}
\frac{d~ \vec{v}}{d ~t} &=& -~ \frac{\mu}{r^{2}} ~ \vec{e}_{R} ~+~
	  \beta ~ \frac{\mu}{r^{2}} ~
	  \left \{  \left ( 1~-~
	  \frac{\vec{v} \cdot \vec{e}_{R}}{c} \right ) ~ \vec{e}_{R}
	  ~-~ \frac{\vec{v}}{c} \right \} ~+~
\nonumber \\
& &  \eta ~ \frac{\beta}{\bar{Q}'_{1}}  ~ \frac{\mu}{r^{2}} ~\frac{u}{c} ~
     \left \{ \left [ \left ( 1~-~ \frac{\vec{v} \cdot \vec{e}_{R}}{u}
     \right ) ~ \vec{e}_{R} ~-~ \frac{\vec{v}}{u} \right ] ~+~	\right .
\nonumber \\
& & \left . \left ( \frac{\vec{v} \cdot \vec{e}_{R}}{u} \right ) ~
    \frac{\vec{v}}{u} ~+~ \frac{1}{2} ~ \left [
    \left ( \frac{\vec{v}}{u} \right ) ^{2} ~-~ \left (
    \frac{\vec{v} \cdot \vec{e}_{R}}{u} \right ) ^{2} \right ]
    \vec{e}_{T} \right \} ~.
\end{eqnarray}
The first term in Eq. (174) is Newtonian gravity
for two-body problem, the second term is the Poynting-Robertson effect
in flat spacetime, the third term corresponds to the action of the solar
wind up to the second order in $v/u$.

Eq. (174) may be written in the following form:
\begin{eqnarray}\label{175}
\frac{d~ \vec{v}}{d ~t} &=& -~ \frac{\mu}{r^{2}} ~ \vec{e}_{R} ~+~
      \beta \left ( 1 + \frac{\eta}{\bar{Q}'_{1}}
      \frac{u}{c} \right ) ~ \frac{\mu}{r^{2}} ~ \vec{e}_{R} ~-~
      \beta ~\left ( 1 + \frac{\eta}{\bar{Q}'_{1}} \right ) ~
      \frac{\mu}{r^{2}} \left (
      \frac{\vec{v} \cdot \vec{e}_{R}}{c} ~ \vec{e}_{R}
      + \frac{\vec{v}}{c} \right ) ~+~
\nonumber \\
& &  \eta~ \frac{\beta}{\bar{Q}'_{1}} ~ \frac{\mu}{r^{2}} ~\left \{
     \frac{\vec{v} \cdot \vec{e}_{R}}{c} ~
     \frac{\vec{v}}{u} ~+~ \frac{1}{2} ~ \left [
     \frac{\vec{v} ^{2}}{c ~u} ~-~
     \frac{\left ( \vec{v} \cdot \vec{e}_{R} \right ) ^{2}}{c ~u} \right ]
     \vec{e}_{T} \right \} ~.
\end{eqnarray}

\subsection{Secular changes of orbital elements -- radiation pressure
as a part of central acceleration}
Neglecting solar wind pressure term
$\eta ( \beta / \bar{Q}'_{1} ) ( \mu / r^{2} ) ( u / c )$, we can rewrite Eq.
(175) to the following form:
\begin{eqnarray}\label{176}
\frac{d~ \vec{v}}{d ~t} &=& -~ \frac{\mu \left ( 1 - \beta \right )}{r^{2}} ~
      \vec{e}_{R} ~-~ \beta ~\frac{\mu}{r^{2}} ~
      \left ( 1 + \frac{\eta}{\bar{Q}'_{1}}  \right ) \left (
      \frac{\vec{v} \cdot \vec{e}_{R}}{c} ~ \vec{e}_{R}
      + \frac{\vec{v}}{c} \right ) ~+~
\nonumber \\
& &  \eta~ \frac{\beta}{\bar{Q}'_{1}} ~ \frac{\mu}{r^{2}} ~\left \{
     \frac{\vec{v} \cdot \vec{e}_{R}}{c} ~
     \frac{\vec{v}}{u} ~+~ \frac{1}{2} ~ \left [
     \frac{\vec{v} ^{2}}{c ~u} ~-~
     \frac{\left ( \vec{v} \cdot \vec{e}_{R} \right ) ^{2}}{c ~u} \right ]
     \vec{e}_{T} \right \} ~,
\end{eqnarray}
where the second term causes deceleration of the particle's motion.
We use $-~\mu ~( 1~-~\beta )~\vec{e}_{R}~/~r^{2}$ as a central
acceleration determining osculating orbital elements; $\beta$ is considered to
be a constant during the motion. On the basis of Eq. (176), we can
immediately write for components of disturbing acceleration
to Keplerian motion
\begin{eqnarray}\label{177}
F_{\beta ~R} &=& -~2~ \beta
      \left ( 1 + \frac{\eta}{\bar{Q}'_{1}} \right ) ~
      \frac{\mu}{r^{2}} ~\frac{v_{\beta~R}}{c} ~+~
      \eta~ \frac{\beta}{\bar{Q}'_{1}} ~\frac{\mu}{r^{2}} ~
      \frac{v_{\beta~R} ^{2}}{c~u} ~,
\nonumber \\
F_{\beta ~T} &=& -~ \beta
      \left ( 1 + \frac{\eta}{\bar{Q}'_{1}} \right ) ~
      \frac{\mu}{r^{2}} ~\frac{v_{\beta~T}}{c} ~+~
      \eta~ \frac{\beta}{\bar{Q}'_{1}} ~\frac{\mu}{r^{2}} ~
      \left \{ \frac{v_{\beta~R} ~v_{\beta~T}}{c~u} ~+~ \frac{1}{2}~
      \frac{v_{\beta~T} ^{2}}{c~u} \right \} ~,
\nonumber \\
F_{\beta ~N} &=& 0 ~,
\end{eqnarray}
where $F_{\beta ~R}$, $F_{\beta ~T}$ and $F_{\beta ~N}$ are radial, transversal
and normal components of perturbation acceleration, and
the two-body problem yields
\begin{eqnarray}\label{178}
v_{\beta~R} &=& \sqrt{\frac{\mu ~( 1 - \beta )}{p_{\beta}}} ~
	e_{\beta} \sin f_{\beta} ~,~
\nonumber \\
v_{\beta~T} &=& \sqrt{\frac{\mu ~( 1 - \beta )}{p_{\beta}}} ~
\left ( 1 + e_{\beta} \cos f_{\beta} \right ) ~,
\end{eqnarray}
where $e_{\beta}$ is osculating eccentricity of the orbit,
$f_{\beta}$ is true anomaly,  $p_{\beta} = a_{\beta} ( 1 - e_{\beta}^{2} )$,
$a_{\beta}$ is semi-major axis.
The important fact that perturbation acceleration is proportional to
$v/c$ ($\ll$ 1) ensures small changes of orbital elements
during a time interval $T$ ($T$ is time interval
between passages through two following pericenters / perihelia). Thus,
we may take a time average in an analytical way
\begin{eqnarray}\label{179}
\langle g \rangle &\equiv& \frac{1}{T}	\int_{0}^{T} g(t) dt =
\frac{\sqrt{\mu ~( 1 - \beta )}}{a_{\beta}^{3/2}} ~
 \frac{1}{2 \pi}  \int_{0}^{2 \pi} g(f_{\beta})
\left ( \frac{df_{\beta}}{dt} \right )^{-1} df_{\beta}
\nonumber \\
&=& \frac{\sqrt{\mu ~( 1 - \beta )}}{a_{\beta}^{3/2}} ~ \frac{1}{2 \pi} \int_{0}^{2 \pi}
g(f_{\beta}) ~\frac{r^{2}}{\sqrt{\mu ~( 1 - \beta ) ~p_{\beta}}}~ df_{\beta}
\nonumber \\
&=& \frac{1}{a_{\beta}^{2}~ \sqrt{1 - e_{\beta}^{2}}} ~\frac{1}{2 \pi} ~
  \int_{0}^{2 \pi} ~ g(f_{\beta}) ~r^{2} ~ df_{\beta} ~,
\end{eqnarray}
assuming non-pseudo-circular orbits and the fact that orbital elements exhibit
only small changes during the time interval $T$;
the second and the third Kepler's laws were used:
$r^{2} ~df_{\beta}/dt = \sqrt{\mu ( 1 - \beta ) p_{\beta}}$ --
conservation of angular momentum,
$a_{\beta}^{3}/T^{2} = \mu ( 1 - \beta ) / (4 \pi^{2})$.

Relevant perturbation equations of celestial mechanics yield for osculating
orbital elements ($\omega_{\beta}$ --
longitude of pericenter / perihelion; $\Theta_{\beta}$
is the position angle of the particle on the orbit, when measured
from the ascending node in the direction of the particle's motion,
$\Theta_{\beta} = \omega_{\beta} + f_{\beta}$):
\begin{eqnarray}\label{180}
\frac{d a_{\beta}}{d t} &=& \frac{2~a_{\beta}}{1~-~e_{\beta}^{2}} ~
      \sqrt{\frac{p_{\beta}}{\mu \left ( 1 ~-~ \beta \right )}} ~
      \left \{
      F_{\beta ~R} ~e_{\beta}~ \sin f_{\beta} +
      F_{\beta ~T} \left ( 1~+~e_{\beta}~ \cos f_{\beta} \right ) \right \} ~,
\nonumber \\
\frac{d e_{\beta}}{d t} &=&
      \sqrt{\frac{p_{\beta}}{\mu \left ( 1 ~-~ \beta \right )}} ~ \left \{
      F_{\beta ~R} ~ \sin f_{\beta} +
      F_{\beta ~T} \left [ \cos f_{\beta} ~+~
     \frac{e_{\beta} +	\cos f_{\beta}}{1 + e_{\beta} \cos f_{\beta}}
	  \right ] \right \} ~,
\nonumber \\
\frac{d \omega_{\beta}}{d t} &=& -~ \frac{1}{e_{\beta}} ~
      \sqrt{\frac{p_{\beta}}{\mu \left ( 1 ~-~ \beta \right )}} ~ \left \{
      F_{\beta ~R} \cos f_{\beta} - F_{\beta ~T}
      \frac{2 + e_{\beta} \cos f_{\beta}}{1 + e_{\beta} \cos f_{\beta}}
      \sin f_{\beta} \right \} ~,
\end{eqnarray}
where $r = p_{\beta} / (1 + e_{\beta} \cos f_{\beta})$.

Putting Eq. (177) and (178) into Eq. (180), making procedure of averaging
of the type given by Eq. (179), we finally receive
\begin{eqnarray}\label{181}
\langle \frac{d a_{\beta}}{dt} \rangle &=& -~ \beta ~\frac{\mu}{c}~
	  \frac{2  + 3 e_{\beta}^{2}}{a_{\beta} \left ( 1~-~e_{\beta}^{2}
	  \right )^{3/2}} \left \{ 1 + \frac{\eta}{\bar{Q}'_{1}} \left [ 1 -
	  \sqrt{\frac{\mu \left ( 1 - \beta \right )}{a_{\beta}
	  \left ( 1 - e_{\beta}^{2} \right )}} ~
	  \frac{1}{2~ u} \right ] \right \} ~,
\nonumber \\
\langle \frac{d e_{\beta}}{dt} \rangle &=& -~ \beta ~\frac{\mu}{c}~
    \frac{5 e_{\beta} / 2}{a_{\beta} ^{2} \sqrt{1 - e_{\beta}^{2}}}
    \left \{ 1 + \frac{\eta}{\bar{Q}'_{1}} \left [ 1 -
	  \sqrt{\frac{\mu \left ( 1 - \beta \right )}{a_{\beta}
	  \left ( 1 - e_{\beta}^{2} \right )}} ~
	  \frac{1}{2~ u} \right ] \right \} ~,
\nonumber \\
\langle \frac{d\omega_{\beta}}{dt} \rangle  &=&
 \frac{3 \mu ^{3/2}}{c^{2} a_{\beta} ^{5/2} \left ( 1 - e_{\beta} ^{2} \right )}
 ~\frac{\beta \left ( 1 - \beta \right )}{\left ( 1 - \beta \right ) ^{1/2}}
 ~\frac{1}{3} ~\frac{\eta}{\bar{Q}'_{1}} ~\frac{c}{u} ~.
\end{eqnarray}
Initial conditions are given by Eq. (59), or, Eqs. (60) -- (62).

Similarly, for the quantity $p_{\beta} = a_{\beta}$
$\left ( 1 - e_{\beta}^{2} \right )$, one can easily obtain
\begin{eqnarray}\label{182}
\langle \frac{d p_{\beta}}{dt} \rangle
&=& - ~2~ \beta ~\frac{\mu}{c} ~\frac{\left (
    1 - e_{\beta}^{2} \right ) ^{3/2}}{p_{\beta}}
    \left \{ 1 + \frac{\eta}{\bar{Q}'_{1}} \left [ 1 -
	  \sqrt{\frac{\mu \left ( 1 - \beta \right )}{p_{\beta}}} ~
	  \frac{1}{2~ u} \right ] \right \} ~.
\end{eqnarray}

We come to the conclusion that
the following relation results from Eqs. (181) and (182):
\begin{eqnarray}\label{183}
p_{\beta}  ~e_{\beta} ^{-4/5} =
p_{\beta ~in}  ~e_{\beta ~in} ^{-4/5} ~,
\end{eqnarray}
as for secular changes of $p_{\beta}$ and $e_{\beta}$,
for the simple case of equation of motion of an interplanetary dust particle
under the action of solar wind represented by Eq. (173) and for the P-R effect.
Thus, Eq. (182) can be solved as a separate equation
\begin{eqnarray}\label{184}
\langle \frac{d p_{\beta}}{dt} \rangle
&=& - ~2 \beta \frac{\mu}{c} \frac{\left [
    1 - e_{\beta ~in}^{2} \left ( p_{\beta} / p_{\beta ~in} \right ) ^{5/2}
    \right ] ^{3/2}}{p_{\beta}}
    \left \{ 1 + \frac{\eta}{\bar{Q}'_{1}} \left [ 1 - \frac{
	  \sqrt{\mu \left ( 1 - \beta \right ) / p_{\beta}}}{2~ u}
	  \right ] \right \} ~,
\end{eqnarray}
and, analogously,
\begin{eqnarray}\label{185}
\langle \frac{d e_{\beta}}{dt} \rangle &=& -~ \beta \frac{\mu}{c}
    \frac{5}{2}
    \frac{e_{\beta in} ^{8/5}}{p_{\beta in} ^{2}}
    \frac{\left ( 1 - e_{\beta}^{2} \right ) ^{3/2}}{e_{\beta}^{3/5}}
    \left \{ 1 + \frac{\eta}{\bar{Q}'_{1}} \left [ 1 -
	  \sqrt{\frac{\mu \left ( 1 - \beta \right )}{p_{\beta in}}}
	  \frac{1}{2~ u}
	  \left ( \frac{e_{\beta in}}{e_{\beta}} \right ) ^{2/5}
	  \right ] \right \} ~.
\end{eqnarray}

\subsection{Secular changes of orbital elements -- gravitation as a central
acceleration}
We can immediately write, on the basis of Eqs. (103), (172) and (181):
\begin{eqnarray}\label{186}
\frac{d a_{\beta}}{dt} &=& -~ \beta ~
	  \left ( 1 + \frac{\eta}{\bar{Q}'_{1}} \right )
	  \frac{\mu}{c} ~
	  \frac{2  + 3 e_{\beta}^{2}}{a_{\beta} \left ( 1~-~e_{\beta}^{2}
	  \right )^{3/2}} ~,
\nonumber \\
\frac{d e_{\beta}}{dt} &=& -~ \beta ~
	  \left ( 1 + \frac{\eta}{\bar{Q}'_{1}} \right )
	  \frac{\mu}{c} ~\frac{5}{2} ~
    \frac{e_{\beta}}{a_{\beta} ^{2} \sqrt{1 - e_{\beta}^{2}}} ~,
\end{eqnarray}
\begin{eqnarray}\label{187}
a &=& a_{\beta} \left ( 1 - e_{\beta}^{2} \right ) ^{3/2} ~
     \frac{1}{2 \pi} ~ \int_{0}^{2 \pi}
    \frac{\left [ 1 + \beta \left ( 1 + e_{\beta}^{2} + 2 e_{\beta}
    \cos x \right ) / \left ( 1 - e_{\beta}^{2} \right ) \right ]^{-1}}
    {\left ( 1 + e_{\beta} \cos x \right )^{2}}
    ~dx ~,
\nonumber \\
e &=& \left ( 1 - e_{\beta}^{2} \right ) ^{3/2} ~
     \frac{1}{2 \pi} ~ \int_{0}^{2 \pi}
\frac{\sqrt{\left ( 1 - \beta \right ) ^{2}  e_{\beta}^{2} + \beta ^{2} -
    2 \beta \left ( 1 - \beta \right )	e_{\beta}
    \cos x}}{\left ( 1 + e_{\beta} \cos x \right )^{2}}
    ~dx ~,
\end{eqnarray}
and, for secular change of longitude of pericenter/perihelion we have
\begin{eqnarray}\label{188}
\frac{d \omega}{d t} &=&
    \frac{\sqrt{\mu \left ( 1 - \beta \right )}}{a_{\beta} ^{3/2}}
    \left \{ \vartheta _{H} \left ( \beta - \frac{e_{\beta}}{1 - e_{\beta}} \right )
    + \frac{\beta \left ( 1 + \eta / \bar{Q}'_{1} \right )}{
    \sqrt{1 - \beta}} \frac{\sqrt{\mu}}{c}
    \frac{M_{\omega 1} + M_{\omega 2}}{\sqrt{a_{\beta}
    \left ( 1 - e_{\beta}^{2} \right )}}  \right \} ~,
\nonumber \\
M_{\omega 1} &=&
    \left ( 1 - e_{\beta}^{2} \right ) ^{3/2} \left [
    5~ I_{3} \left ( e_{\beta} \right ) ~-~
    2~ I_{2} \left ( e_{\beta} \right ) \right ] \vartheta _{H}
    \left ( \beta - \frac{e_{\beta}}{1 - e_{\beta}} \right ) ~,
\nonumber \\
M_{\omega 2} &=& \frac{5}{4} \left \{ I_{1} \left ( \xi \right ) -
    \left [ \frac{\beta ^{2} - \left ( 1 - \beta \right ) ^{2}
    e_{\beta}^{2}}{\beta ^{2} + \left ( 1 - \beta \right ) ^{2}
    e_{\beta}^{2}} \right ] ^{2}
    I_{2} \left ( \xi \right ) \right \} ~,
\nonumber \\
\xi &=& -~ \frac{2 \beta \left ( 1 - \beta \right ) e_{\beta}}{
\left ( 1 - \beta \right ) ^{2} e_{\beta}^{2} + \beta ^{2}} ~,
\nonumber \\
I_{\alpha} \left ( \varepsilon \right ) &=& \frac{1}{2 \pi} \int_{0}^{2 \pi}
       \frac{x}{\left ( 1 + \varepsilon \cos x \right ) ^{\alpha}} ~ d x ~,
       ~~~ \alpha = 1, 2, 3 ~,
\end{eqnarray}
where $\vartheta _{H} ( x ) =$ 1 if $x >$ 0,
$\vartheta _{H} ( x ) =$ 0 if $x <$ 0 (Heaviside's step function); it is
assumed that $\beta \ne e_{\beta} / ( 1 - e_{\beta} )$.

Initial conditions are given by: \\
i) Eq. (59), or, Eqs. (60) -- (62) for the set of Eqs. (186) -- (187), and, \\
ii) Eq. (59), or, Eqs. (60) -- (62) and
Eq. (72), or, Eq. (73) for the set of Eqs. (186) and (188):
Eqs. (72) -- (73) are required for initial value of $\omega$.

\subsection{Solar wind -- discussion}
The results presented in this section hold for the most simple approximation
of the solar wind action -- only radial component of the solar wind particles
is considered (Eq. (173) is taken as an approximation to more general equation
of motion presented in Kla\v{c}ka and Saniga 1993; see also Leinert and Gr\H{u}n
1990). Moreover, solar wind causes
decrease of particle's mass and the secular change of particle's mass $m$
(present in $\beta$, see Eq. (43)) is given as $dm / dt = - K A'_{eff}$ $/$
$( a_{\beta}^{2} \sqrt{1 - e_{\beta}^{2}} )$, where $K$ is a constant depending
on the material properties of the particle and $A'_{eff}$ is
the proper effective cross sectional area of the particle. If $A'_{eff}$
(area toward the Sun) is changing during the particle's motion,
one has to use $dm / dt = - K A'_{eff} ( 1 - v_{{\beta} ~R} / u ) / r^{2}$
and no averaging is possible -- considerations made in section 16 hold
for the case $A'_{eff} \equiv A'$, where $A'$ was defined above Eq. (16),
and, moreover, $A'_{eff}$ does not change during the particle's motion
(e. g., spherical particle).
Real velocity vector of solar wind particles is nonradial and the
nonradial component increases with decreasing distance from the Sun
(e. g., Stix 2002). As a consequence, real solar wind effect may
cause acceleration of meteoroids in small distances from the Sun,
instead of their deceleration.

Finally, solar wind causes also charging of meteoroids and Lorentz force
has to be taken into account in the case of submicron grains.

\section{Summary and conclusions}
The paper derives and presents relativistically covariant equation of motion
for dust particle under the action of electromagnetic radiation -- see
Eq. (40).
As for most frequent applications to systems in the universe
(e. g., meteoroids in the Solar System, dust particles in circumstellar disks),
equation of motion in the form of Eqs. (41) and (42) are sufficient:
application of Eq. (41) (for the case $F'_{ej} = 0$, $j =$ 1, 2, 3 and
under some assumption about particle's rotation) may be
found in Kocifaj {\it et al.} (2000), Kla\v{c}ka and Kocifaj (2001).
Some other accelerations may be added to the right-hand side of
Eq. (41) -- e. g., gravitational perturbations of planets, solar wind effect
(see Eq. (173) or some more precise form of equation of motion)
or some other nongravitational accelerations.

Special attention was devoted to the Poynting-Robertson effect, since this
effect is standardly used in Solar System studies. We have derived secular
orbital evolution for the P-R effect up to the second order in $v/c$.
General equation of motion for interaction between particle and incident
electromagnetic radiation shows that radiation cannot be considered as a part
of central acceleration. The central acceleration has to contain only gravity
of the central body (star/Sun); moreover, this corresponds to the physical
situation when radiation effect is considered as a disturbing effect.
In order to compare secular changes of semi-major axis and eccentricity
for real cosmic dust particle and the P-R effect,
the paper derives and presents also secular changes of these orbital
elements for the P-R effect: see Eq. (103) in section 6.2.3 and
Eq. (172) in section 8 -- advancement of pericenter/perihelion exists
even in the first order of $v/c$.

Solar wind effect is also considered in section 10. Solar wind,
in its simplest approximation, produces
secular changes of the orbital elements analogous to the P-R effect,
as for the first order in $v/c$ -- see first two equations in Eq. (181)
and Eq. (182). However, second order in solar wind effect
produces more significant changes of orbits than it is in the case of the
P-R effect, if radiation pressure is a part of central acceleration
 -- compare Eqs. (136), (146), (153) and (181).
Section 10.2 presents secular changes of orbital elements for the P-R effect and
the most simple approximation of the solar wind effect when solar gravitation
alone is considered to be a central acceleration.
in its simplest  produces

Application to larger bodies, e. g., asteroids, may be found in
Kla\v{c}ka (2000c) -- some kind of thermal emission has to be added
(quantities $Q'_{ej}$, $j =$ 1, 2, 3 present in Eqs. (40) -- (42)
have to be calculated).

\section*{Appendix A: Another formulation of the equation of motion}

(Reference to equation of number (j) of this appendix is denoted as Eq. (A~j).
Reference to equation of number (i) of the main text is denoted as Eq. (i).)

\setcounter{equation}{0}

\subsection*{Proper reference frame of the particle -- stationary particle}
The equation of motion of the particle in its proper frame of
reference is taken in the form
\begin{eqnarray}\label{A1}
\frac{d~ E'}{d~ \tau} &=& 0 ~,
\nonumber \\
\frac{d~ \vec{p'}}{d~ \tau} &=& \frac{1}{c} ~ S'~ \left ( C ' ~
		\vec{e}'_{1} \right )
		~+~ \sum_{j=1}^{3} F'_{e j} \vec{e}'_{j} ~,
\end{eqnarray}
where $E'$ is particle's energy, $\vec{p'}$ its momentum, $\tau$
is proper time, $S'$ is the flux density of radiation energy
(energy flow through unit area perpendicular to the ray per unit
time), $C'$ is the radiation pressure cross section 3 $\times$ 3
matrix, unit vector $\vec{e}'_{1}$ is directed along the path of
the incident radiation (it is supposed that beam of photons
propagate in parallel lines) and its orientation corresponds to
the orientation of light propagation; $\vec{F}'_{e}$ is emission
component of the radiation force acting on the particle (see Eq. (6)).

\subsection*{Stationary frame of reference}
Our aim is to derive equation of motion for the particle in the
stationary frame of reference.

\subsection*{Covariant equation of motion -- first case}
Let the components of the pressure cross section 3 $\times$ 3
matrix $C'$ be an orthonormal basis $\vec{e}'_{b1}$,
$\vec{e}'_{b2}$, $\vec{e}'_{b3}$. We may then write
\begin{equation}\label{A2}
\vec{e}'_{n} = \sum_{k=1}^{3} ~\left ( \vec{e}'_{bk} \cdot \vec{e}'_{n} \right )
	   \vec{e}'_{bk} ~, ~~ n = 1, 2, 3 ~,
\end{equation}
and
\begin{equation}\label{A3}
\vec{e}'_{bn} = \sum_{k=1}^{3} ~\left ( \vec{e}'_{bn} \cdot \vec{e}'_{k} \right )
	   \vec{e}'_{k} ~, ~~ n = 1, 2, 3 ~,
\end{equation}
where $\vec{e}'_{1}$, $\vec{e}'_{2}$ and $\vec{e}'_{3}$ form an
orthonormal basis. Similarly
\begin{equation}\label{A4}
C' ~\vec{e}_{1}' = \sum_{k=1}^{3} ~\left ( \vec{e}^{'T}_{bk} ~ C' ~
	   \vec{e}'_{1} \right ) ~ \vec{e}'_{bk} ~.
\end{equation}
On the basis of Eqs. (A1), (A2), (A3) and (A4), one obtains
\begin{eqnarray}\label{A5}
\frac{d~ \vec{p'}}{d~ \tau} &=& \frac{S'}{c} ~ \sum_{k=1}^{3}
     ~ \left \{ \left (
     \vec{e}^{'T}_{bk} ~ C' ~ \vec{e}'_{1} \right ) ~
     \sum_{j=1}^{3} ~\left ( \vec{e}'_{bk} \cdot \vec{e}'_{j} \right )
	   \vec{e}'_{j} \right \}
		~+~ \sum_{j=1}^{3} F'_{e j} \vec{e}'_{j} =
\nonumber \\
&=& \frac{S'}{c} ~ \sum_{j=1}^{3} ~\left \{ \sum_{k=1}^{3} ~ \left (
     \vec{e}'_{bk} \cdot \vec{e}'_{j} \right ) ~
     \left ( \vec{e}^{'T}_{bk} ~ C' ~ \vec{e}'_{1} \right ) ~
     \right \} ~ \vec{e}'_{j}
		~+~ \sum_{j=1}^{3} F'_{e j} \vec{e}'_{j} =
\nonumber \\
&=& \frac{S'}{c} ~ \sum_{j=1}^{3} ~\left \{ \sum_{k=1}^{3} ~
     \sum_{l=1}^{3} ~ \left ( \vec{e}'_{bk} \cdot \vec{e}'_{j} \right ) ~
     \left ( \vec{e}'_{bl} \cdot \vec{e}'_{1} \right ) ~
     \left ( \vec{e}^{'T}_{bk} ~ C' ~ \vec{e}'_{bl} \right ) ~
     \right \} ~ \vec{e}'_{j}
		~+~ \sum_{j=1}^{3} F'_{e j} \vec{e}'_{j} \equiv
\nonumber \\
&\equiv& \frac{S'}{c} ~ \sum_{j=1}^{3} ~\left \{ \bar{Q}'_{j} ~ A' \right \} ~
     \vec{e}'_{j}
		~+~ \sum_{j=1}^{3} F'_{e j} \vec{e}'_{j} ~,
\end{eqnarray}
which corresponds Eq. (37).
Thus, the covariant form is represented by Eq. (38).

More straightforward consideration: Let the components of the
matrix $C'$ be given in the orthonormal basis $\vec{e}'_{1}$,
$\vec{e}'_{2}$ and $\vec{e}'_{3}$ -- $C'_{kl} = \vec{e}_{k}^{'T} ~
C'~\vec{e}'_{l}$, $k$, $l$ $=$ 1 to 3. Then
\begin{equation}\label{A6}
C' ~ \vec{e}'_{1} = \sum_{j=1}^{3} \left ( \vec{e}_{j}^{'T} ~ C'~
	    \vec{e}'_{1} \right ) ~ \vec{e}'_{j} ~.
\end{equation}
Substitutions $\bar{Q}'_{j} ~A' \equiv \vec{e}_{j}^{'T} ~ C'~\vec{e}'_{1}$,
$j =$ 1, 2, 3 immediately yield Eq. (37) which has already been rewritten to
the covariant form represented by Eq. (38).

\subsection*{Covariant equation of motion -- second case}
We want to derive an equation of motion for the particle in the
frame of reference in which particle moves with actual velocity $\vec{v}$.
We will use the fact that we know this equation in the proper frame of
reference -- see Eq. (A1).

Let us have a four-vector $A^{\mu} = ( A^{0}, \vec{A} )$, where
$A^{0}$ is its time component and $\vec{A}$ is its spatial component.
Since generalized special Lorentz transformations do not form a group
(in general, composition of two generalized special Lorentz transformations is
not a generalized special Lorentz transformation), we will consider
more general Lorentz transformation. This Lorentz transformation can be
written as
\begin{equation}\label{A7}
A^{' \mu} = \Lambda^{\mu}_{~~\nu} ~ A^{\nu} ~,
\end{equation}
where summation over repeated indices is supposed (and also in all the following
equations) -- 0, 1, 2, 3 for Greek letters and 1, 2, 3 for Latin letters --
and its inverse
\begin{equation}\label{A8}
A^{\mu} = \Lambda_{\alpha}^{~~\mu} ~A^{' \alpha} ~.
\end{equation}
Important property of the Lorentz transformation
\begin{equation}\label{A9}
A^{''} = \Lambda ~ A
\end{equation}
is, that it can be composed of the following two transformations:
\begin{eqnarray}\label{A10}
A^{'} = L ~ A ~,
\nonumber \\
A^{''} = R ~ A ' ~,
\end{eqnarray}
where $L$ corresponds to generalized special Lorentz transformation
and $R$ represents rotation in 3-dimensional space:
\begin{equation}\label{A11}
A^{' \mu} = L^{\mu}_{~~\nu} ~ A^{\nu} ~,
\end{equation}
and (see Eq. (13))
\begin{eqnarray}\label{A12}
L^{0}_{~~0} &=& \gamma	~,
\nonumber \\
L^{0}_{~~i} &=& L^{i}_{~~0} = -~ \gamma ~( \vec{v} / c )_{i} ~,
		      ~~ i = 1, 2, 3 ~,
\nonumber \\
L^{i}_{~~j} &=& \delta_{i j} ~+~ ( \gamma ~-~ 1 ) ~( \vec{v} )_{i}
	      ~ ( \vec{v} )_{j} ~/~ \vec{v}^{2}   ~,
	     ~~ i = 1, 2, 3 ~,~~ j = 1, 2, 3 ~,
\end{eqnarray}
where $\delta_{i j} =$ 1 if $i = j$ and
$\delta_{i j} =$ 0 if $i \ne j$,
\begin{equation}\label{A13}
A^{\mu} = L_{\alpha}^{~~\mu} ~A^{' \alpha} ~,
\end{equation}
where (see Eq. (14))
\begin{eqnarray}\label{A14}
L_{\alpha}^{~~\beta} &=& \eta _{\alpha ~ \rho} ~ \eta ^{\beta ~ \gamma} ~
		  L^{\rho}_{~~\gamma} ~,
\nonumber \\
\eta _{\alpha ~ \beta} &=& diag ( +~1, -~1, -~1, -~1 )
\end{eqnarray}
and
\begin{eqnarray}\label{A15}
L_{0}^{~~0} &=& \gamma ~,
\nonumber \\
L_{0}^{~~i} &=& L_{i}^{~~0} = \gamma ~( \vec{v} / c )_{i} ~,
		~~ i = 1, 2, 3 ~,
\nonumber \\
L_{i}^{~~j} &=& \delta_{i j} ~+~ ( \gamma ~-~ 1 ) ~( \vec{v} )_{i}
	      ~ ( \vec{v} )_{j} ~/~ \vec{v}^{2}   ~,
	     ~~ i = 1, 2, 3 ~,~~ j = 1, 2, 3 ~,
\end{eqnarray}
\begin{eqnarray}\label{A16}
R^{0}_{~~0} &=& 1 ~,
\nonumber \\
R^{0}_{~~i} &=& R^{i}_{~~0} = 0 ~,
		      ~~ i = 1, 2, 3 ~,
\nonumber \\
R^{i}_{~~j} &=& r_{i j} ~, ~~ i = 1, 2, 3 ~,~~ j = 1, 2, 3 ~,
\end{eqnarray}
where $r_{i j}$ can be expressed in terms of Euler angles, and, moreover,
orthogonality conditions are fulfilled:
\begin{eqnarray}\label{A17}
\sum_{i = 1}^{3} r_{i j} ~r_{i k} = \delta_{j k} ~,
				    ~~ j = 1, 2, 3 ~,~~ k = 1, 2, 3  ~,
\nonumber \\
\sum_{j = 1}^{3} r_{i j} ~r_{k j} = \delta_{i k} ~,
				    ~~ i = 1, 2, 3 ~,~~ k = 1, 2, 3  ~.
\end{eqnarray}

On the basis of Eqs. (A9) and (A10) we can write
\begin{equation}\label{A18}
\Lambda^{\mu}_{~~\nu} = R^{\mu}_{~~\varrho}~ L^{\varrho}_{~~\nu} ~.
\end{equation}
Eqs. (A12), (A16) and (A18) yield
\begin{eqnarray}\label{A19}
\Lambda^{0}_{~~0} &=& L^{0}_{~~0} = \gamma  ~,
\nonumber \\
\Lambda^{0}_{~~i} &=& L^{0}_{~~i} = -~ \gamma ~( \vec{v} / c )_{i} ~,
		      ~~ i = 1, 2, 3 ~.
\end{eqnarray}

Finally, requirement $A^{\mu} A_{\mu} = A^{' \mu} A' _{\mu}$ yields,
\begin{equation}\label{A20}
\Lambda^{\mu}_{~~\nu} ~ \Lambda_{\mu}^{~~\varrho} = \delta^{\varrho}_{\nu} ~,
\end{equation}
where $\delta^{\varrho}_{\nu} = 1$ if $\varrho = \nu$ and
$\delta^{\varrho}_{\nu} = 0$ if $\varrho \ne \nu$; Eq. (A7) was also used:
$A^{' \mu} = \Lambda^{\mu}_{~~\nu} ~ A^{\nu}$ ~,
$A'_{\mu} = \Lambda_{\mu}^{~~\nu} ~ A_{\nu}$ ~.

\subsection*{Incoming radiation}
Applying Eqs. (A7) and (A19) to quantity $( E_{i} / c, \vec{p}_{i} )$
(four-momentum per unit time -- proper time is a scalar quantity) and
taking into account also the fact that
$\vec{p}_{i} = E_{i} / c ~\vec{e}_{1}$,
we can write
\begin{eqnarray}\label{A21}
E_{i} '  &=& E_{i}  ~w_{1} ~,
\end{eqnarray}
where
\begin{equation}\label{A22}
w_{1} \equiv \gamma ~ ( 1 ~-~ \vec{v} \cdot \vec{e}_{1} / c ) ~.
\end{equation}

Using the fact that $p^{\mu} = ( h ~\nu , h ~\nu ~ \vec{e}_{1} )$
for photons, we have
\begin{eqnarray}\label{A23}
\nu ' &=& \nu  ~w_{1} ~.
\end{eqnarray}

We have four-vector
$p_{i}^{\mu} = ( E_{i} / c, \vec{p}_{i} ) = ( 1, \vec{e}_{1} ) E_{i} / c$
$= ( 1 / w_{1}, \vec{e}_{1} / w_{1} ) ~ w_{1} ~ E_{i} / c$
$\equiv$ $b_{1}^{\mu} ~ w_{1} ~ E_{i} / c$. We have found a new four-vector
$b_{1}^{\mu}$, which is given as
$b_{1}^{' \mu} = \Lambda^{\mu}_{~~\nu} ~ b_{1}^{\nu}$
in the proper frame of reference of the particle:
$b_{1}^{' \mu} = ( 1, \vec{e} '_{1} )$. The transformation of
space components between $b_{1}^{\mu}$ and $b_{1}^{' \mu}$ corresponds to
aberration of light.

For monochromatic radiation the flux density of radiation energy
becomes
\begin{equation}\label{A24}
S' = n' ~h~ \nu ' ~ c ~; ~~~ S = n ~h~ \nu  ~ c ~,
\end{equation}
where $n$ and $n'$ are concentrations of photons (photon number
densities) in the corresponding reference frames. We also have
continuity equation
\begin{equation}\label{A25}
\partial _{\mu} ~ j^{\mu} = 0 ~, ~~~j^{\mu} = ( c~n, c~n~ \vec{e}_{1} ) ~,
\end{equation}
with current density $j^{\mu}$. Application of Eqs. (A7) and (A19) then
yields
\begin{equation}\label{A26}
n' = w_{1} ~ n ~.
\end{equation}
Using Eqs. (A23), (A24) and (A26) we finally obtain
\begin{equation}\label{A27}
S' = w_{1}^{2} ~ S ~.
\end{equation}
Eqs. (A21) and (A27) then together give $E_{i} = w_{1} ~ S
~A'$, $\vec{p}_{i} = w_{1} ~ S ~A'~\vec{e}_{1} / c$.

\subsection*{Covariant equation of motion}
Inspiration comes from the fact that space components of four-momentum
are written as a product with unit vector $\vec{e}'_{1}$. We know that
this unit vector can be generalized to a four-vector
\begin{eqnarray}\label{A28}
b_{1}^{\mu} &=& \left ( 1 ~/~ w_{1}, \vec{e}_{1} ~/~w_{1} \right )
\nonumber \\
w_{1} &\equiv& \gamma ~ ( 1 ~-~ \vec{v} \cdot \vec{e}_{1} / c ) ~.
\end{eqnarray}
Moreover, we know that $w_{1}^{2} ~S ~/~ c$ is a scalar quantity -- invariant
of the Lorentz transformation (see Eqs. (A27)).

The idea is to write covariant equation of motion in the form
\begin{equation}\label{A29}
\frac{d~ p^{\mu}}{d~ \tau} = \frac{w_{1}^{2}~S}{c} ~ G^{\mu~\nu} ~b_{1~ \nu}
		 ~+~ \frac{1}{c} \sum_{j=1}^{3} F'_{e j} \left (
		 c ~ b_{j}^{\mu} - u^{\mu} \right ) ~,
\end{equation}
if the result for the emission component of the radiation force acting
on the particle was added (see sections 2 and 3 in the main text).
The only problem is to find components of the tensor of the second rank
$G^{\mu~\nu}$.

In order to find $G^{\mu~\nu}$, we will proceed in two steps. At first, we
will rewrite Eq. (A29) in the proper frame of reference of the particle.
Comparison with Eq. (A1) will yield components of $G^{' ~\mu~\nu}$.
The second step is transformation from $G^{' ~\mu~\nu}$ to $G^{\mu~\nu}$.

In the proper frame of reference, Eq. (A29) yields
\begin{eqnarray}\label{A30}
\frac{d~ E'}{d~ \tau} &=& \frac{w_{1}^{2}~S}{c} ~\left \{ G^{'~0~0} ~-~
\sum_{j=1}^{3} G^{'~0~j} ~ \left ( \vec{e}'_{1} \right ) _{j} \right \} ~,
\nonumber \\
\frac{d~ \left ( \vec{p'} \right ) _{k}}{d~ \tau} &=&  \frac{w_{1}^{2}~S}{c} ~
\left \{ G^{'~k~0} ~-~
\sum_{j=1}^{3} G^{'~k~j} ~ \left ( \vec{e}'_{1} \right ) _{j} \right \}
		~+~ \sum_{j=1}^{3} F'_{e j} ~
		\left ( \vec{e}'_{j}  \right ) _{k}~,
\end{eqnarray}
where the term $1/w_{1}'$ in brackets is omitted due to the simple
fact that it equals 1 in the proper frame of reference. Comparison
with Eq. (A1) yields
\begin{eqnarray}\label{A31}
G^{'~0~0} = G^{'~0~j} = G^{'~j~0} = 0 ~, ~~ j = 1, 2, 3 ~,
\nonumber \\
G^{'~k~j} = -~ C '_{k~j} ~, ~~j = 1, 2, 3 ~, ~~k = 1, 2, 3 ~.
\end{eqnarray}

In order to find $G^{\mu~\nu}$, we have to use the Lorentz
transformation
\begin{equation}\label{A32}
G^{\mu ~\nu} = \Lambda_{\alpha}^{~~\mu} ~ \Lambda_{\beta}^{~~\nu} ~
	   G^{'~\alpha~\beta} ~,
\end{equation}
where
\begin{eqnarray}\label{A33}
\Lambda_{\alpha}^{~~\beta} &=& \eta _{\alpha ~ \rho} ~ \eta ^{\beta ~ \gamma} ~
		  \Lambda^{\rho}_{~~\gamma} ~,
\nonumber \\
\eta _{\alpha ~ \beta} &=& diag ( +~1, -~1, -~1, -~1 ) ~.
\end{eqnarray}

Usage of the generalized special Lorentz transformation (Eqs. (A12) or (A15))
yields the results of Kimura {\it et al.} (2002) -- the authors take the space
tensor $C'$ as a scalar quantity under Lorentz transformation.
However, generalized special Lorentz transformation has not to be used.
The reason is that any body in torque-free, accelerated
motion undergoes rotation. This effect is known as the Thomas precession
(see e. g., Robertson and Noonan 1968, pp. 66 -- 69). Thus, the
process of complete relativistic derivation and the corresponding result
presented by Kimura {\it et al.} (2002) is incorrect. We can mention that
the last statement is evident already at first glance: Kimura {\it et al.} (2002)
take the radiation pressure cross section 3 $\times$ 3 matrix -- space tensor --
as a relativistically invariant quantity.

\subsection*{Consistency of the covariant formulations}
We have obtained equation of motion in the form of Eq. (A29). Another form of
covariant equation is presented in Eq. (35) (Eq. (38)).
Are these equations consistent?

Eq. (38) is covariant equation of motion in the form corresponding to
($\beta ^{\mu} \equiv u^{\mu} / c$)
\begin{equation}\label{A34}
\left ( \frac{d~ p^{\mu}}{d~ \tau} \right ) _{I} = \frac{w_{1}^{2}~S~A'}{c} ~
       \sum_{j=1}^{3} \bar{Q}'_{j} \left (
       b_{j}^{\mu} ~-~ \beta^{\mu} \right )
		 ~+~ \frac{1}{c} \sum_{j=1}^{3} F'_{e j} \left (
		 c ~ b_{j}^{\mu} - u^{\mu} \right ) ~.
\end{equation}
This appendix has discussed an equation of motion of the form
\begin{equation}\label{A35}
\left ( \frac{d~ p^{\mu}}{d~ \tau} \right ) _{II} = \frac{w_{1}^{2}~S}{c} ~
		 G^{\mu~\nu} ~ b_{1~ \nu}
		 ~+~ \frac{1}{c} \sum_{j=1}^{3} F'_{e j} \left (
		 c ~ b_{j}^{\mu} - u^{\mu} \right ) ~.
\end{equation}
We want to show that Eqs. (A34) and (A35) are equivalent, i. e., that
$\left ( d~ p^{\mu} ~/~ d~ \tau \right ) _{I}$ $=$
$\left ( d~ p^{\mu} ~/~ d~ \tau \right ) _{II}$.

We will not write the terms $F'_{e j}$, in what follows.

Multiplication of Eq. (A34) by four-vector $b_{k~\mu}$ (and summation over
$\mu$) yields
\begin{equation}\label{A36}
\left ( \frac{d~ p^{\mu}}{d~ \tau} \right ) _{I} ~b_{k~ \mu}  =
    \frac{w_{1}^{2}~S~A'}{c} ~\sum_{j=1}^{3} \bar{Q}'_{j} \left (
    b_{j}^{\mu}~b_{k~ \mu}  ~-~ \beta^{\mu}~ b_{k~ \mu}  \right )
		~~, ~k = 1, 2, 3 ~.
\end{equation}
In calculations of $\beta^{\mu} ~b_{k~ \mu}$ we will use the fact that it
represents scalar product of two four-vectors. Thus, its value is independent
on the frame of reference. For the proper frame of reference
\begin{equation}\label{A37}
\beta^{\mu} ~b_{k~ \mu} = 1 ~~, ~k = 1, 2, 3 ~.
\end{equation}
It can be easily verified, that
\begin{equation}\label{A38}
b_{j}^{\mu}~b_{k~ \mu} = 1 ~-~ \vec{e}_{j} ' \cdot \vec{e}_{k} ' =
	      1 ~-~ \delta_{j~k} ~~, ~j = 1, 2, 3 ~~, ~k = 1, 2, 3 ~,
\end{equation}
since in the optics of scattering processes it is assumed (defined) that unit
vectors $\vec{e}_{1}'$, $\vec{e}_{2}'$ and $\vec{e}_{3}'$ are orthogonal. Thus,
we obtain (inserting results of Eqs. (A37) and (A38) into Eq. (A36))
\begin{equation}\label{A39}
\left ( \frac{d~ p^{\mu}}{d~ \tau} \right ) _{I} ~b_{k~ \mu}  = -~
	   \frac{w_{1}^{2}~S~A'}{c} ~\bar{Q}'_{k} ~, ~~ k = 1, 2, 3 ~.
\end{equation}

Multiplication of Eq. (A35) by four-vector $b_{k~\mu}$ (and summation over
$\mu$) yields
\begin{equation}\label{A40}
\left ( \frac{d~ p^{\mu}}{d~ \tau} \right ) _{II} ~b_{k~ \mu}  =
	   \frac{w_{1}^{2}~S}{c} ~G^{\mu~\nu} ~ b_{1~\nu} ~b_{k~ \mu}
		~, ~~k = 1, 2, 3 ~.
\end{equation}
Again, the value is independent on the frame of reference.
For the proper frame of reference Eq. (A31) yields
\begin{eqnarray}\label{A41}
\left ( \frac{d~ p^{\mu}}{d~ \tau} \right ) _{II} ~b_{k~ \mu}  &=& ~-
	   \frac{w_{1}^{2}~S}{c} ~
     C'_{~j~i} ~  ( \vec{e}'_{1} )_{i} ~ ( \vec{e}_{k} ')_{j}
\nonumber \\
&\equiv&  -~ \frac{w_{1}^{2}~S}{c} ~
     ( \vec{e}_{k} ')^{T} ~ ( C'~ \vec{e}'_{1} ) ~, ~~ k = 1, 2, 3 ~.
\end{eqnarray}
It was already shown (see Eq. (A6) and the text below it)
that right-hand sides of Eqs. (A39) and (A41) are identical.
Thus, also left-hand sides of Eqs. (A39) and (A40) are identical:
\begin{equation}\label{A42}
\left ( \frac{d~ p^{\mu}}{d~ \tau} \right ) _{I} ~b_{k~ \mu} =
\left ( \frac{d~ p^{\mu}}{d~ \tau} \right ) _{II} ~b_{k~ \mu} ~, ~~ k = 1, 2, 3 ~.
\end{equation}
If we take into account that four-vectors $b_{k}^{\mu}$
may be taken in various ways, Eq. (A42) yields
\begin{equation}\label{A43}
\left ( \frac{d~ p^{\mu}}{d~ \tau} \right ) _{I} =
\left ( \frac{d~ p^{\mu}}{d~ \tau} \right ) _{II}  ~.
\end{equation}
Thus, Eqs. (A34) and (A35) are equivalent, Q. E. D..

\section*{Appendix B: Einstein's example}

(Reference to equation of number (j) of this appendix is denoted as Eq. (B~j).
Reference to equation of number (i) of the main text is denoted as Eq. (i).)

\setcounter{equation}{0}

Let us consider a plane mirror moving (at a given moment) along
x-axis (system S) with velocity $\vec{v} = ( v, 0, 0 )$, $v > 0$;
the mirror is perpendicular to the x-axis (the plane of the mirror is
parallel to the yz-plane). A beam of incident (hitting) photons is
characterized by unit vector
$\vec{S} ' = ( \cos \theta ', \sin \theta ', 0 )$
in the proper frame (primed quantities) of the mirror. Reflected beam
is described by the unit vector
$\vec{e} ' = ( -~ \cos \theta ', \sin \theta ', 0 )$
(in the proper frame S').

The problem is: Find equation of motion of the mirror in the frame
of reference S.

\subsection*{Solution 1: trivial manner}
Consider one photon (frequency $f'$) in the proper frame of the mirror.
Since the directions (and orientations) of the incident and outgoing photons are
characterized by
\begin{eqnarray}\label{B1}
\vec{S} ' &=& \left ( + ~ \cos \theta ', \sin \theta ', 0 \right ) ~,
\nonumber \\
\vec{e} ' &=& \left ( - ~ \cos \theta ', \sin \theta ', 0  \right ) ~,
\end{eqnarray}
we can immediately write
\begin{eqnarray}\label{B2}
p_{i}^{' \mu} &=& \frac{h~f'}{c} ~
		  \left ( 1, +~ \cos \theta ', \sin \theta ', 0 \right ) ~,
\nonumber \\
p_{o}^{' \mu} &=& \frac{h~f'}{c} ~
		  \left ( 1, -~ \cos \theta ', \sin \theta ', 0 \right ) ~,
\end{eqnarray}
for the four-momentum of the photon before interaction with the mirror
and after the interaction.

As a consequence, the mirror obtains four-momentum
\begin{equation}\label{B3}
p^{' \mu} = p_{i}^{' \mu} ~-~ p_{o}^{' \mu} = \frac{h~f'}{c} ~
		  \left ( 0, 2~ \cos \theta ', 0, 0 \right ) ~.
\end{equation}

Application of the special Lorentz transformation to Eq. (B3) yields
\begin{equation}\label{B4}
p^{\mu} = \frac{h~f'}{c} ~2 ~ \gamma ~\left ( \cos \theta ' \right ) ~
	  \left ( \beta, 1, 0, 0 \right ) ~,
\end{equation}
where, as standardly abbreviated,
\begin{equation}\label{B5}
\gamma = 1 ~/~ \sqrt{1 ~-~ \beta ^{2}}~~; ~~~\beta = v ~/~c ~.
\end{equation}

On the basis of Eq. (B4), we can immediately write equation of motion
of the mirror
\begin{equation}\label{B6}
\frac{d p^{\mu}}{d \tau} = \frac{E_{i}'}{c} ~ 2 ~ \gamma ~
			   \left ( \cos \theta ' \right ) ~
			   \left ( \beta, 1, 0, 0 \right ) ~,
\end{equation}
where $E_{i}'$ is the total energy (per unit time) of the incident radiation
measured in the proper frame of reference.

\subsection*{Solution 2: application of general theory presented in Sec. 3
of the main text}
We have to choose orthonormal vectors in the systems S':
we will use $\vec{e}_{1} ' \equiv \vec{S} '$
and $\vec{e}_{2} '$ and one can easily find
\begin{eqnarray}\label{B7}
\vec{e}_{1} ' &\equiv& \vec{S} ' = \left ( + ~ \cos \theta ', \sin \theta ', 0 \right ) ~,
\nonumber \\
\vec{e}_{2} ' &=& \left ( - ~ \sin \theta ', \cos \theta ', 0  \right ) ~.
\end{eqnarray}
We have to write ($Q_{3} ' = 0$)
\begin{equation}\label{B8}
\vec{p} ' =  \frac{h~f'}{c} ~ \left ( Q_{1}' ~\vec{e}_{1} '
	    ~+~ Q_{2} ' ~\vec{e}_{2} ' \right ) ~.
\end{equation}
On the basis of Eqs. (B3), (B7) and (B8) we have ($Q_{3} ' =$ 0)
\begin{equation}\label{B9}
Q_{1} ' = 2 ~( \cos \theta ' ) ^{2} ~, ~~
Q_{2} ' = - ~2 ~( \sin \theta ') ~( \cos \theta ' )  ~.
\end{equation}
Other prescription yields (see Eq. (34))
\begin{eqnarray}\label{B10}
b_{1}^{0} &=& \gamma ~ ( 1 + \vec{v} \cdot \vec{e}_{1} ' / c ) =
	      \gamma ( 1 + \beta  \cos \theta ' ) ~,
\nonumber \\
\vec{b}_{1} &=& \vec{e}_{1} ' +
	    \left [ ( \gamma - 1 )  \vec{v} \cdot \vec{e}_{1} ' / \vec{v}^{2}
	    + \gamma / c \right ]  \vec{v} =
      \left ( \gamma \cos \theta ' + \gamma  \beta, \sin \theta ', 0 \right ) ~,
\end{eqnarray}
\begin{eqnarray}\label{B11}
b_{2}^{0} &=& \gamma ~ ( 1 + \vec{v} \cdot \vec{e}_{2} ' / c ) =
	      \gamma ( 1 - \beta  \sin \theta ' ) ~,
\nonumber \\
\vec{b}_{2} &=& \vec{e}_{2} ' +
	    \left [ ( \gamma - 1 ) \vec{v} \cdot \vec{e}_{2} ' / \vec{v}^{2}
	    + \gamma / c \right ]  \vec{v} =
       \left ( - \gamma \sin \theta ' + \gamma  \beta, \cos \theta ', 0 \right ) ~.
\end{eqnarray}
Inserting Eqs. (B9) -- (B11) (and $F'_{ej} =$ 0 for $j =$ 1, 2, 3)
into Eqs. (29) -- (30), one obtains
\begin{eqnarray}\label{B12}
\frac{d p^{\mu}}{d \tau} &=& \frac{E_{i}'}{c} ~\left \{
	      \left [ 2 ( \cos \theta ' ) ^{2}  \right ]
	      \left ( b_{1}^{\mu} - \beta^{\mu} \right ) +
	      \left [ - 2 ( \sin \theta ') ( \cos \theta ' )
	      \right ] ~ \left ( b_{2}^{\mu} - \beta^{\mu}
	      \right ) \right \}
\nonumber \\
			 &=& \frac{E_{i}'}{c}  2  \gamma
			     \left ( \cos \theta ' \right )
			     \left ( \beta, 1, 0, 0 \right ) ~.
\end{eqnarray}

Unit vectors $\vec{e}_{1} ' \equiv \vec{S} '$ and $\vec{e}_{2} '$ are used.
They are orthonormal in the system S'. However, corresponding vectors are
not orthogonal in the system S.
\begin{eqnarray}\label{B13}
\vec{e}_{1} &=& \frac{1}{w '}  \left \{ \vec{e}_{1} ' +
	    \left [ ( \gamma - 1 ) ~ \vec{v} \cdot \vec{e}_{1} ' / \vec{v}^{2}
	    + \gamma / c \right ]  \vec{v} \right \} ~,
\nonumber \\
w ' &=& \gamma  \left ( 1 + \vec{v} \cdot \vec{e}_{1} ' / c \right ) ~,
\end{eqnarray}
and analogous equation holds for vector $\vec{e}_{2}$.
Inserting Eqs. (B7), one obtains:
\begin{eqnarray}\label{B14}
\vec{e}_{1} &=& \left \{
	    \frac{\cos \theta ' + \beta}{1 + \beta  \cos \theta '},
	    \frac{\sin \theta '}{\gamma
	    \left ( 1 + \beta  \cos \theta '\right )}, 0 \right \} ~,
\nonumber \\
\vec{e}_{2} &=& \left \{
	    \frac{-~\sin \theta ' + \beta}{1 - \beta  \sin \theta '},
	    \frac{\cos \theta '}{\gamma
	    \left ( 1 - \beta  \sin \theta ' \right )}, 0 \right \} ~.
\end{eqnarray}
It can be easily verified that scalar product of these two vectors is nonzero,
in general.

\subsection*{Comparison with Einstein's result}
Inserting $E'_{i} = w^{2} ~S~ A'_{mirror} ~\cos \theta '$ into Eq. (B6)
(or Eqs. (B12)) and using Eq. (B14)
($\vec{e}_{1} \equiv ( \cos \theta,  \sin \theta, 0 )$)
for the purpose of obtaining
$\cos \theta ' = ( \cos \theta - \beta) / (1 - \beta ~\cos \theta)$, one
easily obtains: \\
i) $dE/d \tau = 2~\gamma^{3}~ S~A'_{mirror}~(\cos \theta - \beta )^{2} ~ \beta$;
using definition of radiation pressure $dE/dt \equiv P~v~A'_{mirror}$, we have
$P = 2~(S~/~c)~( \cos \theta - \beta)^{2} ~/~( 1 - \beta ^{2})$, or, \\
ii) $dp/d \tau = 2~\gamma^{3}~ (S~A'_{mirror}~/~c) ~(\cos \theta - \beta )^{2}$;
using definition of radiation pressure $P \equiv (dp/dt)~/~A'_{mirror}$, we have
$P = 2~(S~/~c)~( \cos \theta - \beta)^{2} ~/~( 1 - \beta ^{2})$.

Result for $P$ is consistent with the result presented in
Einstein (1905).

\section*{Appendix C: Equation of motion presented by Kimura {\it et al.} (2002)}

(Reference to equation of number (i) of this appendix is denoted as Eq. (C~i).
Reference to equation of number (j) of the appendix B is denoted as Eq. (B~j).
Reference to equation of number (k) of the main text is denoted as Eq. (k).)

\setcounter{equation}{0}

It is said that the paper by Kimura {\it et al.} (2002) is relevant;
irrelevant are papers by Kla\v{c}ka (2000a, 2000b, 2000c), or derivations and
results presented in Sec. 3 of this paper. Thus, it is important to clarify the
situation by presentation of detailed arguments.

Kimura {\it et al.} (2002) derive and present equation of motion
of the form
\begin{eqnarray}\label{C1}
\frac{d~ \vec{v}}{d ~t} &=&  \frac{S ~A'}{m~c} ~ \times \vec{\zeta} ~,
\nonumber \\
\vec{\zeta} &=& Q_{1} ' ~ \left [ \left ( 1~-~ \vec{v} \cdot \vec{k}_{1} / c \right ) ~
	\vec{k}_{1} ~-~ \vec{v} / c \right ]  ~+~
\nonumber \\
& &	Q_{2} ' ~ \left [ \left ( 1~-~2~ \vec{v} \cdot \vec{k}_{1} / c \right ) ~
	\vec{k}_{2} ~+~ \left ( \vec{v} \cdot \vec{k}_{2} / c \right ) ~
	\vec{k}_{1} \right ]  ~+~
\nonumber \\
& &	Q_{3} ' ~  \left ( 1~-~2~ \vec{v} \cdot \vec{k}_{1} / c \right ) ~
	\vec{k}_{3} ~,
\nonumber \\
\vec{e}'_{1} &=& ( 1 ~+~ \vec{v} \cdot \vec{k}_{1} / c ) ~ \vec{k}_{1} ~-~
	  \vec{v} / c ~,
\nonumber \\
\vec{e}'_{2} &=& \vec{k}_{2} ~+~ \left (
	  \vec{v} \cdot \vec{k}_{2} / c \right ) ~ \vec{k}_{1} ~,
\nonumber \\
\vec{e}'_{3} &=& \vec{k}_{3} ~,
\nonumber \\
\vec{e}'_{i} \cdot \vec{e}'_{j} &=& \delta _{ij} ~, ~~~
		   i, j \in \left \{ 1, 2, 3 \right \} ~,
\nonumber \\
\vec{k}_{i} \cdot \vec{k}_{j} &=& \delta _{ij} ~, ~~~
		   i, j \in \left \{ 1, 2, 3 \right \} ~,
\nonumber \\
\vec{v} \cdot \vec{k}_{3} &=& 0 ~,
\end{eqnarray}
(see Eqs. (1), (2), (3), (10), (14) in Kimura {\it et al.} 2002) and it is
supposed that $\vec{F}'_{e} =$ 0.

There have appeared two independent suggestions: \\
i) Eq. (C1) is correct and Eq. (36) is incorrect. \\
ii) Eq. (C1) is equivalent to Eq. (36). As a proof, the following argument
is presented. Equations of motion presented by Kla\v{c}ka (2000a, 2000b, 2000c)
and Eq. (36) (for the case $\vec{F}'_{e} =$ 0) and Kimura {\it et al.} (2002)
are equivalent and all the difference comes from usage of different basic vectors.
The set of non-orthogonal unit vectors $\{ \vec{e}_{j} ; j = 1, 2, 3 \}$
(see Eq. (36)) is replaced by new set of orthonormal
vectors $\{ \vec{k}_{j} ; j = 1, 2, 3 \}$ in the way
\begin{eqnarray}\label{C2}
\vec{e}_{1} &=& \vec{k}_{1} ~,
\nonumber \\
\vec{e}_{2} &=& \vec{k}_{2} + ( \vec{v} \cdot \vec{k}_{2} / c ) ~
( \vec{k}_{1} - \vec{k}_{2} ) + \vec{v} / c ~,
\nonumber \\
\vec{e}_{3} &=& \vec{k}_{3} - ( \vec{v} \cdot \vec{k}_{3} / c ) ~
\vec{k}_{3} + \vec{v} / c ~,
\nonumber \\
\vec{v} \cdot \vec{k}_{3} &=& 0 ~.
\end{eqnarray}
Eq. (C2) yields
$\vec{e}'_{1} = ( 1 + \vec{v} \cdot \vec{k}_{1} / c ) \vec{k}_{1} - \vec{v} / c$,
$\vec{e} '_{2} = \vec{k}_{2} + ( \vec{v} \cdot \vec{k}_{2} / c ) \vec{k}_{1}$
and $\vec{e} '_{3} = \vec{k}_{3}$, which is equivalent to Eq. (C1).

\subsection*{Kimura {\it et al.} (2002) and Einstein's example}
Let us consider a plane mirror moving (at a given moment) along
x-axis (system S) with velocity $\vec{v} = ( v, 0, 0 )$, $v > 0$;
the mirror is perpendicular to the x-axis (the plane of the mirror is
parallel to the yz-plane). A beam of incident (hitting) photons is
characterized by unit vector
$\vec{S} ' = ( \cos \theta ', \sin \theta ', 0 )$
in the proper frame (primed quantities) of the mirror. Reflected beam
is described by the unit vector
$\vec{e} ' = ( -~ \cos \theta ', \sin \theta ', 0 )$
(in the proper frame S').

We are interested in application of Kimura {\it et al.} (2002) general
equation of motion to this example.

On the basis of considerations presented in Appendix B, Eqs. (B7) and (B9)
immediately yield
\begin{eqnarray}\label{C3}
\vec{e}_{1} ' &=& \left ( + ~ \cos \theta ', \sin \theta ', 0 \right ) ~,
\nonumber \\
\vec{e}_{2} ' &=& \left ( - ~ \sin \theta ', \cos \theta ', 0  \right ) ~.
\nonumber \\
Q_{1} ' &=& 2 ~( \cos \theta ' ) ^{2} ~, ~~
Q_{2} ' = - ~2 ~( \sin \theta ') ~( \cos \theta ' )  ~,~~ Q_{3} ' = 0 ~.
\end{eqnarray}
Eq. (C1) yields for orthonormal vectors
\begin{eqnarray}\label{C4}
\vec{k}_{1} &=& ( 1 ~-~ \vec{v} \cdot \vec{e}'_{1} / c ) ~ \vec{e}'_{1} ~+~
	  \vec{v} / c ~,
\nonumber \\
\vec{k}_{2} &=& \vec{e}'_{2} ~-~ \left (
	  \vec{v} \cdot \vec{e}'_{2} / c \right ) ~ \vec{e}'_{1} ~.
\end{eqnarray}
Using Eqs. (C3) and (C4),
\begin{eqnarray}\label{C5}
\vec{k}_{1} &=& \left \{ \cos \theta ' ~+~ \frac{v}{c} ~
		\left ( \sin \theta ' \right ) ^{2} , \sin \theta ' ~-~
		\frac{v}{c} ~ \cos \theta ' ~ \sin \theta ' ,
		0 \right \} ~,
\nonumber \\
\vec{k}_{2} &=& \left \{ -~ \sin \theta ' ~+~ \frac{v}{c} ~
		\cos \theta ' ~ \sin \theta ' , \cos \theta ' ~+~ \frac{v}{c}
		~ \left ( \sin \theta ' \right ) ^{2} ,
		0 \right \} ~.
\end{eqnarray}

Inserting Eqs. (C3) and (C5) into Eqs. (C1), one easily obtains
\begin{eqnarray}\label{C6}
\frac{d \vec{v}}{d t} &=& \frac{S~ A'_{mirror} ~\cos \theta '}{m~c} ~\left \{
	   2 ~\cos \theta ' ~-~ \frac{v}{c} \left [ 1 ~+~ 2 ~
	   \left ( \cos \theta ' \right ) ^{2} \right ] ,
	   0, 0 \right \} ~.
\end{eqnarray}

Using the relation
$E'_{i} = S~ ( 1 ~-~ 2 ~v ~\cos \theta ' / c )$ $A'_{mirror} ~\cos \theta '$
in Eqs. (B6) or (B12), the correct result can be written in the form
\begin{eqnarray}\label{C7}
\frac{d \vec{v}}{d t} &=& \frac{S~ A'_{mirror} ~\cos \theta '}{m~c} ~\left \{
       2~ \left ( 1 ~-~ 2 ~ \frac{v}{c} ~\cos \theta ' \right ) ~\cos \theta ' ,
       0, 0 \right \} ~.
\end{eqnarray}

Eq. (C6) was obtained on the basis of equation of motion presented
by Kimura {\it et al.} (2002), while the correct result corresponds to
Eq. (C7). Thus, Kimura {\it et al.} (2002) are incorrect -- their general
equation of motion yields result not consistent with Einstein's result
(Einstein 1905).

\subsection*{Physics of transformation represented by Eq. (C2)}
Transformation $\left \{ \vec{e}'_{i} ; i = 1, 2, 3 \right \}$
$\longrightarrow$ $\left \{ \vec{e}_{i} ; i = 1, 2, 3 \right \}$ from
frame of reference $S'$ to frame of reference $S$ corresponds to
aberration of light (see Eq. (36) or heuristic derivation Sec. 3.4).
In the given frame of reference $S$ we may want to define two sets of
unit vectors $\left \{ \vec{e}_{i} ; i = 1, 2, 3 \right \}$ and
$\left \{ \vec{l}_{i} ; i = 1, 2, 3 \right \}$, where
$\vec{e}_{1} = \vec{l}_{1}$
and $\vec{e}_{i} \cdot \vec{e}_{j} \ne \delta_{ij}$,
$\vec{l}_{i} \cdot \vec{l}_{j} = \delta_{ij}$,
$i, j \in \left \{ 1, 2, 3 \right \}$.
Relation between
$\left \{ \vec{l}_{i} ; i = 1, 2, 3 \right \}$ and
$\left \{ \vec{e}_{i} ; i = 1, 2, 3 \right \}$ corresponds to pure spatial
rotation (geometry) and it can be described as
\begin{eqnarray}\label{C8}
\vec{e}_{1} &=& \vec{l}_{1} ~,
\nonumber \\
\vec{e}_{2} &=& \xi_{1} ~\vec{l}_{1} ~+~ \xi_{2} ~\vec{l}_{2} ~+~
		\xi_{3} ~\vec{l}_{3}~,
\nonumber \\
\vec{e}_{3} &=& \eta_{1} ~\vec{l}_{1} ~+~ \eta_{2} ~\vec{l}_{2} ~+~
		\eta_{3} ~\vec{l}_{3} ~.
\end{eqnarray}
The conditions
\begin{eqnarray}\label{C9}
\vec{l}_{i} \cdot \vec{l}_{j} &=& \delta_{ij} ~, ~~~
i, j \in \left \{ 1, 2, 3 \right \}
\end{eqnarray}
lead to
\begin{eqnarray}\label{C10}
\sum_{i=1}^{3} \xi_{i} ^{2} &=& 1 ~, ~~~~~
\sum_{i=1}^{3} \eta_{i} ^{2} = 1 ~,
\nonumber \\
\xi_{1} &=& \vec{e}_{1} \cdot \vec{e}_{2}  ~, ~~~
\eta_{1} = \vec{e}_{1} \cdot \vec{e}_{3}  ~,
\nonumber \\
\sum_{i=1}^{3} \xi_{i} \eta_{i} &=& \vec{e}_{2} \cdot \vec{e}_{3}  ~.
\end{eqnarray}

The important property is that both sets of unit vectors
$\left \{ \vec{l}_{i} ; i = 1, 2, 3 \right \}$ and
$\left \{ \vec{e}_{i} ; i = 1, 2, 3 \right \}$ are defined in the same frame
frame of reference $S$. As for transformation defined by Eq. (C2), the sets
of unit vectors $\left \{ \vec{e}_{i} ; i = 1, 2, 3 \right \}$ and
$\left \{ \vec{k}_{i} ; i = 1, 2, 3 \right \}$ are also defined in the
system $S$. But what about translational velocity terms $+ ~ \vec{v} / c$
in Eq. (C2)? Any velocity term $+ ~ \vec{v} / c$ in
transformation corresponds to transformation from one (local) inertial
frame of reference to another one. Thus, the conclusion is evident:
Eq. (C2) is of no physical sense.


\begin{thebibliography}{}
\bibitem{}Balek V., Kla\v{c}ka J., 2002. Perihelion motion due to the
Poynting-Robertson effect. http://xxx.lanl.gov/abs/astro-ph/0207496
(complete and different derivation: to be submitted to
{\it General Relativity and Gravitation})
\bibitem{}Breiter S., Jackson A. A., 1998. Unified analytical solutions to
two-body problems with drag.
{\it Mon. Not. R. Astron. Soc.} {\bf 299}, 237-243.
\bibitem{}Brouwer D., Clemence G. M., 1961. Celestial Mechanics.
Academic Press, New York and London, 598 pp.
\bibitem{}Burns J. A., Lamy P. L., Soter S., 1979. Radiation forces on small
particles in the Solar System. {\it Icarus} {\bf 40}, 1-48.
\bibitem{}Einstein A., 1905. Zur Elektrodynamik der bewegter K\H{o}rper.
{\it Annalen der Physik} {\bf 17}, 891-920.
\bibitem{}Gajdo\v{s}\'{\i}k M., Kla\v{c}ka J., 1999. Cometary dust trails
and ejection velocities. In: Evolution and Source Regions of Asteroids and
Comets, Proc. IAU Coll. 173, J. Svore\v{n}, E. M. Pittich, and H. Rickman (eds.),
Astron. Inst. Slovak Acad. Sci., Tatransk\'{a} Lomnica, pp. 239-242.
\bibitem{}Harwit M., 1963. Origins of the Zodiacal dust cloud.
{\it J. Geophys. Res.} {\bf 86}, 2171-2180.
\bibitem{}Harwit M., 1988. Astrophysical Concepts, 2-nd edition.
Springer-Verlag, New York-Berlin-Heidelberg.
\bibitem{}Kimura H., Okamoto H., Mukai T., 2002. Radiation Pressure
and the Poynting-Robertson Effect for Fluffy Dust Particles. {\it
Icarus} {\bf 157}, 349-361.
\bibitem{}Kla\v{c}ka J., 1992a. Poynting-Robertson effect. I. Equation of
motion. {\it Earth, Moon, and Planets} {\bf 59}, 41-59.
\bibitem{}Kla\v{c}ka J., 1992b. Poynting-Robertson effect. II. Perturbation
equations. {\it Earth, Moon, and Planets} {\bf 59}, 211-218.
\bibitem{}Kla\v{c}ka J., 1994a. Interplanetary dust particles and solar
radiation. {\it Earth, Moon, and Planets} {\bf 64}, 125-132.
\bibitem{}Kla\v{c}ka J., 1994b. Radial forces and orbital elements. In:
{\it Dynamics and Astrometry of Natural and Artificial Celestial Bodies},
K. Kurzynska, F. Barlier, P. K. Seidelmann and	I. Wytrzyszczak (eds.),
Astronomical Observatory of A. Mickiewicz University, Poznan, pp. 181-185.
\bibitem{}Kla\v{c}ka J., 2000a. Electromagnetic radiation and motion of real
particle. \\
http://xxx.lanl.gov/abs/astro-ph/0008510
\bibitem{}Kla\v{c}ka J., 2000b. Aberration of light and motion of real
particle. \\
http://xxx.lanl.gov/abs/astro-ph/0009108
\bibitem{}Kla\v{c}ka J., 2000c. Solar radiation and asteroidal motion,
http://xxx.lanl.gov/abs/astro-ph/0009109
\bibitem{}Kla\v{c}ka J., 2001. On the Poynting-Robertson effect and
analytical solutions. In: Dynamics of Natural and Artificial Celestial Bodies,
H. Pretka-Ziomek, E. Wnuk, P. K. Seidelmann and D. Richardson
(eds.), Kluwer Academic Publishers, Dordrecht, pp. 215-217.
(astro-ph/0004181)
\bibitem{}Kla\v{c}ka J., Kaufmannov\'{a} J., 1992. Poynting-Robertson
effect: `circular' orbit. {\it Earth, Moon, and Planets} {\bf 59}, 97-102.
\bibitem{}Kla\v{c}ka J., Kaufmannov\'{a} J., 1993. Poynting-Robertson effect
and small eccentric orbits. {\it Earth, Moon, and Planets} {\bf 63}, 271-274.
\bibitem{}Kla\v{c}ka J., Kocifaj M. 1994. Electromagnetic radiation and
equation of motion for a dust particle.
In: Dynamics and Astrometry of Natural and
Artificial Celestial Bodies, K. Kurzy\'{n}ska, F. Barlier, P. K. Seidelmann
and I. Wytrzyszczak (eds.), Astronomical Observatory of A. Mickiewicz
University, Pozna\'{n}, Poland, 187-190.
\bibitem{}Kla\v{c}ka J., Kocifaj M., 2001. Motion of nonspherical dust
particle under the action of electromagnetic radiation,
{\it J. Quant. Spectrosc. Radiat. Transfer} {\bf 70/4-6}, 595-610.
\bibitem{}Kla\v{c}ka J., Saniga M., 1993. Interplanetary dust particles and
solar wind. {\it Earth, Moon, and Planets} {\bf 60}, 23-29.
\bibitem{}Kocifaj M., Kla\v{c}ka J., Kundrac\'{\i}k F., 2000.
Motion of realistically shaped cosmic dust particle in Solar System.
In: {\it Light Scattering by Nonspherical Particles: Halifax Contributions},
G. Videen, Q. Fu, and P. Ch\'{y}lek (eds.), Army Research Laboratory,
Adelphi Maryland, pp. 257-261.
\bibitem{}Landau L. D., Lifshitz E. M., 1975. The Classical Theory of Fields,
4th ed., Pergamon Press, New York (In Russian: Teoria polia. Nauka, Moscow,
1973).
\bibitem{}Leinert Ch., Gr\H{u}n E., 1990. Interplanetary dust. In: {\it
Physics of the Inner Heliosphere I}, R. Schwenn and E. Marsch (eds.),
Springer-Verlag, Berlin Heidelberg, pp. 207-275.
\bibitem{}Lyttleton R. A., 1976. Effects of solar radiation on the orbits of
small particles. {\it Astrophysics and Space Science} {\bf 44}, 119-140.
\bibitem{}Mediavilla E., Buitrago J., 1989. The Poynting-Robertson effect and
its generalization to the case of an extended source in rotation.
{\it Eur. J. Phys.} {\bf 10}, 127-132.
\bibitem{}Mignard F., 1992. On the radiation forces. In: {\it Interrelations
Between Physics and Dynamics for Minor Bodies in the Solar System}, D. Benest
and C. Froeschle (eds.), Fronti\`{e}res, pp. 419-451.
\bibitem{}Mishchenko M., 2001. Radiation force caused by scattering, absorption,
and emission of light by nonspherical particles.
{\it J. Quant. Spectrosc. Radiat. Transfer} {\bf 70}, 811-816.
\bibitem{}Mishchenko M., Travis L. D., Lacis A. A., 2002. Scattering, Absorption
and Emission of Light by Small Particles. Cambridge University Press,
Cambridge (UK), 445 pp.
\bibitem{}Poynting J. M., 1904. Radiation in the Solar System: Its effect on
temperature and its pressure on small bodies. {\it Philosophical Transactions
of the Royal Society of London} {\bf Series A 202}, 525-552.
\bibitem{}Robertson H. P., 1937. Dynamical effects of radiation in the Solar
System. {\it Mon. Not. R. Astron. Soc.} {\bf 97}, 423-438.
\bibitem{}Robertson H. P., Noonan T. W., 1968. Relativity and Cosmology.
Saunders, Philadelphia.
\bibitem{}Srikanth, 1999. Physical interpretation of the Poynting-Robertson
effect. {\it Icarus} {\bf 140}, 231-234.
\bibitem{}Stix M., 2002. The Sun. Springer-Verlag, Berlin Heidelberg, 490 pp.
\bibitem{}Williams I. P., 2002. The effect of radiation on the motion of
meteoroids. In: {\it Optics of Cosmic Dust}, G. Videen and
M. Kocifaj (eds.), Kluwer Academic Publishers, Dordrecht, pp. 283-300.
\bibitem{}Wyatt S. P., Whipple F. L., 1950. The Poynting-Robertson effect on
meteor orbits. {\it Astrophys. J.} {\bf 111}, 558-565.
\end{thebibliography}
\end{document}